\documentclass[aps,pra,amsmath,amssymb,reprint,floatfix,superscriptaddress]{revtex4-1}

\usepackage[utf8]{inputenc}
\usepackage[T1]{fontenc}
\usepackage[english]{babel}
\usepackage[dvipsnames,svgnames,x11names]{xcolor}
\usepackage{mathtools,dsfont,braket}
\usepackage{tensor}
\usepackage{nicefrac}
\usepackage[version=4]{mhchem}
\usepackage{siunitx}

\usepackage[caption=false,labelfont=bf,position=top]{subfig}
\usepackage{booktabs}
\usepackage{multirow}
\usepackage{scrextend}
\usepackage{graphicx}
\usepackage{dcolumn}
\usepackage{bm}
\usepackage{microtype}
\usepackage{listings,xcolor}
\usepackage{tikz}
\usepackage{tikz-3dplot}
\usepackage{tikz-cd}
\usetikzlibrary{
  arrows,
  arrows.meta,
  calc,
  fit,
  patterns,
  matrix,
  shapes.arrows,
  decorations.pathmorphing,
  decorations.pathreplacing,
  decorations.markings,
  intersections,
}
\usepackage{lipsum}
\usepackage{xr}
\newcommand*\alignmultr[2][]{%
\begin{tabular}{@{}S[#1]@{}}
  #2
\end{tabular}}

\definecolor{uniddblue}{RGB}{0,65,145}   
\definecolor{uninred}{RGB}{255,0,0}      
\definecolor{unigray}{RGB}{22,22,24}     
\definecolor{unidblue}{RGB}{0,143,255}   
\definecolor{unibblue}{RGB}{176,173,5}   
\definecolor{unired}{RGB}{238,77,46}     

\begin{document}

\preprint{Version1}

\title{Towards photoassociation processes of ultracold rubidium trimers} 

\author{Jan Schnabel}
\email[]{schnabel@theochem.uni-stuttgart.de}
\affiliation{Institute for Theoretical Chemistry and Center for Integrated Quantum Science and Technology
  (IQ\textsuperscript{ST}), University of Stuttgart,
  70569 Stuttgart, Germany}

\author{Tobias Kampschulte}
\author{Simon Rupp}
\author{Johannes {Hecker Denschlag}}
\affiliation{Institut f{\"u}r Quantenmaterie and Center for Integrated Quantum Science and Technology
  (IQ\textsuperscript{ST}), Universit{\"a}t Ulm, 89069 Ulm, Germany}

\author{Andreas K{\"o}hn}
\email[]{koehn@theochem.uni-stuttgart.de}
\affiliation{Institute for Theoretical Chemistry and Center for Integrated Quantum Science and Technology
  (IQ\textsuperscript{ST}), University of Stuttgart,
  70569 Stuttgart, Germany}

\date{\today}

\begin{abstract}
We theoretically investigate the prospects for photoassociation (PA) of
\ce{Rb3}, in particular at close range. We provide an overview of accessible
states and possible transitions. The major focus is placed on the calculation of
equilibrium structures, the survey of spin-orbit effects and the investigation of
transition dipole moments. Furthermore we discuss Franck-Condon overlaps and
special aspects of trimers including the (pseudo) Jahn-Teller effect and the
resulting topology of adiabatic potential-energy surfaces. With this we
identify concrete and suitable PA transitions to potentially produce long-lived trimer
bound states. Calculations are performed using the multireference
configuration-interaction method together with a large-core effective
core potential and a core-polarization potential with a large
uncontracted even-tempered basis set.
\end{abstract}


\maketitle 

\section{Introduction}

Ultracold molecules offer great opportunities for research and applications,
as they can be prepared in precisely defined quantum states
\cite{Quemener2012,Bohn2017,Balakrishnan2016,Krems2008,Carr2009,Doyle2004}. Besides
studying the molecular properties with high precision, collisions and chemical
reactions can then be investigated in the quantum regime where only a single
partial wave contributes. Furthermore, cold molecules have a number of
applications, ranging from metrology to quantum sensors, to quantum simulation
and computation~\cite{Bohn2017,Carr2009}. In recent years a number of ways to
produce cold molecules have been developed ranging from buffer gas cooling,
slowing and filtering, laser cooling, to associating ultracold atoms. The
coolest temperatures and the highest control in preparing the molecular
quantum state have been typically achieved by associating ultracold
atoms~\cite{Quemener2012,Balakrishnan2016,Krems2008}. In this way a variety of
different ultracold diatomic molecules has been produced, typically consisting
of alkali-atoms, such as \ce{Li2}, \ce{Na2}, \ce{K2}, \ce{Rb2}, \ce{Cs2},
\ce{NaRb}, \ce{RbCs}, \ce{RbK}, \ce{NaK}, \ce{LiNa}, \ce{LiK}, \ce{LiRb},
\ce{LiCs}, \ce{NaCs}, but there are also other compounds, such as \ce{LiYb},
\ce{RbYb}, see,
e.g. Refs.~\cite{Bohn2017,Balakrishnan2016,Krems2008,Doyle2004} and references
therein. Possible methods for the molecule production are e.g., three-body
recombination \cite{Burt1997,Stamper1998,Greene2017}, photoassociation
\cite{Ulmanis2012,Jones2006}, and sweeping over a Feshbach resonance
\cite{Koehler2006,Chin2010}.

Alkali metal dimer systems have also been studied theoretically in great detail.
Accurate potential energy curves (PECs), dipole moments and spin-orbit
interactions can be obtained via several \emph{ab initio}
methods~\cite{Dulieu2009}. Among others, the Fourier Grid Hamiltonian
method~\cite{FGH1,FGH2} or the discrete variable representation (DVR)
method~\cite{DVR} were used to analyze the level structure of the well-known
coupled
$\mathrm{A}\,\tensor*[^1]{\Sigma}{_u^+}$--$\mathrm{b}\,\tensor*[^3]{\Pi}{_u}$
manifold in homonuclear alkali metal dimers.

Producing and understanding ultracold alkali trimers (i.e., e.g.:
$\mathrm{X}_2\mathrm{Y}$, $\mathrm{X}_3$, etc. with
$\mathrm{X},\mathrm{Y}\in\lbrace\ce{Li}, \ce{Na}, \ce{K}, \ce{Rb},
\ce{Cs}\rbrace$) clearly is a next milestone. Alkali trimers are much more
complex and challenging as compared to alkali dimers, both from the
theoretical and the experimental point of view.  One aspect of the
complexity of an alkali trimer is that many of its levels are prone to quick
decay due to fast internal relaxation and dissociation mechanisms. This makes
it challenging to prepare and manipulate the trimer on the quantum
level. Indeed, detailed and highly resolved spectroscopy on free trimer
molecules is generally still lacking. Ultracold trimers have not been
produced yet, apart from the extremely weakly-bound Efimov states
\cite{Greene2017,Ferlaino2011}, which are fast-decaying three-body states of
resonantly interacting atoms.  Alkali trimers at mK-temperatures, however,
have been produced in experiments using supersonic beam expansion of \ce{Ar}
seeded with, e.g. sodium atoms, as in
Refs.~\cite{Na3ArBeam1986,ErnstArBeam1995,ErnstArBeam1997}, or in experiments
with alkali metal clusters formed on helium
droplets~\cite{NaglTrimesHeDroplets,AuboeckRb3K3QuartetsHeDroplets,NaglHeteroHomoHighSpinTrimersHeDroplets,HauserPCCP1,HauserPCCP2}.
Theoretical interest in alkali metal clusters already goes back to the 80s and
90s, with a number of pioneering
works~\cite{Davidson1978,Martins1983,Thompson1985Li3Na3K3,Cocchini1988Na3,Spiegelmann1988NanKn,Meiswinkel1990,TheoreticalStudAbsSpectNa3B1997}
giving insights into the electronic properties of alkali trimers, the
corresponding ground state potential energy surfaces (PESs) and the occurring
Jahn-Teller (JT) effect. 
Yet, these studies were restricted to light alkali metal species,
i.e. \ce{Li}, \ce{Na} and \ce{K}. Later, following the success of the helium
droplet method, theoretical investigations of alkali trimer systems were
reappearing -- now also containing heavier elements such as
\ce{Rb}~\cite{HauserPotassiumDoublets,HauserJTK3Rb3,HauserJTK3Rb3Doublets,HauserJTNa3,HauserBook2009,HauserPhD,SoldanPESRb3}.
In these works, the main focus was on selected JT states and the reproduction
of special transitions and spectra measured with \ce{He}-droplet spectroscopy.
In recent years, the advent of experiments studying ultracold collisions
between an alkali atom and an alkali dimer also triggered further calculations
of ground state alkali trimer PESs. For an overview see,
e.g. \cite{Quemener2012}.

\begin{figure}[tb]
  \includegraphics[width=.7\columnwidth]{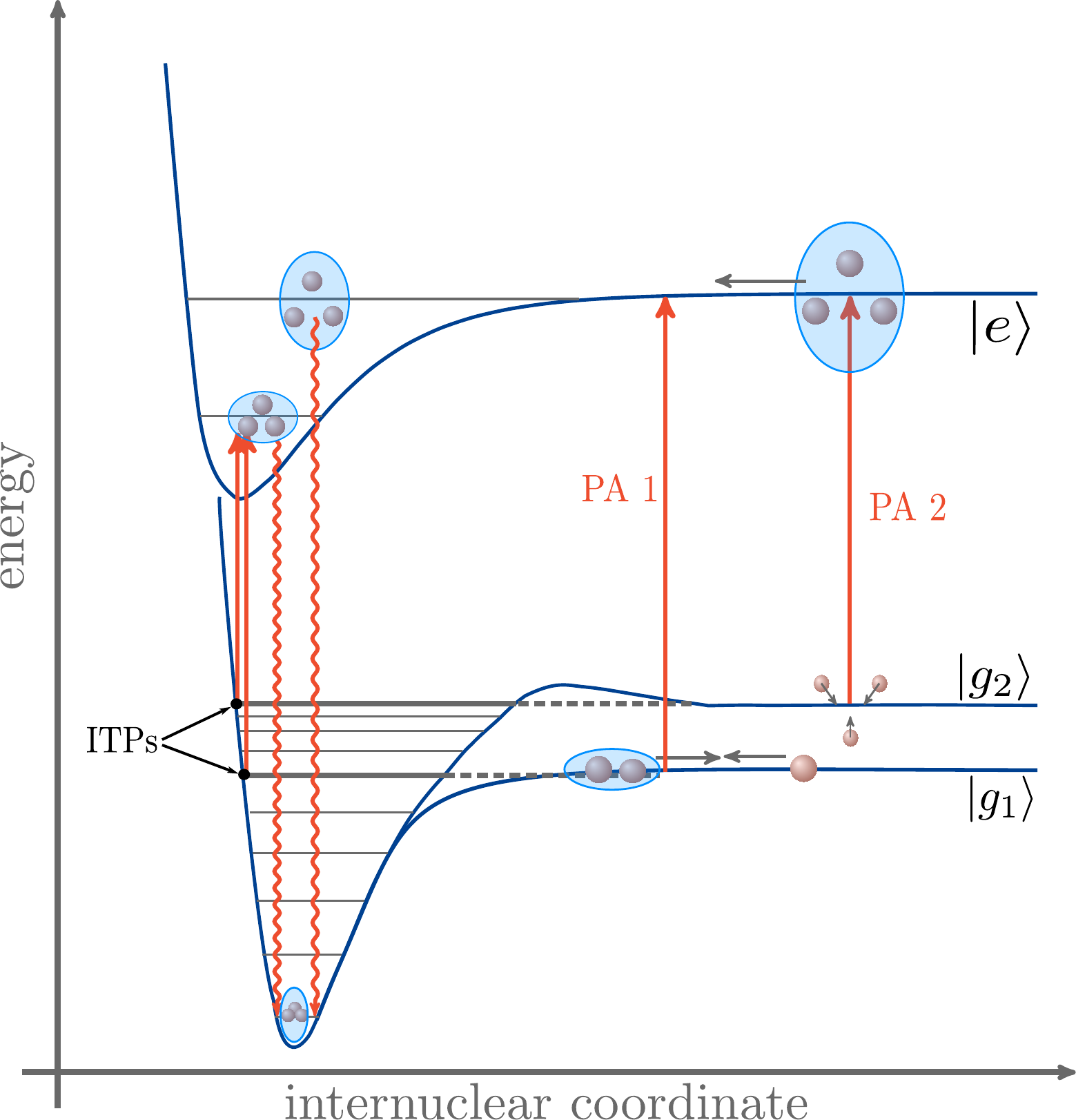}
  \caption{\label{fig:photoassociation-scheme}(Color online) Strongly simplified illustration
    of the two different photoassociation (PA) schemes for the production of
    \ce{Rb3} species. PA 1 photoassociates a \ce{Rb2} dimer with a free
    \ce{Rb} atom starting from the asymptote $\ket{g_1}$. PA 2 photoassociates
    three \ce{Rb} atoms starting from the asymptote $\ket{g_2}$. Both PA
    processes can be also realized (in principle) starting from an inner
    turning point (ITP), as discussed in Sec.~\ref{subsec:ConfigSpace} and
    indicated on the very left. In both cases the excited trimer (more weakly
    bound for PA starting from $\ket{g_1}$ or $\ket{g_2}$) can radiatively
    decay to the ground state.}
\end{figure}

A promising approach for preparing isolated trimer molecules in precisely
defined quantum states is photoassociation (PA) which so far has been used for
creating dimers in the ultracold regime \cite{Ulmanis2012,Jones2006}.
In a PA process a colliding atom pair in the electronic ground state and a
laser photon is transferred into a well-defined bound, electronically excited
state $\ket{e}$~\cite{Jones2006}. From there, the excited molecule can
spontaneously decay into a number of ro-vibrational, long-lived levels of the
molecular ground state manifold $\ket{g}$. In analogy, one can in principle
think of two possible PA schemes for the production of trimer molecules,
cf. Fig.~\ref{fig:photoassociation-scheme}:
\begin{itemize}
  \item[(i)] A dimer molecule and a free ground-state atom are
    photoassociated ($\equiv$ PA 1). This is shown in
    Fig.~\ref{fig:photoassociation-scheme}. The laser photon drives a
    transition from the asymptote $\ket{g_1}$ to an electronically excited
    bound state $\ket{e}$ of the trimer complex. From there it can
    spontaneously relax to the ground state.
  \item[(ii)] Three colliding free atoms are photoassociated ($\equiv$ PA 2).
    As shown in Fig.~\ref{fig:photoassociation-scheme}, the photon drives the
    transition now from the asymptote $\ket{g_2}$ to the excited trimer state.
\end{itemize}
The PA can in principle take place at long range (large internuclear
distances) or at short range (small internuclear distances).
Photoassociation at large distances was recently discussed theoretically in
\cite{Rios2015}. Here, we therefore rather focus on trimer photoassociation at
short distances. Recent theoretical work in
Ref.~\cite{PhysRevLett.108.263001}, however, suggests that the simultaneous
collision of three atoms is strongly suppressed due to an effective repulsive
barrier in the short-range of the three-body potential, rendering the
realization of PA2 at short range less likely. For PA1, however, such a
restriction is not expected. Working out concrete schemes for trimer PA
requires detailed knowledge of the involved trimer states and the optical
transitions between them. With the present work we provide a broad overview of
states in terms of energy levels and the topology of potential energy surfaces
(PESs). Previous theoretical studies on alkali
trimers~\cite{HauserPCCP1,HauserPCCP2,Davidson1978,Martins1983,Thompson1985Li3Na3K3,Cocchini1988Na3,Spiegelmann1988NanKn,Meiswinkel1990,TheoreticalStudAbsSpectNa3B1997,Mukherjee2014,HauserPotassiumDoublets,HauserJTK3Rb3,HauserJTK3Rb3Doublets,HauserJTNa3,HauserBook2009,HauserPhD}
were essentially restricted to either the doublet or quartet ground state or
they investigated selected JT distorted excited states. Furthermore, we
calculate the electronic dipole transition matrix elements between states. We
discuss special aspects of trimers including different coordinate systems, the
(pseudo) Jahn-Teller effect, the Renner-Teller effect for linear
configurations, as well as accidental degeneracies. Finally, we suggest
specific PA transitions and investigate coupling effects in terms of
spin-orbit interaction. Our work is intended as a basis for further detailed
investigations of PA, which at the next stage will require the simulation of
nuclear dynamics.

This work is organized as follows. Section~\ref{sec:GeneralOverview} briefly
introduces the computational aspects and convenient coordinate systems for
trimers. Hereafter we discuss major topological features of the corresponding
PESs by means of special cuts and comment on the (pseudo) Jahn-Teller (and
Renner-Teller) effect. Here, we additionally provide an overview of the
expected quartet and doublet equilibrium states of the trimer system within a
certain energy range and comment on spin-orbit coupling effects and estimate
their magnitude. In Sec.~\ref{sec:PAStatesIdentify} we analyze the excited
electronic states with regard to their applicability in PA processes. We find
that they can be reached conveniently via the inner-turning points on the
quartet ground state PES. We identify one component of the
$1{\,}\tensor*[^{4}]{\mathrm{E}}{^{\prime\prime}}$ Jahn-Teller pair as a
promising candidate for PA experiments. We thoroughly investigate its
suitability as a target state by studying electronic transition dipole
strengths with the quartet ground state, spin-orbit coupling and further
mixing effects with other states in its close proximity, as well as its
distance from conical intersections (COINs). Finally, we summarize the main points of
this work in Sec.~\ref{sec:Summary} and give an outlook to ongoing work.

\section{\label{sec:GeneralOverview}General Overview of the Rubidium Trimer System}
\subsection{\label{subsec:CompAspects}Computational Aspects}
\begin{table*}[tb]
  \centering
  \caption{\label{tab:Rb2Data}Comparison of experimental (references given in
    square brackets) and calculated values for some spectroscopic constants of
    a few \ce{Rb2} states. $D_e$ is the dissociation energy, $R_e$ the
    equilibrium distance, and $T_e$ the electronic term energy. Calculations
    are performed at MRCI(ECP+CPP)/15s12p7d5f3g level of theory. We also
    report differences $\Delta$ between theory and experiment as well as the
    mean difference $\bar{\Delta}$ and the absolute mean difference
    $\bar{\Delta}_{\text{abs}}$ for the given set of states.}
  \begin{tabular}{l @{\hskip 10pt}S @{\hskip 10pt}S @{\hskip 10pt}S @{\hskip
        8pt}S @{\hskip 10pt}S @{\hskip 10pt}S @{\hskip 10pt}S @{\hskip 4pt}S @{\hskip 4pt}S}
    \toprule\toprule
    \multirow{2}{*}{State} & 
    \multicolumn{3}{c}{$D_e\,[\mathrm{cm}^{-1}]$} &
    \multicolumn{3}{c}{$R_e\,[\si{\angstrom}]$} &
    \multicolumn{3}{c}{$T_e\,[\mathrm{cm}^{-1}]$} \\
    \cmidrule(r){2-4} \cmidrule(r){5-7} \cmidrule(r){8-10} 
    & \multicolumn{1}{c}{\small this work} & \multicolumn{1}{c}{\small exp.} &
    \multicolumn{1}{c}{\small $\Delta$} &
    \multicolumn{1}{c}{\small this work} & \multicolumn{1}{c}{\small exp.} &
    \multicolumn{1}{c}{\small $\Delta$} & 
    \multicolumn{1}{c}{\small this work} & \multicolumn{1}{c}{\small exp.} &
    \multicolumn{1}{c}{\small $\Delta$} \\
    \midrule
    {$\mathrm{X}{\,}\tensor*[^{1}]{\Sigma}{_g}\;$~\cite{Strauss2010}} & 4116 & 3993.593 & 122 & 4.1689
    & 4.2099 & -0.0410 & 0 & 0 & 0 \\
    {$\mathrm{a}{\,}\tensor*[^{3}]{\Sigma}{_u}\;$~\cite{Strauss2010}} & 250 & 241.503 & 8 & 6.0065 &
    6.0940 & -0.0875 & 3866 & {--} & {--} \\
    {$\mathrm{b}{\,}\tensor*[^{3}]{\Pi}{_u}\;$ ~\cite{Salami2009}} & 7218 & 7039 & 179 & 4.1537 &
    4.1329 & 0.0208 & 9632 & 9601 & 31 \\
    {$\mathrm{A}{\,}\tensor*[^{1}]{\Sigma}{_u}\;$~\cite{Salami2009}} & 6071 & 5981 & 90 & 4.8637 &
    4.8737 & -0.0100 & 10778 & 10750 & 28 \\
    {$(2){\,}\tensor*[^{1}]{\Sigma}{_g}\;$~\cite{Drews2017,Amiot1987}} & 3140 & 2963 & 177 & 5.4081 & 5.4399 &
    -0.0318 & 13709 & 13602 & 107 \\
    {$(1){\,}\tensor*[^{1}]{\Pi}{_u}\;$~\cite{Amiot1990}} & 2150 & 1907 & 243 & 4.5203 & {--} & {--} & 
    14700 & 14666 & 34 \\
    {$(1){\,}\tensor*[^{1}]{\Pi}{_g}\;$~\cite{Amiot1986}} & 1246 & 1290 & -44 & 5.4225 & 5.4188 & 0.0037 &
    15604 & 15510 & 94 \\
    \midrule
    ~ & ~ & $\bar{\Delta}:$ & 111 & ~ & ~ & -0.0243 & ~ & ~ &
    59 \\
    ~ & ~ & $\bar{\Delta}_{\text{abs}}:$ & 123 & ~ & ~ & 0.0325 & ~ & ~ & 59\\
    \bottomrule\bottomrule
  \end{tabular}
\end{table*}
Since investigating PA processes of \ce{Rb3} requires an extensive survey of a
large number of expected states and transitions in the \ce{Rb3} system, a
pragmatic but considerably accurate computational approach has to be
applied. In this work, we are using a large-core effective core potential
(ECP) in combination with a core polarization potential (CPP) as it has been
developed in Ref.~\cite{LargeCoreECP} with a large [15s12p7d5f3g]
(uncontracted and even-tempered) basis set (UET15) -- see supplementary
material~\cite{Supplementary} for details. In doing so merely the valence electron of \ce{Rb} is
treated explicitly while the remaining 36 electrons are described by the
ECP. The CPP accounts for dynamic polarization of the core electrons by the
valence electrons. All doublet and quartet states of \ce{Rb3} within a certain
energy range were computed using the internally contracted multireference
configuration interaction (MRCI) method~\cite{MRCI1,MRCI2,MRCI3,MRCI4,MRCI5}.
As we are only dealing with an effective three-electron system, the MRCI
method has no problem with the separability of the wavefunction. This means
that the PESs are entirely well defined and show correct dissociation behavior
into three non-interacting \ce{Rb} atoms. All calculations are performed using
the \textsc{molpro} 2018.2 program package~\cite{MOLPRO_brief}.

The pragmatic ECP$+$CPP approximation is sufficient for gaining a reliable
understanding of the physics of the system, as shown in the following, while
saving tremendously on computational costs. By construction, the ECP
reproduces the experimentally determined atomic energy levels up to the
${}^2\mathrm{F}$ state~\cite{LargeCoreECP}. The results in
Tab.~\ref{tab:Rb2Data} illustrate the expected accuracy for molecular systems
-- here in terms of benchmark calculations for spectroscopic constants of
selected singlet and triplet states of \ce{Rb2} in comparison to experimental
results. The calculations do not account for spin orbit coupling effects,
which are only rather small perturbations in most cases. This we will also
show in the present study for \ce{Rb3}. The mean differences $\bar{\Delta}$
reported in Tab.~\ref{tab:Rb2Data} show a systematic overestimation of the
binding energies by $100$ to $250\,\mathrm{cm}^{-1}$, while equilibrium
distances are typically underestimated by $0.01$ to
$0.04\,\si{\angstrom}$. This over- and underestimation is a well-known bias
introduced by the large-core ECPs due to the approximate description of the
repulsion of the core
electrons~\cite{CoreCoreRepulsionDiatomics,SoldanCoreRepulsion}. For the
electronic term energies $T_e$, the errors are on the order of $30$ to
$100\,\mathrm{cm}^{-1}$.  Since the \ce{Rb3} system forms three Rb-Rb bonds,
we can estimate the accuracy of our \emph{ab initio} method from the above
mean errors by $\approx\pm 300\,\mathrm{cm}^{-1}$. For bond lengths the same
accuracy as for \ce{Rb2} is expected (about 1 per cent of the total predicted
distance).  According to the above experience, binding energies are probably
mostly overestimated while bond lengths are underestimated.  While these
deviations seem large from a spectroscopist's point of view, we note that
these deviations are already within the regime of accurate quantum chemical
methods, typically defined by the 'chemical accuracy' level of $\approx\pm
1\,\mathrm{kcal}/\mathrm{mol}\approx\pm 350\,\mathrm{cm}^{-1}$ for
energies. Increasing this accuracy is possible but requires steeply increasing
computational resources, while our present approach only requires
approximately 40 minutes on 8 cores for solving for 27 electronic states at a
given \ce{Rb3} geometry, thus allowing to explore the configurational space
efficiently.

\subsection{\label{subsec:Coords}Coordinates}
\begin{figure*}[tb]
  \centering
  \begin{minipage}[t]{.3267\textwidth}
    \subfloat[]{\includegraphics[width=.64\linewidth]{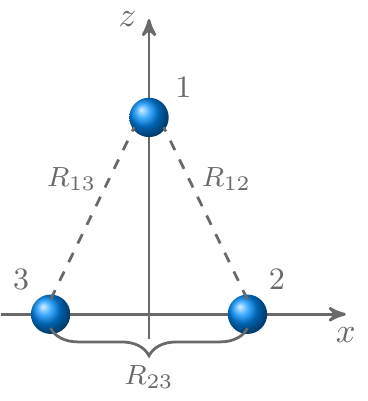}}
  \end{minipage}
  \hspace*{-1cm}
  \begin{minipage}[t]{.3267\textwidth}
    \subfloat[]{\includegraphics[width=.66\linewidth]{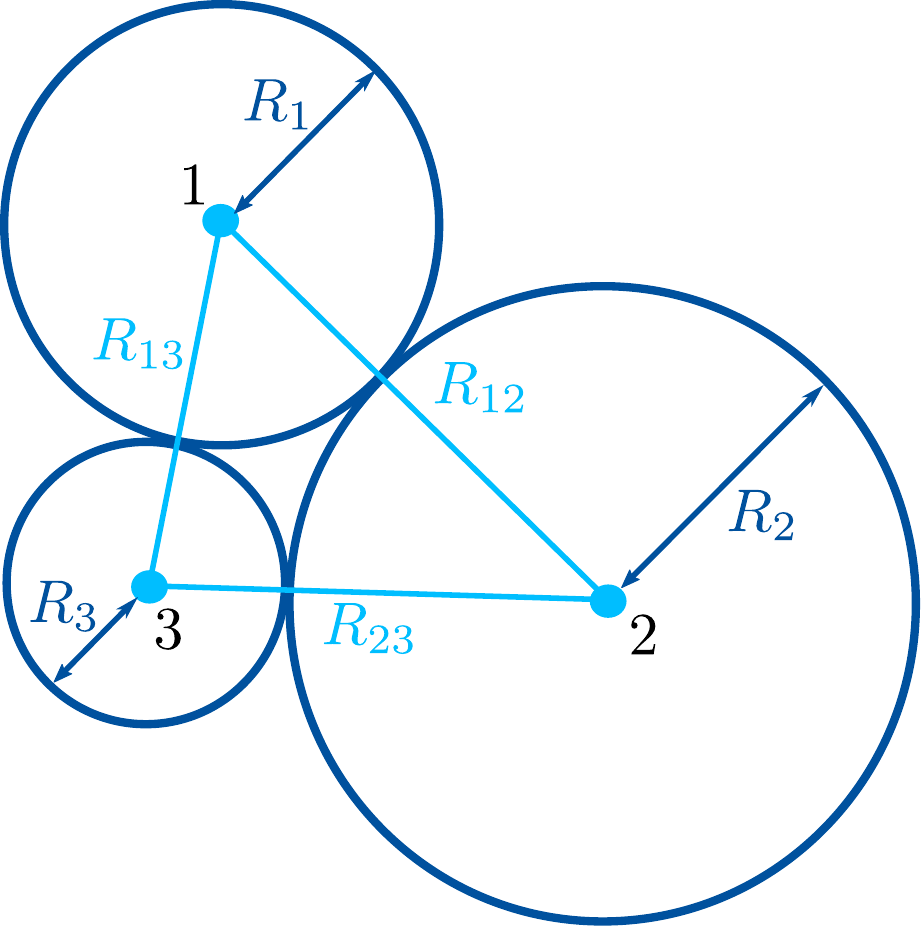}}
  \end{minipage}
  \begin{minipage}[t]{.3267\textwidth}
    \subfloat[]{\includegraphics[width=\linewidth]{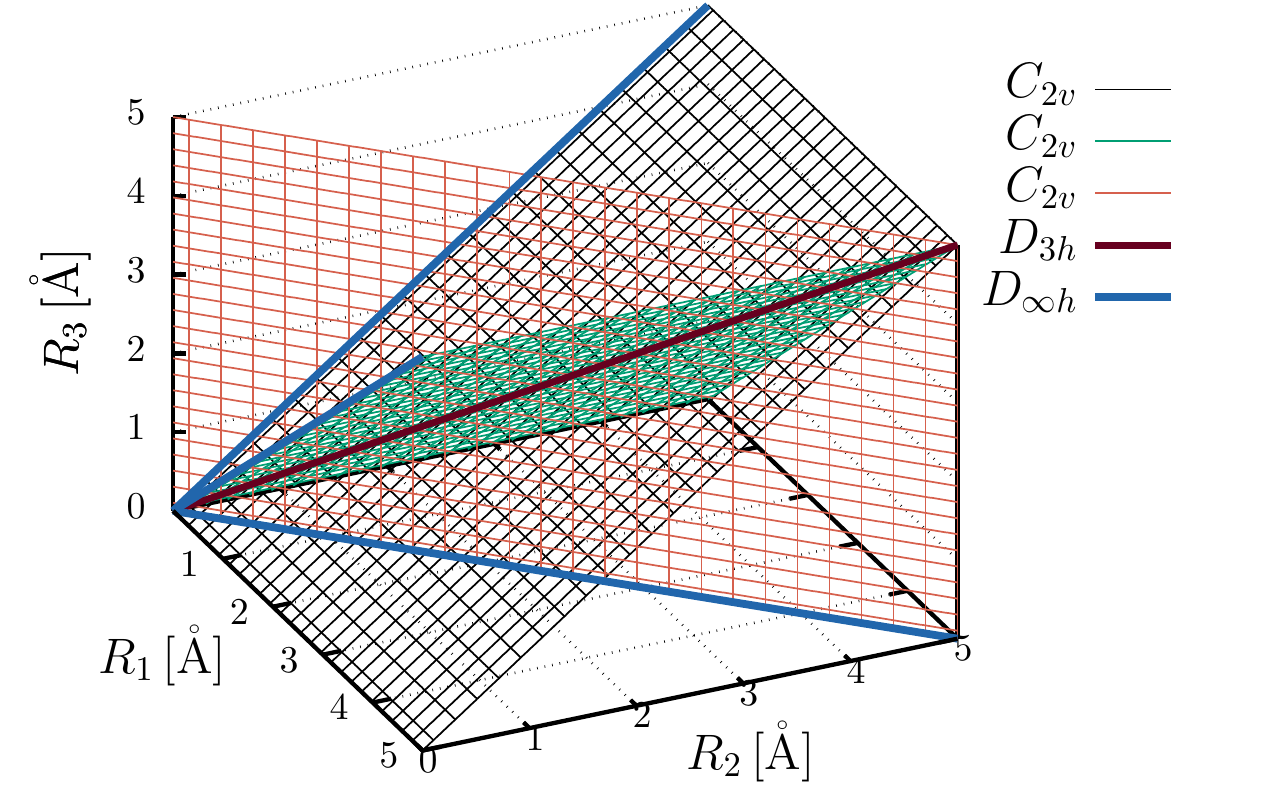}}
  \end{minipage}
  \caption{\label{fig:coords}(Color online) \textbf{(a)} Sketch of the \ce{Rb3} system in
    the $xz$ plane with \emph{internuclear distances}
    $R_{12},R_{23},R_{13}$. \textbf{(b)} Illustration of the geometric
    interpretation of the \emph{perimetric coordinates} for triatomic
    molecules. They represent the radii of three mutual tangent circles
    centered on the nuclei. \textbf{(c)} Division of the positive octant in perimetric
    coordinates to show special configuration subspaces of triatomic systems
    (i.e. $D_{\infty h}$, $D_{3h}$ and $C_{2v}$).}
\end{figure*}
The atoms of non-linear triatomic molecular systems always define a plane,
which we choose, without loss of generality, as the $xz$ plane, see
Fig.~\ref{fig:coords}~(a).  The system has three internal degrees of freedom,
with the only exception of linear geometries for which the system has a fourth
degree of freedom. There are many coordinate systems available for properly
studying the physics of the system -- like the well-known Jacobi and
hyperspherical coordinates, see, e.g. Ref.~\cite{PhysRevA.65.042725} and
references therein. In general every coordinate system has its strengths and
weaknesses and the choice strongly depends on what one wants to analyze. In
this work we are making use of three different coordinate systems which are
introduced in the following.

It is straightforward to use \emph{internuclear distances} as shown in
Fig.~\ref{fig:coords}~(a). However, not every triple of numbers
$(R_{12},R_{23},R_{13})$ obeys the triangular condition and defines a possible
molecular configuration. It is convenient to employ \emph{perimetric
  coordinates}~\cite{PhysRev.51.855,PhysRev.112.1649,PhysRev.115.1216,PhysRev.126.1057,PhysRev.127.509,PhysRevA.4.516,PhysRevA.4.885},
as used by Davidson in his analysis of \ce{H3}~\cite{Davidson1977}. Given the
set of internal coordinates $\lbrace R_{12},R_{23},R_{13}\rbrace$, the
perimetric coordinates can be expressed as
\begin{subequations}
  \label{eq:PekerisCoordinates}
  \begin{align}
    R_1 &= \frac{1}{2}\left(R_{12} + R_{13} - R_{23}\right) \\ 
    R_2 &= \frac{1}{2}\left(R_{12} + R_{23} - R_{13}\right) \\ 
    R_3 &= \frac{1}{2}\left(R_{13} + R_{23} - R_{12}\right)\,.
  \end{align}
\end{subequations}
The perimetric coordinates are the radii of mutually tangent circles centered on each
nucleus (as shown in Fig.~\ref{fig:coords}~(b)). The general
topology of this coordinate system reveals some properties:
\begin{enumerate}
  \item[a)] Every triple of numbers ($R_1,R_2,R_3$) in the positive octant
    (cf. Fig.~\ref{fig:coords}~(c)) gives a unique molecular conformation
    (modulo permutational inversion); i.e. the coordinates satisfy the
    triangular inequality
  \item[b)] Internuclear distances are given as the sum of the corresponding
    perimetric coordinates (e.g. $R_{12}=R_1+R_2$)
  \item[c)] Linear molecules are found at the three equivalent boundary planes
    of the positive octant, where one of the perimetric coordinates is zero
    (e.g. for $R_1=0$, $R_2=R_{12}$ and $R_3=R_{13}$).
  \item[d)] The dissociation limits (atom $+$ dimer) are obtained by one of
    the coordinates being large
    (e.g. $\mathrm{atom}_1+\mathrm{dimer}_{23}\Leftrightarrow R_1\to\infty$).
\end{enumerate}
The positive octant contains further special positions (i.e. configurations
higher than $C_s$ symmetry) summarized in Fig.~\ref{fig:coords}~(c). Linear
molecules of $D_{\infty h}$ symmetry are found on three equivalent diagonals
of the boundary planes. Equilateral triangular configurations ($D_{3h}$
symmetry) correspond to the space diagonal while isosceles triangles
(i.e. $C_{2v}$ symmetry) are, due to the permutational symmetry, represented
via one of the three equivalent space diagonal surfaces (note that strictly
speaking there are six equivalent subspaces of $C_{2v}$ configurations since
it is not defined if the atoms are labelled clockwise or counterclockwise --
transition to the inverted structure takes place over a linear one
($R_i=0$)). In the context of this work the perimetric coordinates are a
powerful tool for investigating the configuration space
(cf. Sec.~\ref{subsec:ConfigSpace}) of equilibrium states of \ce{Rb3} helping
to identify appropriate states for PA processes.

Since homonuclear (alkali metal) triatomics are prominent systems showing the
Jahn-Teller (JT)
effect~\cite{BersukerReview,HauserPotassiumDoublets,HauserJTK3Rb3,HauserJTK3Rb3Doublets,HauserJTNa3,RochaC3}
it is also useful to introduce the \emph{(symmetry-adapted) JT coordinates} to
characterize the corresponding major topological features (COIN seam,
mexican-hat like PES and triply degenerate COINs for pseudo
Jahn-Teller (PJT) interactions) near $D_{3h}$ equilateral triangular
conformations. Given the internuclear distances $(R_{12},R_{23},R_{13})$ they
are defined by~\cite{RochaC3}
\begin{align}
  \label{QCoords}
  \begin{pmatrix}
    Q_1\\
    Q_2 \\
    Q_3 
  \end{pmatrix}
  &=
  \begin{tikzpicture}[baseline=-0.3ex]
    \matrix[matrix of math nodes,
      left delimiter=(,
      right delimiter=),
      nodes in empty cells,
      ampersand replacement=\&] (m)
           {
             ~ \& ~ \& ~ \& ~ \& \\
             ~ \& ~ \& ~ \& ~ \& \\
             ~ \& ~ \& ~ \& ~ \& \\
             ~ \& ~ \& ~ \& ~ \& ~ \\
             ~ \& ~ \& ~ \& ~ \& \\
             ~ \& ~ \& ~ \& ~ \& \\
             ~ \& ~ \& ~ \& ~ \& \\
             ~ \& ~ \& ~ \& ~ \& \\
             ~ \& ~ \& ~ \& ~ \& ~ \\
             ~ \& ~ \& ~ \& ~ \& \\
             ~ \& ~ \& ~ \& ~ \& \\
             ~ \& ~ \& ~ \& ~ \& \\
             ~ \& ~ \& ~ \& ~ \& \\
             ~ \& ~ \& ~ \& ~ \& ~ \\
             ~ \& ~ \& ~ \& ~ \& \\
           };
           \draw[thick,color=gray] (m-2-3.north) -- (m-4-4.south) --
           (m-4-2.south) -- cycle;
           \draw[thick,->] (m-2-3.north) -- (m-1-3.north);
           \draw[thick,->] (m-4-4.south) -- (m-5-4.south east);
           \draw[thick,->] (m-4-2.south) -- (m-5-1.south east);
           \shade[ball color=unidblue] (m-2-3.north) circle [radius=.075];
           \shade[ball color=unidblue] (m-4-4.south) circle [radius=.075];
           \shade[ball color=unidblue] (m-4-2.south) circle [radius=.075];
           \draw[thick,color=gray] (m-7-3.north) -- (m-9-4.south) --
           (m-9-2.south) -- cycle;
           \shade[ball color=unidblue] (m-7-3.north) circle [radius=.075];
           \shade[ball color=unidblue] (m-9-4.south) circle [radius=.075];
           \shade[ball color=unidblue] (m-9-2.south) circle [radius=.075];
           \draw[thick,->] (m-7-3.north) -- (m-7-4.north);
           \draw[thick,->] (m-9-4.south) -- (m-8-3.east);
           \draw[thick,->] (m-9-2.south) -- (m-10-1.south east);
           \draw[thick,color=gray] (m-12-3.north) -- (m-14-4.south) --
           (m-14-2.south) -- cycle;
           \shade[ball color=unidblue] (m-12-3.north) circle [radius=.075];
           \shade[ball color=unidblue] (m-14-4.south) circle [radius=.075];
           \shade[ball color=unidblue] (m-14-2.south) circle [radius=.075];
           \draw[thick,->] (m-12-3.north) -- (m-11-3.north);
           \draw[thick,->] (m-14-4.south) -- (m-15-3.south east);
           \draw[thick,->] (m-14-2.south) -- (m-15-3.south west);
  \end{tikzpicture} \notag \\
  &= 
  \begin{pmatrix}
    \sqrt{\nicefrac{1}{3}} & \sqrt{\nicefrac{1}{3}} &
    \sqrt{\nicefrac{1}{3}} \\
    -\sqrt{\nicefrac{1}{2}} & \sqrt{\nicefrac{1}{2}} & 0 \\
    -\sqrt{\nicefrac{1}{6}} &
    -\sqrt{\nicefrac{1}{6}} & \sqrt{\nicefrac{2}{3}} 
  \end{pmatrix}
  \cdot
  \begin{pmatrix}
    R_{12} \\
    R_{13} \\
    R_{23}
  \end{pmatrix}\,.
\end{align}
They describe the planar vibrational modes of the system where $Q_1$ is
associated with the breathing mode (preserving $D_{3h}$ geometry), $Q_2$ with
the asymmetric stretch mode (distorting the equilateral triangle into a $C_s$
configuration) and $Q_3$ with the symmetric stretch mode (taking the system
into a $C_{2v}$ conformation). Note that this only holds for $D_{3h}$
symmetry, in the subspace of lower symmetry, e.g. $C_{2v}$, the actual modes
are mixtures of $Q_1$ and $Q_3$. Using this set of coordinates, e.g.  Hauser
et al., studied several aspects of the JT effect in \ce{K3} and
\ce{Rb3}~\cite{HauserPotassiumDoublets,HauserJTK3Rb3,HauserJTK3Rb3Doublets} by
$C_{2v}$ cuts (one- and two-dimensional) through the PESs of both species. We
are applying these coordinates for investigating the
$1{\,}\tensor*[^{4}]{\mathrm{E}}{^{\prime\prime}}$ state in the context of PA
experiments in Sec.~\ref{subsec:1Epp}.

\subsection{\label{subsec:PESCuts}Special Cuts through the PESs}
\begin{figure*}[tb]
  \centering
  \begin{minipage}[t]{.49\textwidth}
    \includegraphics[width=\linewidth]{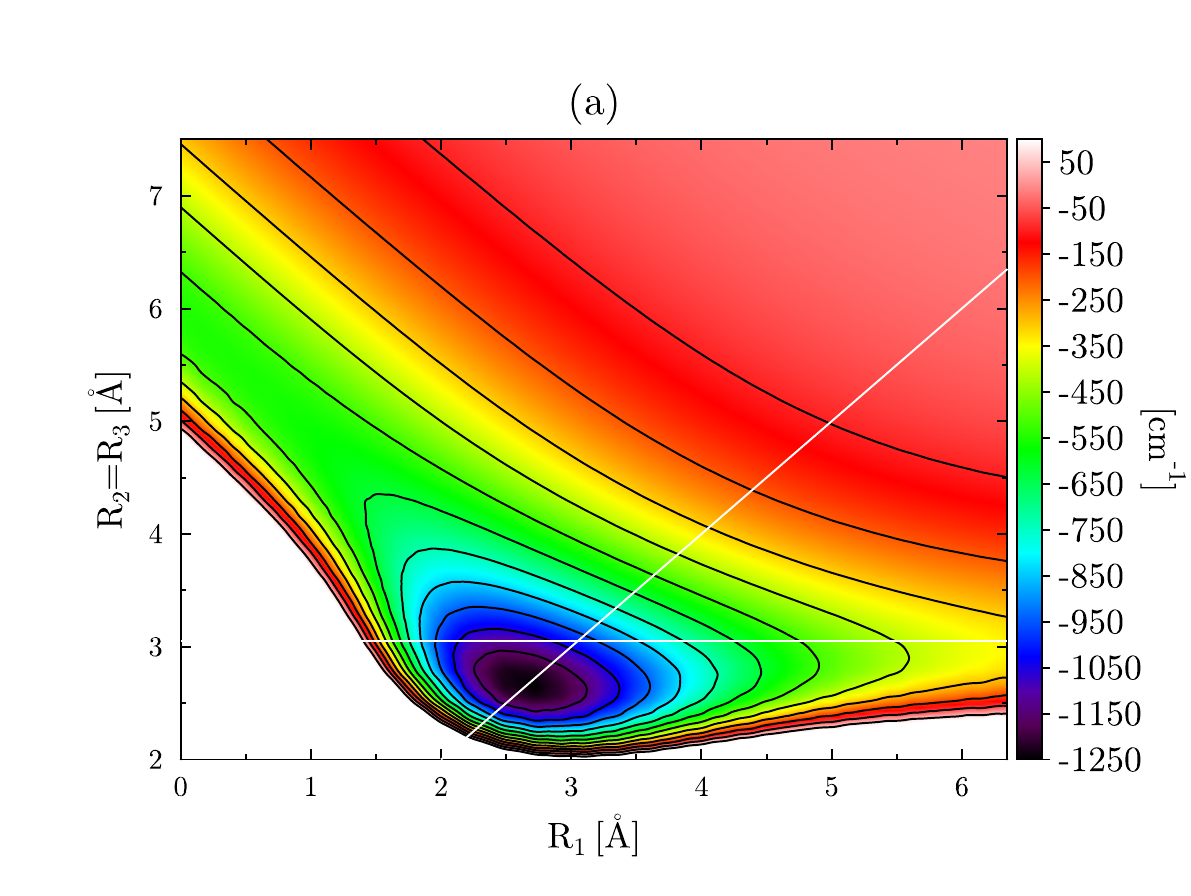}
  \end{minipage}
  \begin{minipage}[t]{.49\textwidth}
    \includegraphics[width=\linewidth]{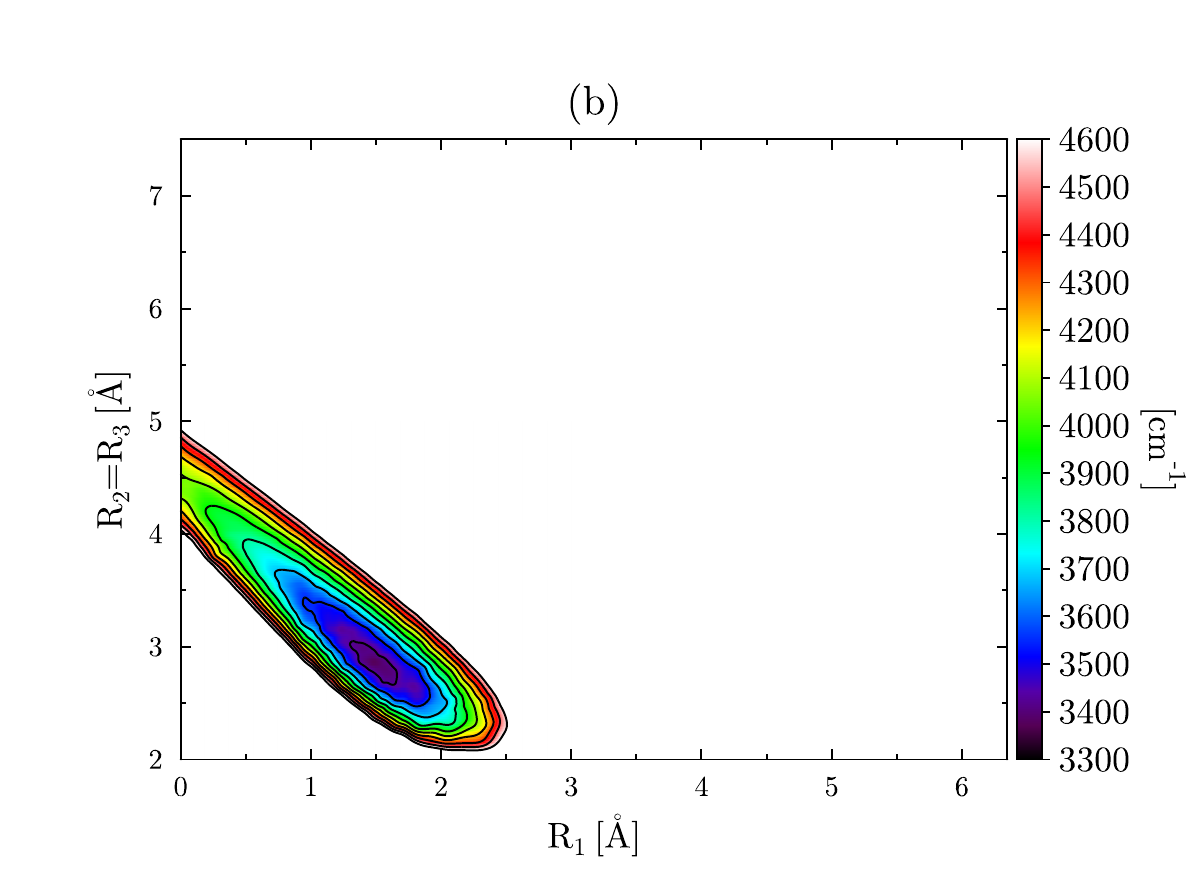}
  \end{minipage}
  \caption{\label{fig:2DCuts1B1and1A2}(Color online) Two-dimensional contour plots of the
    quartet ground state $1{\,}\tensor*[^{4}]{\mathrm{B}}{_1}$ in \textbf{(a)}
    and the first quartet excited state $1{\,}\tensor*[^{4}]{\mathrm{A}}{_2}$
    in \textbf{(b)} PESs in the subspace $R_1$ and $R_2=R_3$ of perimetric
    coordinates, i.e. along one space diagonal surface in
    Fig.~\ref{fig:coords}~(c). The space diagonal (i.e. $D_{3h}$
    configurations) corresponds to the diagonal line shown in white, while the
    horizontal line (both in (a)) represents the special one-dimensional
    $C_{2v}$-cut for $R_{23}=\SI{6.094}{\angstrom}$. The wavy character in
    \textbf{(b)} is due to the underlying spline interpolation of the
    corresponding \emph{ab-initio} data.}
\end{figure*}
To get an idea of the system's physics, in particular the occurring
coupling and crossing effects, we start with analyzing special cuts through
the PESs of both doublet and quartet manifolds. For this we restrict our
investigations to the $C_{2v}$ subspace since it turns out that all
equilibrium structures show at least $C_{2v}$ symmetry. Therefore we are
labeling the resulting electronic states according to the irreducible
representations (IRREPs) of this point group. Given the choice of coordinates
shown in Fig.~\ref{fig:coords}~(a), $\mathrm{A}_1$ and $\mathrm{B}_1$ states
are symmetric, and $A_2$ and $B_2$ states are antisymmetric with respect to
reflection of the electronic coordinates at the molecular plane.
Figure~\ref{fig:2DCuts1B1and1A2} gives a first impression of the topology of
the potential energy landscapes for the quartet ground state
($1{\,}\tensor*[^4]{\mathrm{B}}{_1}$) and the first excited quartet state
($1{\,}\tensor*[^4]{\mathrm{A}}{_2}$) in terms of two-dimensional cuts for
$C_{2v}$-symmetric nuclear configurations. These correspond to one of the
space diagonal surfaces shown in Fig.~\ref{fig:coords}~(c). The quartet ground
state ($1{\,}\tensor*[^{4}]{\mathrm{B}}{_1}$) in
Fig.~\ref{fig:2DCuts1B1and1A2}~(a) is well isolated from excited quartet
states (i.e. crossings with other states only appear at energies high above
the minimum and the dissociation limit of this state) with the global minimum
occurring at equilateral triangular ($D_{3h}$) geometry
\cite{SoldanPESRb3}. At a symmetric linear geometry (for this cut at $R_1=0$) we obtain
a saddle point marking the transition to the inverted structure. Moreover, we
note that the PES of the $1{\,}\tensor*[^{4}]{\mathrm{B}}{_1}$ is rather
shallow. These properties have also been pointed out by Sold\'{a}n in his work
concerning the quartet ground state of \ce{Rb3} in
Ref.~\cite{SoldanPESRb3}. Figure~\ref{fig:2DCuts1B1and1A2}~(b) shows the first
excited quartet state ($1{\,}\tensor*[^{4}]{\mathrm{A}}{_2}$ in $C_{2v}$) with
the global minimum occurring at isosceles triangular (i.e. $C_{2v}$)
geometry. This is due to the JT effect forming a twofold degenerate
$\mathrm{E}^{\prime\prime}$ state (with $1{\,}\tensor*[^4]{\mathrm{B}}{_2}$)
at $D_{3h}$ geometries (this will be discussed in more detail in
Sec.~\ref{subsec:1Epp}). This PES rises significantly steeper than the shallow
quartet ground state PES.

\begin{figure*}[tb]
  \centering
  \begin{minipage}[t]{.49\textwidth}
    \includegraphics[width=.91\columnwidth]{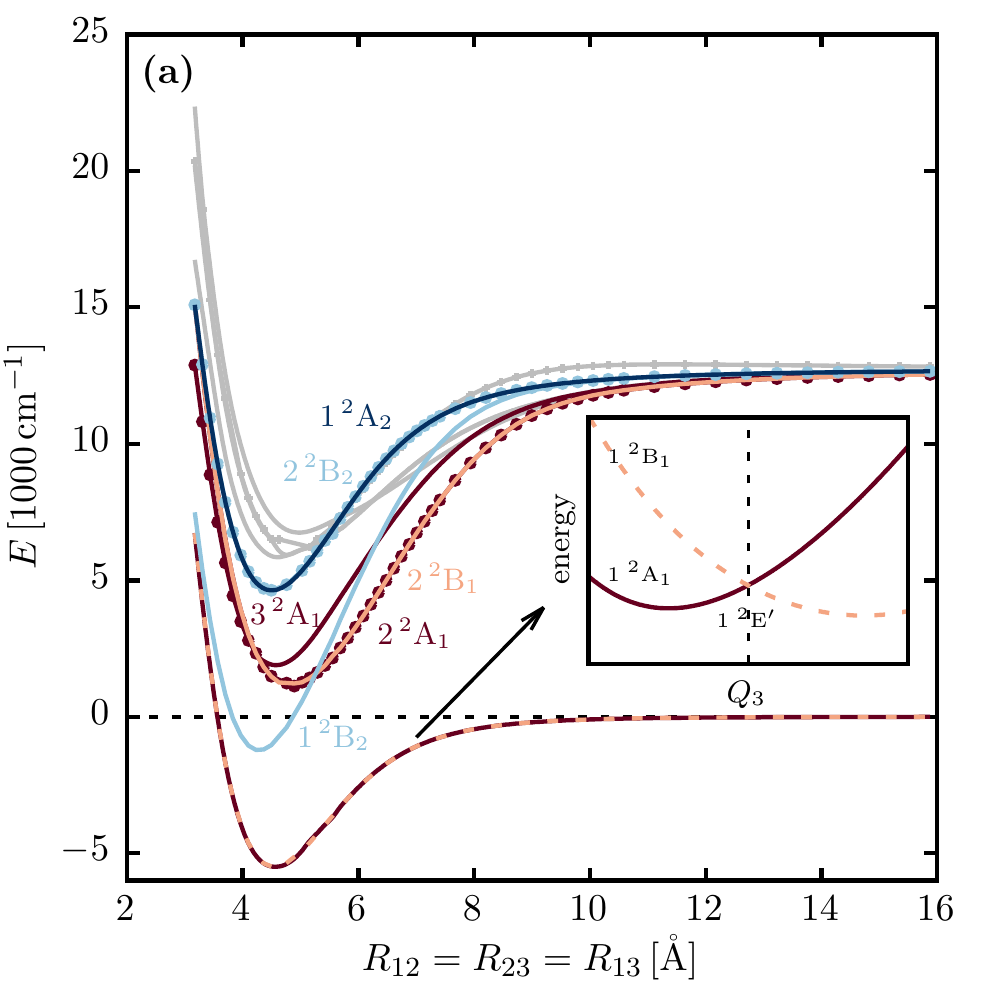}
  \end{minipage}
  \hspace{\fill}
  \begin{minipage}[t]{.49\textwidth}
    \includegraphics[width=.91\columnwidth]{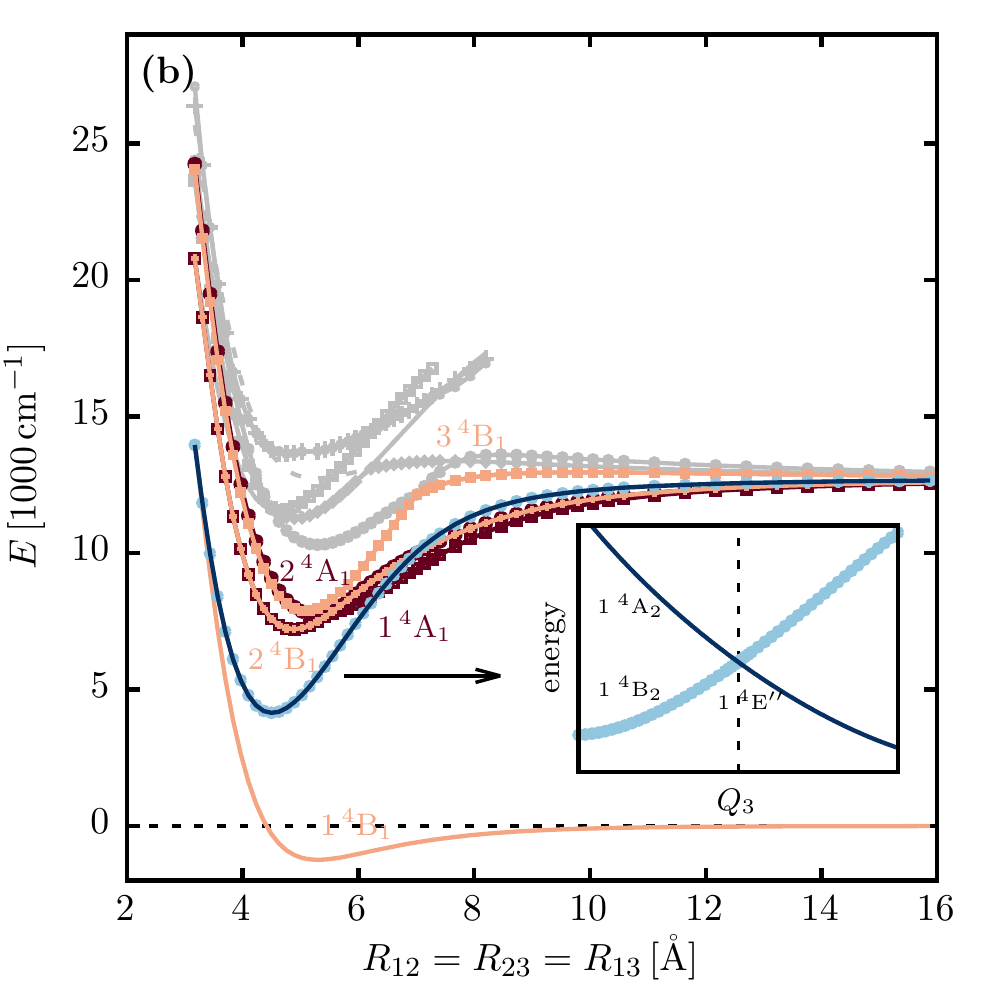}
  \end{minipage}
  \begin{minipage}[t]{.49\textwidth}
    \includegraphics[width=.91\columnwidth]{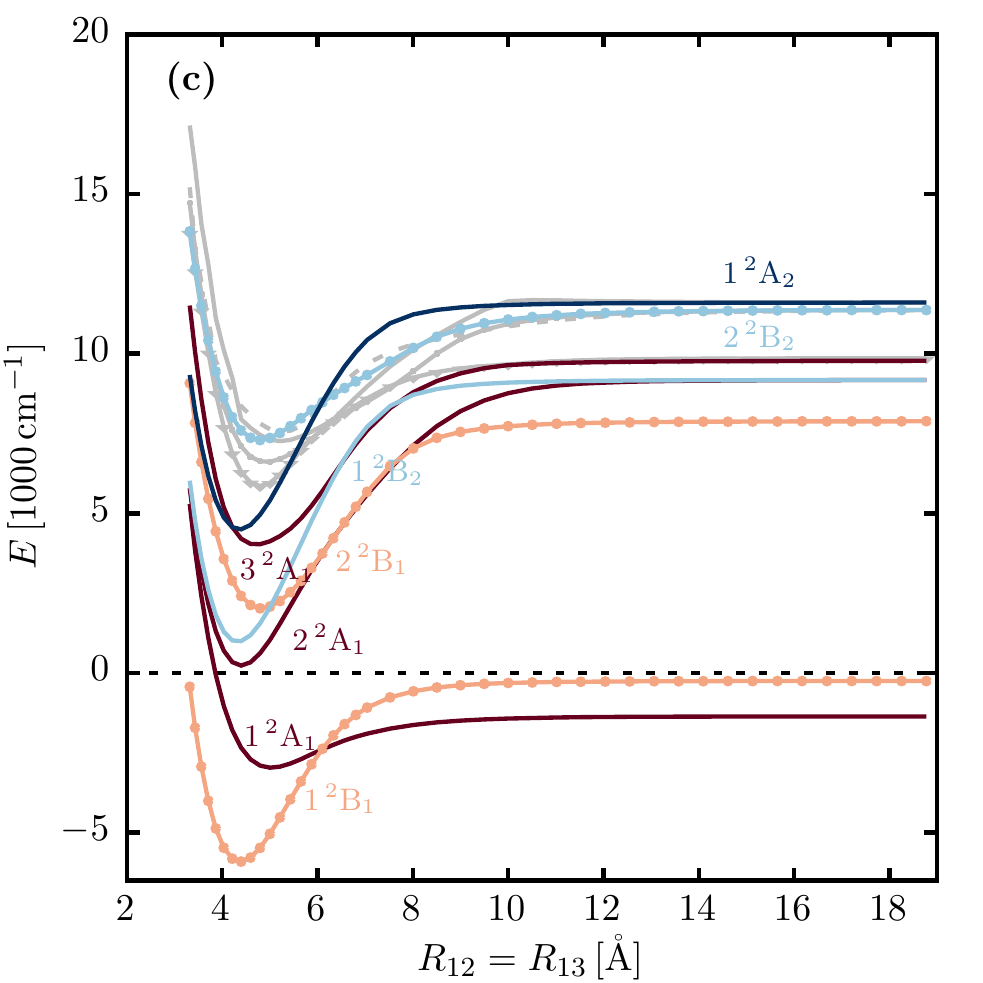}
  \end{minipage}
  \begin{minipage}[t]{.49\textwidth}
    \includegraphics[width=.91\columnwidth]{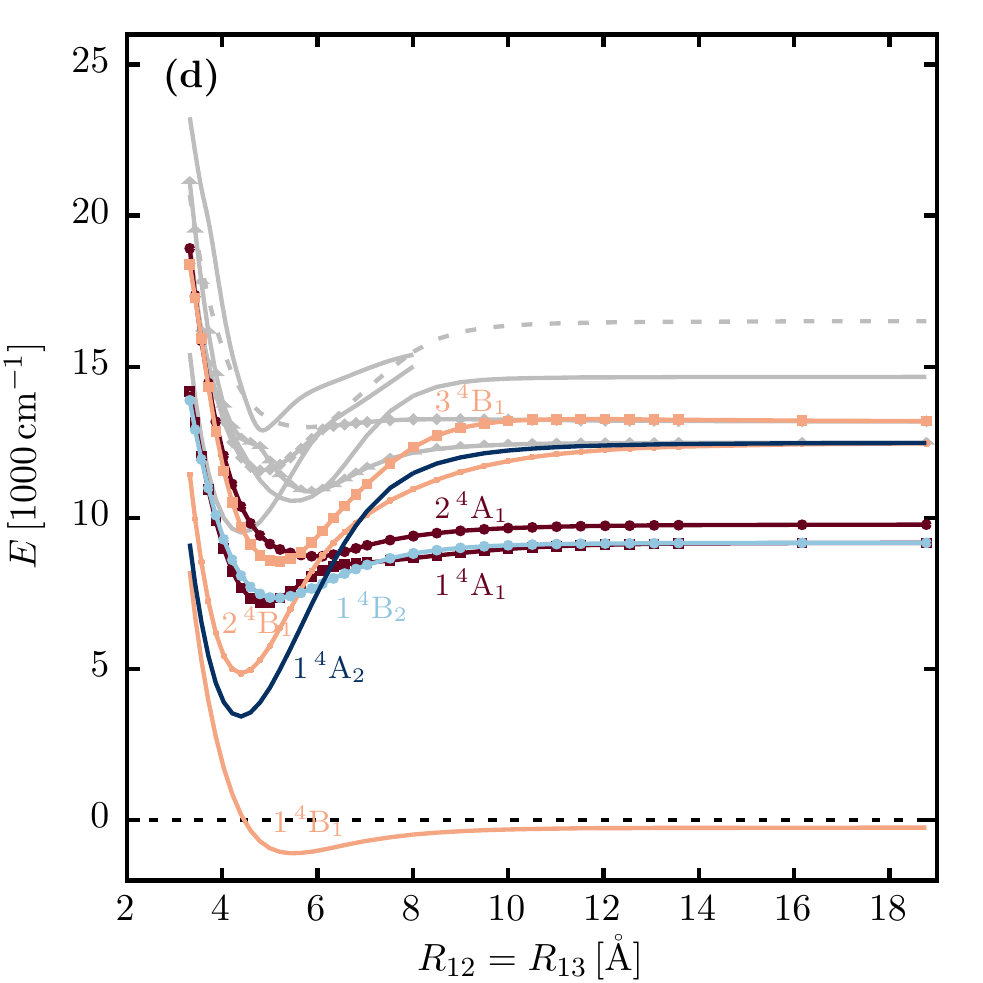}
  \end{minipage}
  \caption{\label{fig:D3hScan}(Color online) \textbf{Upper panel:} One-dimensional cut
    through the potential energy surfaces (PESs) along the space diagonal in
    the perimetric coordinate space shown Fig.~\ref{fig:coords}~(c) and in
    terms of the diagonal shown Fig.~\ref{fig:2DCuts1B1and1A2}~(a)
    (i.e. $D_{3h}$ scan maintaining the equilateral triangular
    configuration). Doublet states are shown in \textbf{(a)} and quartet
    states in \textbf{(b)}, respectively. The low-lying states discussed in the
    text are highlighted, the presence of further states is indicated by the gray lines.
    \textbf{Lower panel:} One-dimensional cut through the PESs along one
    special direction on one space diagonal surface shown in
    Fig.~\ref{fig:coords}~(c). This $C_{2v}$ scan  corresponds to a fixed distance
    $R_{23}=R_e(a{\,}^3\Sigma_u)=\SI{6.094}{\angstrom}$. Doublets are shown in \textbf{(c)} and
    quartets in \textbf{(d)}, respectively.}
\end{figure*}

The presence of (pseudo) Jahn-Teller interactions can be also observed from
one-dimensional scans along the $D_{3h}$ subspace, i.e. along the diagonal
shown in white in Fig.~\ref{fig:2DCuts1B1and1A2}~(a). The resulting potential
energy curves (PECs) are shown in Fig.~\ref{fig:D3hScan} in the space of
internal coordinates ($R_{12} = R_{23} = R_{13}$).
Equilateral triangular configurations of homonuclear triatomics display
$D_{3h}$ symmetry and allow for two-fold degenerate, so called $\mathrm{E}$
terms (cf. Tab.~\ref{S-tab:D3h} in the supplementary material~\cite{Supplementary}). According to
the Jahn-Teller
theorem~\cite{BersukerReview,RelExEJT,VibJTExe,relJTReview,POLUYANOV2008125,RochaC3},
the PES of at least one of these degenerate states has no extremum at this
high-symmetry point. Thus, the system lowers its symmetry to lift the
degeneracy, here branching off into $\mathrm{A}_1+\mathrm{B}_1$ states for
$\mathrm{E}^\prime$ or into $\mathrm{B}_2+\mathrm{A}_2$ states for
$\mathrm{E}^{\prime\prime}$, respectively. This is accompanied by an energy
lowering and the formation of a COIN at the point of
degeneracy. This is also indicated by the insets shown in
Fig.~\ref{fig:D3hScan}. Potential energy curves which are degenerate over the
whole range shown in Fig.~\ref{fig:D3hScan} are actually one-dimensional
COIN seams in the three-dimensional configuration space.

The doublet ground states of alkali trimers show their global minimum at
obtuse isosceles triangular geometries due to the JT effect (studied
theoretically for \ce{Li3} in Refs.~\cite{Martins1983,Thompson1985Li3Na3K3},
for \ce{Na3} in
Refs.~\cite{Martins1983,Davidson1978,Thompson1985Li3Na3K3,Cocchini1988Na3},
for \ce{K3} in
Refs.~\cite{Martins1983,Thompson1985Li3Na3K3,HauserPotassiumDoublets,Mukherjee2014}
and for \ce{Rb3} in
Refs.~\cite{HauserJTK3Rb3Doublets,HauserBook2009,HauserPhD}). This finding is also
illustrated by the corresponding PECs in Fig.~\ref{fig:D3hScan}~(a) (and by
the alternative one-dimensional cuts in Fig.~\ref{S-fig:JT-scans-QD} in the
supplementary material~\cite{Supplementary}). A further peculiarity -- well-known as the pseudo
Jahn-Teller (PJT) effect -- is formed, e.g., by the triple of states $\lbrace
2{\,}\tensor*[^{2}]{\mathrm{A}}{_1}, 2{\,}\tensor*[^{2}]{\mathrm{B}}{_1},
3{\,}\tensor*[^{2}]{\mathrm{A}}{_1}\rbrace$, where
$2{\,}\tensor*[^{2}]{\mathrm{A}}{_1}$ and
$2{\,}\tensor*[^{2}]{\mathrm{B}}{_1}$ are degenerate components of the
$2{\,}\tensor*[^{2}]{\mathrm{E}}{^\prime}$ term (for $D_{3h}$ configurations)
and the $3{\,}\tensor*[^{2}]{\mathrm{A}}{_1}$ state is nearby in energy (near
degeneracy) -- cf. Fig.~\ref{fig:D3hScan}~(a) and
Fig.~\ref{S-fig:JT-scans-QD}. Consequently, all three states can mix for
$C_{2v}$ configurations which is described within the theory of pseudo
JT-coupling (cf. e.g. Refs.~\cite{PJTEvsRTE,PJTEReview}). It follows that due
to the third state which is close in energy the COIN seam of the doubly
degenerate JT state at high-symmetry geometries vanishes. Only at a \emph{single}
point in the $D_{3h}$ subspace all three states become degenerate forming a
\emph{triply} degenerate COIN point~\cite{Cocchini1988Na3}. This intersection
is analogous to the JT one, but it is not required by symmetry (accidental
degeneracy). All of this is essential to fully understand the well-known
experimentally observed $\mathrm{B}$ band in alkali metal
triatomics~\cite{BersukerReview}.

In contrast to the doublet ground state, the quartet ground state is free of
JT distortions with its global minimum at $D_{3h}$ configuration. The first
pair of excited quartet states, however, is degenerate along a one-dimensional
COIN seam in the $D_{3h}$ configuration space, and spans a
$1{\,}\tensor*[^{4}]{\mathrm{E}}{^{\prime\prime}}$ term, which splits into
$1{\,}\tensor*[^{4}]{\mathrm{A}}{_2}$ and
$1{\,}\tensor*[^{4}]{\mathrm{B}}{_2}$ states when the symmetry is lowered.
Besides those states which are exactly degenerate, there are also a number of
nearly degenerate states. In particular, there are quadruple
interactions~\cite{HauserPhD} present within the subset $\mathcal{Q}$ of
quartet states
\begin{align}
  \label{eq:Q}
  \mathcal{Q} &= \lbrace 1{\,}\tensor*[^{4}]{\mathrm{A}}{_1},
  2{\,}\tensor*[^{4}]{\mathrm{A}}{_1}, 2{\,}\tensor*[^{4}]{\mathrm{B}}{_1},
  3{\,}\tensor*[^{4}]{\mathrm{B}}{_1}\rbrace\,.
\end{align}
This peculiarity can be seen in Fig.~\ref{fig:D3hScan}~(b) (and
Fig.~\ref{S-fig:JT-scans-QD}~(b) in the supplementary material~\cite{Supplementary}), where those
states are almost degenerate in a region reaching from
$\approx\SI{5.0}{\angstrom}$ to $\approx\SI{7.0}{\angstrom}$. More details on
all JT and PJT pairs within this energy range can be found in the
supplementary material~\cite{Supplementary} in Tab.~\ref{S-tab:PJTStates}.

The other one-dimensional cut indicated in Fig.~\ref{fig:2DCuts1B1and1A2}~(a)
corresponds to a collision trajectory between a \ce{Rb2} molecule and a
\ce{Rb} atom. For this cut, we fixed the distance $R_{23}$ to the equilibrium
distance of the lowest triplet state ($a{\,}^3\Sigma_u$) of \ce{Rb2}. The
resulting cuts in Figs.~\ref{fig:D3hScan}~(c) and~(d) give a first impression
of the states possibly involved in a PA 1 scheme. Moreover, this graph shows
one dimension of the 2D branching space (formally spanned by $Q_2$ and $Q_3$,
cf. Sec.~\ref{subsec:Coords}) where the degeneracies from high-symmetry
configurations (here found at the point
$R_{12}=R_{13}=R_{23}=\SI{6.094}{\angstrom}$) are lifted. This gives a notion
of the topology of the full 3D potential energy landscape. The density of
states increases for higher energies for both doublet and quartet manifolds
and decreases the chance for finding sufficiently long-lived target states for
PA experiments. Therefore and due to the fact that the doublet ground state
has a rather complex behavior due to JT distortions we focus on quartet
states.

Linear configurations of the trimer system are subject to Renner-Teller (RT)
or combined PJT plus RT interactions. A detailed analysis of this, however, is
beyond the scope of the present work. Nevertheless, a comment can be found in
the supplementary material~\cite{Supplementary}.

\begin{figure*}[tb]
  \centering
  \includegraphics[width=.7\textwidth]{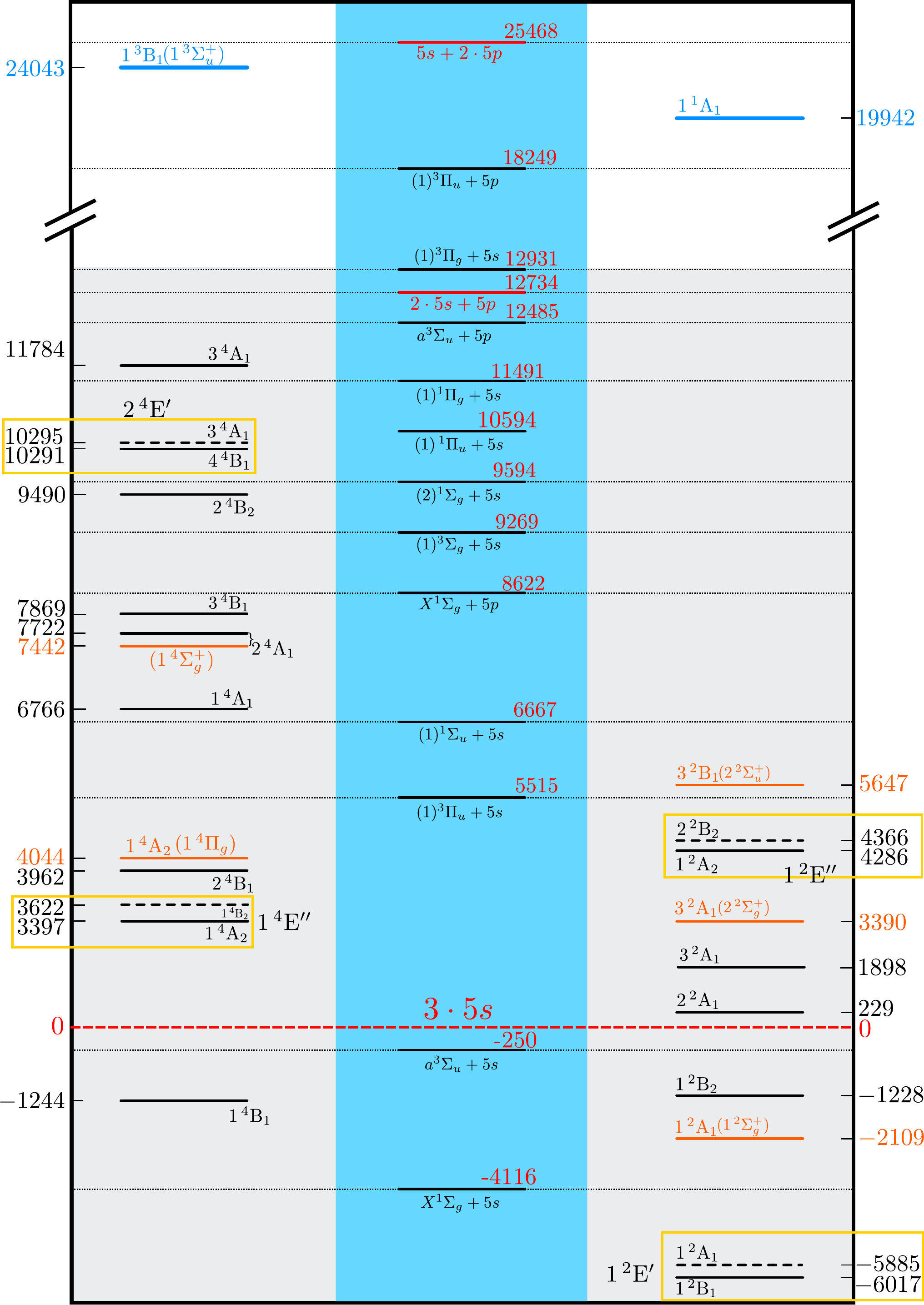}
  \caption{\label{fig:termscheme}(Color online) Energy level diagram of the extremal points
    of \textbf{doublet (right)} and \textbf{quartet (left)} states of \ce{Rb3}
    optimized at MRCI(ECP+CPP)/UET15 level of theory. The dissociation
    asymptotes into \ce{Rb2}+\ce{Rb} (with the corresponding equilibrium
    energies of the \ce{Rb2} states) or \ce{Rb}+\ce{Rb}+\ce{Rb} are shown in
    the area highlighted in blue. Levels given in \textbf{black} belong to
    \textbf{triangular} equilibrium configurations (i.e. $D_{3h}$ or $C_{2v}$
    symmetry) while levels given in \textbf{orange} represent \textbf{linear}
    equilibrium configurations (all of them $D_{\infty h}$ symmetry).
    \textbf{Yellow boxes} mark Jahn-Teller states, where the respective dashed
    lines correspond to saddle points showing isosceles triangular
    geometry. The ionized $\mathrm{Rb}_3^+$ states are shown at the top in
    terms of blue energy levels. All energies are given relative to the free
    atom-atom-atom limit (i.e. $3\cdot\ce{Rb}\left[5s\right]$).}
\end{figure*}

\subsection{\label{subsec:EqStates}Equilibrium States}
A systematic overview of the energy levels of all doublet and quartet states
of \ce{Rb3} considered in this work is given in Fig.~\ref{fig:termscheme}. All
energy levels refer to the electronic energy at the equilibrium geometry. For
finding the equilibrium states we started from high-symmetry configurations
($D_{3h}$) and proceeded to geometries of lower symmetry ($C_{2v}$). Our
analysis did not show any evidence for equilibrium structures of even lower
symmetry, i.e. $C_s$. We determined all equilibrium states and their
electronic term energies in the energy region up to the $5s+2\cdot 5p$
asymptote. The energies of the \ce{Rb2}+\ce{Rb} or \ce{Rb}+\ce{Rb}+\ce{Rb}
dissociation asymptotes are given in the middle panel. The assignment of the
trimer states to the \ce{Rb2}+\ce{Rb} asymptotes is in general only unique for
the quartet ground state $1{\,}\tensor*[^4]{\mathrm{B}}{_1}$ dissociating into
$a{\,}\tensor*[^3]{\Sigma}{_u}+5s$. For one-dimensional $C_{2v}$ cuts as shown
in Fig.~\ref{fig:D3hScan}~(c) and~(d) we obtain a unique assignment for all
quartet states and some doublet states as well. However, in the general case,
for both doublet and quartet states, all \ce{Rb2}+\ce{Rb} asymptotes
correlating with the respective trimer state symmetry are possible
dissociation channels. Most of the excited states correlate to the $2\cdot 5s
+ 5p$ asymptote. Merely the highly excited quartet states
$2{\,}\tensor*[^4]{\mathrm{B}}{_2}$, $3{\,}\tensor*[^4]{\mathrm{B}}{_2}$ and
$4{\,}\tensor*[^4]{\mathrm{A}}{_1}$ correspond to the $5s+2\cdot 5p$ asymptote
and thus to the $(1)\,\tensor*[^3]{\Pi}{_u}+5p$ dissociation limit. The lowest
doublet JT manifold
$1{\,}\tensor*[^2]{\mathrm{E}}{^\prime}=1{\,}\tensor*[^2]{\mathrm{A}}{_1}+1{\,}\tensor*[^2]{\mathrm{B}}{_1}$
dissociates either to $X{\,}\tensor*[^1]{\Sigma}{_g}+5s$ or to
$a{\,}\tensor*[^3]{\Sigma}{_u}+5s$. The remaining doublet states correspond to
\ce{Rb2}+\ce{Rb} asymptotes below the $2\cdot 5s + 5p$ asymptote where both
singlet and triplet \ce{Rb2} states are possible.
Finally, the top panel of Fig.~\ref{fig:termscheme} also shows the
ionized states of \ce{Rb3+} appearing in either singlet or triplet
configuration. This could be useful if a REMPI~\cite{Wolf921} scheme is used
for the detection of previously generated \ce{Rb3} species. All these results
are listed in Tab.~\ref{tab:Rb3statesTriangular} (for triangular geometries)
and in Tab.~\ref{tab:Rb3statesLinear} (for linear configurations) together
with their corresponding harmonic vibrational frequencies $\tilde{\nu}$.
{\setlength{\tabcolsep}{8.3pt}
\begin{table*}[tb]
  \centering
  \caption{\label{tab:Rb3statesTriangular}Synopsis of \textbf{triangular}
    ($C_{2v}$ and $D_{3h}$) doublet and quartet (ground and excited) states of
    \ce{Rb3} as well as the singlet state of \ce{Rb3+} computed at
    MRCI(ECP+CPP)/UET15 level of theory. Equilibrium structures are given in
    terms of the internal coordinates (perimetric coordinates) introduced in
    Fig.~\ref{fig:coords} and all corresponding energies ($E_{\text{rel}}$)
    are given relative to the $(3\cdot 5s)$--asymptote calculated at the same
    level of theory. The states are labelled according to the $C_{2v}$ IRREPs
    while the corresponding assignment to $D_{3h}$ symmetry is given in
    parenthesis. This complements the results of the energy level diagram of
    Fig.~\ref{fig:termscheme}.}
  \begin{tabular}{lcccccc}
    \toprule\toprule
        \multirow{2}{*}{State ($D_{3h}$)} & {$R_{12},R_{23},R_{13}\,[\mathrm{\mathring{A}}]$} & \multirow{2}{*}{Geometry} & \multirow{2}{*}{$E_{\text{rel}}\,[\mathrm{cm}^{-1}]$} &
        \multirow{2}{*}{$\tilde{\nu}_{D_{3h}}\,[\mathrm{cm}^{-1}]$\footnote{\label{comment}In general
            the assignment is not unique but usually $\tilde{\nu}_{D_{3h}}$ is $Q_1$-like, $\tilde{\nu}_{C_{2v}}$ is $Q_3$-like and $\tilde{\nu}_{C_s}$
            is $Q_2$-like.}} &
        \multirow{2}{*}{$\tilde{\nu}_{C_{2v}}\,[\mathrm{cm}^{-1}]$\footref{comment}} &
        \multirow{2}{*}{$\tilde{\nu}_{C_s}\,[\mathrm{cm}^{-1}]$\footref{comment}} \\
        ~ & {$(R_1, R_2, R_3)$} & ~ & ~ & ~ & ~ & ~ \\
        \midrule
        \multirow{2}{*}{$1{\,}\tensor*[^{4}]{\mathrm{B}}{_1}$ ($1{\,}\tensor*[^4]{\mathrm{A}}{^\prime_2}$)} & $5.311, 5.311, 5.311$ & \multirow{2}{*}{$D_{3h}$} &
        \multirow{2}{*}{\alignmultr[table-format=4.1]{-1244}} &
        \multirow{2}{*}{\alignmultr[table-format=4.1]{23.8}} & 
        \multirow{2}{*}{\alignmultr[table-format=4.1]{23.6}} &
        \multirow{2}{*}{\alignmultr[table-format=4.1]{23.6}} \\
        ~ & $(2.656,2.656,2.656)$ & ~ & ~ & ~ & ~ & ~ \vspace*{.05cm} \\
        \multirow{2}{*}{$1{\,}\tensor*[^{4}]{\mathrm{A}}{_2}$ ($1{\,}\tensor*[^4]{\mathrm{E}}{^{\prime\prime}}$)} & $4.368,5.700,4.368$ & \multirow{2}{*}{$C_{2v}$}
        & \multirow{2}{*}{3397} & 
        \multirow{2}{*}{\alignmultr[table-format=4.1]{50.2}} &
        \multirow{2}{*}{\alignmultr[table-format=4.1]{17.2}} &
        \multirow{2}{*}{\alignmultr[table-format=4.1]{35.8}} \\
        ~ & $(1.518,2.850,2.850)$ & ~ & ~ & ~ & ~ & ~ \vspace*{.05cm}\\
        \multirow{2}{*}{$2{\,}\tensor*[^{4}]{\mathrm{B}}{_1}$ ($1{\,}\tensor*[^4]{\mathrm{E}}{^\prime}$)} & $4.442, 8.179, 4.442$ &
        \multirow{2}{*}{$C_{2v}$} & \multirow{2}{*}{3962} & 
        \multirow{2}{*}{\alignmultr[table-format=4.1]{40.3}} &
        \multirow{2}{*}{\alignmultr[table-format=4.1]{9.6}} &
        \multirow{2}{*}{\alignmultr[table-format=4.1]{40.4}} \\
        ~ & $(0.352,4.090,4.090)$ & ~ & ~ & ~ & ~ & ~ \vspace*{.05cm}\\
        \multirow{2}{*}{$1{\,}\tensor*[^{4}]{\mathrm{A}}{_1}$ ($1{\,}\tensor*[^4]{\mathrm{E}}{^\prime}$)} & $4.993, 8.076, 4.993$
        & \multirow{2}{*}{$C_{2v}$} & \multirow{2}{*}{6766} &
        \multirow{2}{*}{\alignmultr[table-format=4.1]{32.8}} &
        \multirow{2}{*}{\alignmultr[table-format=4.1]{9.8}} &
        \multirow{2}{*}{\alignmultr[table-format=4.1]{50.7}} \\
        ~ & $(0.955,4.038,4.038)$ & ~ & ~ & ~ & ~ & ~ \vspace*{.05cm}\\
        \multirow{2}{*}{$2{\,}\tensor*[^{4}]{\mathrm{A}}{_1}$ ($1{\,}\tensor*[^4]{\mathrm{A}}{^\prime_1}$)} & $5.325, 5.325, 5.325$ &
        \multirow{2}{*}{$D_{3h}$} & \multirow{2}{*}{7722} & 
        \multirow{2}{*}{\alignmultr[table-format=4.1]{32.0}} &
        \multirow{2}{*}{\alignmultr[table-format=4.1]{83.9}} &
        \multirow{2}{*}{\alignmultr[table-format=4.1]{83.9}} \\
        ~ & $(2.663,2.663,2.663)$ & ~ & ~ & ~ & ~ & ~ \vspace*{.05cm}\\
        \multirow{2}{*}{$3{\,}\tensor*[^{4}]{\mathrm{B}}{_1}$ ($2{\,}\tensor*[^4]{\mathrm{A}}{^\prime_2}$)} & $5.084, 5.084, 5.084$ &
        \multirow{2}{*}{$D_{3h}$} & \multirow{2}{*}{7869} &
        \multirow{2}{*}{\alignmultr[table-format=4.1]{43.0}} &
        \multirow{2}{*}{\alignmultr[table-format=4.1]{58.2}} &
        \multirow{2}{*}{\alignmultr[table-format=4.1]{58.2}} \\
        ~ & $(2.542,2.542,2.542)$ & ~ & ~ & ~ & ~ & ~ \vspace*{.05cm} \\
        \multirow{2}{*}{$2{\,}\tensor*[^{4}]{\mathrm{B}}{_2}$ ($2{\,}\tensor*[^4]{\mathrm{E}}{^{\prime\prime}}$)} & $4.443, 6.217, 4.443$ &
        \multirow{2}{*}{$C_{2v}$} & \multirow{2}{*}{9490} & 
        \multirow{2}{*}{\alignmultr[table-format=4.1]{52.0}} &
        \multirow{2}{*}{\alignmultr[table-format=4.1]{42.7}} &
        \multirow{2}{*}{\alignmultr[table-format=4.1]{41.3}} \\
        ~ & $(1.335,3.109,3.109)$ & ~ & ~ & ~ & ~ & ~ \vspace*{.05cm}\\
        \multirow{2}{*}{$4{\,}\tensor*[^{4}]{\mathrm{B}}{_1}$ ($2{\,}\tensor*[^4]{\mathrm{E}}{^\prime}$)} & $5.283, 5.337,
        5.283$ & \multirow{2}{*}{$C_{2v}$} & \multirow{2}{*}{10291} &
        \multirow{2}{*}{\alignmultr[table-format=4.1]{31.5}} &
        \multirow{2}{*}{\alignmultr[table-format=4.1]{37.7}} &
        \multirow{2}{*}{\alignmultr[table-format=4.1]{39.0}} \\
        ~ & $(2.615,2.669,2.669)$ & ~ & ~ & ~ & ~ & ~ \vspace*{.05cm}\\
        \multirow{2}{*}{$3{\,}\tensor*[^{4}]{\mathrm{A}}{_1}$ (upper) ($2{\,}\tensor*[^4]{\mathrm{E}}{^\prime}$)} & $4.687, 7.226, 4.687$ & \multirow{2}{*}{$C_{2v}$} & \multirow{2}{*}{11784} 
        & \multirow{2}{*}{\alignmultr[table-format=4.1]{41.5}} &
        \multirow{2}{*}{\alignmultr[table-format=4.1]{28.2}} &
        \multirow{2}{*}{\alignmultr[table-format=4.1]{27.5}} \\
        ~ & $(1.074,3.613,3.613)$ & ~ & ~ & ~ & ~ & ~ \vspace*{.05cm}\\
        \midrule
        \multirow{2}{*}{$1{\,}\tensor*[^{2}]{\mathrm{B}}{_1}$ ($1{\,}\tensor*[^2]{\mathrm{E}}{^\prime}$)} & $4.379, 5.393, 4.379$ &
        \multirow{2}{*}{$C_{2v}$} & \multirow{2}{*}{-6017} & 
        \multirow{2}{*}{\alignmultr[table-format=4.1]{53.1}} &
        \multirow{2}{*}{\alignmultr[table-format=4.1]{20.6}} &
        \multirow{2}{*}{\alignmultr[table-format=4.1]{33.3}} \\
        ~ & $(1.682,2.697,2.697)$ & ~ & ~ & ~ & ~ & ~ \vspace*{.05cm}\\ 
        \multirow{2}{*}{$1{\,}\tensor*[^{2}]{\mathrm{B}}{_2}$ ($1{\,}\tensor*[^2]{\mathrm{A}}{^{\prime\prime}_2}$)} & $4.276, 4.285, 4.276$ &
        \multirow{2}{*}{$C_{2v}$} & \multirow{2}{*}{-1228} & 
        \multirow{2}{*}{\alignmultr[table-format=4.1]{60.9}} &
        \multirow{2}{*}{\alignmultr[table-format=4.1]{43.9}} &
        \multirow{2}{*}{\alignmultr[table-format=4.1]{44.1}} \\
        ~ & $(2.134,2.143,2.143)$ & ~ & ~ & ~ & ~ & ~ \vspace*{.05cm}\\
        \multirow{2}{*}{$2{\,}\tensor*[^{2}]{\mathrm{A}}{_1}$ ($2{\,}\tensor*[^2]{\mathrm{E}}{^\prime}$)} & $4.398, 6.073, 4.398$ &
        \multirow{2}{*}{ $C_{2v}$} & \multirow{2}{*}{229} & 
        \multirow{2}{*}{\alignmultr[table-format=4.1]{50.4}} &
        \multirow{2}{*}{\alignmultr[table-format=4.1]{26.7}} &
        \multirow{2}{*}{\alignmultr[table-format=4.1]{42.1}} \\
        ~ & $(1.361,3.037,3.037)$ & ~ & ~ & ~ & ~ & ~ \vspace*{.05cm}\\
        \multirow{2}{*}{$3{\,}\tensor*[^{2}]{\mathrm{A}}{_1}$ ($1{\,}\tensor*[^2]{\mathrm{A}}{^\prime_1}$)} & $4.557, 4.557, 4.557$ &
        \multirow{2}{*}{$D_{3h}$} & \multirow{2}{*}{1898}
        & \multirow{2}{*}{\alignmultr[table-format=4.1]{51.9}} & \multirow{2}{*}{\alignmultr[table-format=4.1]{100.6}} &
        \multirow{2}{*}{\alignmultr[table-format=4.1]{100.6}} \\
        ~ & $(2.279,2.279,2.279)$ & ~ & ~ & ~ & ~ & ~ \vspace*{.05cm}\\
        \multirow{2}{*}{$1{\,}\tensor*[^{2}]{\mathrm{A}}{_2}$ ($1{\,}\tensor*[^2]{\mathrm{E}}{^{\prime\prime}}$)} & $4.337, 5.132, 4.337$ &
        \multirow{2}{*}{$C_{2v}$} & \multirow{2}{*}{4286} & 
        \multirow{2}{*}{\alignmultr[table-format=4.1]{52.9}} &
        \multirow{2}{*}{\alignmultr[table-format=4.1]{23.7}} &
        \multirow{2}{*}{\alignmultr[table-format=4.1]{30.2}} \\
        ~ & $(1.771,2.566,2.566)$ & ~ & ~ & ~ & ~ & ~ \vspace*{.05cm}\\
        \midrule\midrule
        \multirow{2}{*}{$1{\,}\tensor*[^{1}]{\mathrm{A}}{_{1}}$ (${\,}\tensor*[^1]{\mathrm{A}}{^\prime_1}$)} & $4.610, 4.610, 4.610$ & \multirow{2}{*}{$D_{3h}$}
        & \multirow{2}{*}{19942} & \multirow{2}{*}{\alignmultr[table-format=4.1]{53.2}} &
        \multirow{2}{*}{\alignmultr[table-format=4.1]{36.6}}&
        \multirow{2}{*}{\alignmultr[table-format=4.1]{36.5}} \\
        ~ & $(2.305,2.305,2.305)$ & ~ & ~ & ~ & ~ & ~ \vspace*{0cm} \\
        \bottomrule\bottomrule
  \end{tabular}
\end{table*}
}
{
\setlength{\tabcolsep}{5pt}
\begin{table*}[tb]
  \centering
  \caption{\label{tab:Rb3statesLinear}Synopsis of \textbf{linear} ($D_{\infty
      h}$) doublet and quartet (ground and excited) states of \ce{Rb3} as well
    as the triplet state of \ce{Rb3+} computed at MRCI(ECP+CPP)/UET15 level of
    theory. Equilibrium structures are given in terms of the internal
    coordinates (perimetric coordinates) introduced in Fig.~\ref{fig:coords}
    and all corresponding energies ($E_{\text{rel}}$) are given relative to
    the $(3\cdot 5s)$--asymptote calculated at the same level of theory. The
    states are labelled according to the $C_{2v}$ IRREPs while the
    corresponding assignment to $D_{\infty h}$ symmetry is given in
    parenthesis. This complements the results of the energy level diagram of
    Fig.~\ref{fig:termscheme}.}
  \begin{tabular}{lcccccc}
    \toprule\toprule
    {\multirow{2}{*}{State ($D_{\infty h}$)}} &
    {$R_{12},R_{23},R_{13}\,[\mathrm{\mathring{A}}]$} &
    {\multirow{2}{*}{$E_{\text{rel}}\,[\mathrm{cm}^{-1}]$}} & {\multirow{2}{*}{$\tilde{\nu}_{\text{symm}}\,[\mathrm{cm}^{-1}]$}} & {\multirow{2}{*}{$\tilde{\nu}_{\text{asymm}}\,[\mathrm{cm}^{-1}]$}} &
    {\multirow{2}{*}{$\tilde{\nu}_{\text{bending}_1}\,[\mathrm{cm}^{-1}]$}} &
    {\multirow{2}{*}{$\tilde{\nu}_{\text{bending}_2}\,[\mathrm{cm}^{-1}]$}} \\
    ~ & $(R_1,R_2,R_3)$ & ~ & ~ & ~ & ~ & ~ \\
    \midrule
    \multirow{2}{*}{$1{\,}\tensor*[^{4}]{\mathrm{A}}{_2}+2{\,}\tensor*[^{4}]{\mathrm{B}}{_1}$\footnote{\label{commentlin}Renner-Teller
        pair with the ${\,}^4\mathrm{B}_1$ state turning out as saddle point
        at this linear configuration}($1{\,}\tensor*[^{4}]{\Pi}{_g}$)} & $4.435, 8.869, 4.435$ & \multirow{2}{*}{4044} & \multirow{2}{*}{\alignmultr[table-format=4.1]{33.7}}
    & \multirow{2}{*}{\alignmultr[table-format=4.1]{41.7}} &
    \multirow{2}{*}{\alignmultr[table-format=4.1]{236.8}} &
    \multirow{2}{*}{\alignmultr[table-format=4.1]{84.7}} \\
    ~ & $(0.000,4.435,4.435)$ & ~ & ~ & ~ & ~ & ~ \vspace*{.05cm} \\
    \multirow{2}{*}{$2{\,}\tensor*[^{4}]{\mathrm{A}}{_1}$
      ($1{\,}\tensor*[^{4}]{\Sigma}{^+_g}$)\footnote{\label{commentPJTRT} As a consequence
        of a combined pseudo Jahn-Teller and Renner-Teller interaction 
        two $\mathrm{A}_1$ states, one of them arising from a $\Pi_u$ state,
        can mix for greater displacements along $D_{\infty h}$
        geometries. This is also the reason for non-degenerate frequencies $\tilde{\nu}_{\text{bending}_{1,2}}$}} & $4.937, 9.874, 4.937$ & \multirow{2}{*}{7442} & \multirow{2}{*}{\alignmultr[table-format=4.1]{398.2}}
    & \multirow{2}{*}{\alignmultr[table-format=4.1]{48.9}} &
    \multirow{2}{*}{\alignmultr[table-format=4.1]{282.6}} &
    \multirow{2}{*}{\alignmultr[table-format=4.1]{282.6}} \\
    ~ & $(0.000,4.937,4.937)$ & ~ & ~ & ~ & ~ & ~ \vspace*{.05cm} \\
    \midrule
    \multirow{2}{*}{$1{\,}\tensor*[^{2}]{\mathrm{A}}{_1}$ ($1{\,}\tensor*[^{2}]{\Sigma}{^+_g}$)} & $4.795, 9.590, 4.795$ & \multirow{2}{*}{-2109} &
    \multirow{2}{*}{\alignmultr[table-format=4.1]{24.9}} &
    \multirow{2}{*}{\alignmultr[table-format=4.1]{60.8}} &
    \multirow{2}{*}{\alignmultr[table-format=4.1]{4.1}} &
    \multirow{2}{*}{\alignmultr[table-format=4.1]{4.1}} \\
    ~ & $(0.000,4.795,4.795)$ & ~ & ~ & ~ & ~ & ~ \vspace*{.05cm} \\
    \multirow{2}{*}{$3{\,}\tensor*[^{2}]{\mathrm{A}}{_1}$
      ($2{\,}\tensor*[^{2}]{\Sigma}{^+_g}$)\footref{commentPJTRT}} & $4.440, 8.880, 4.440$ & \multirow{2}{*}{3390}
    & \multirow{2}{*}{\alignmultr[table-format=4.1]{405.9}} &
    \multirow{2}{*}{\alignmultr[table-format=4.1]{46.9}} &
    \multirow{2}{*}{\alignmultr[table-format=4.1]{313.8}} &
    \multirow{2}{*}{\alignmultr[table-format=4.1]{350.3}} \\
    ~ & $(0.000,4.440,4.440)$ & ~ & ~ & ~ & ~ & ~ \vspace*{.05cm} \\
    \multirow{2}{*}{$3{\,}\tensor*[^{2}]{\mathrm{B}}{_1}$ ($2{\,}\tensor*[^{2}]{\Sigma}{^+_u}$) } & $4.930, 9.860, 4.930$ & \multirow{2}{*}{5647} & \multirow{2}{*}{\alignmultr[table-format=4.1]{27.7}}
    & \multirow{2}{*}{\alignmultr[table-format=4.1]{48.4}} &
    \multirow{2}{*}{\alignmultr[table-format=4.1]{169.6}} &
    \multirow{2}{*}{\alignmultr[table-format=4.1]{169.6}} \\
    ~ & $(0.000,4.930,4.930)$ & ~ & ~ & ~ & ~ & ~ \vspace*{.05cm}\\
    \midrule\midrule
    \multirow{2}{*}{$1{\,}\tensor*[^{3}]{\mathrm{B}}{_{1}}$ ($\tensor*[^{3}]{\Sigma}{^+_u}$)} & $4.875, 9.749,
    4.875$ & \multirow{2}{*}{24043} &
    \multirow{2}{*}{\alignmultr[table-format=4.1]{30.3}} &
    \multirow{2}{*}{\alignmultr[table-format=4.1]{49.7}} &
    \multirow{2}{*}{\alignmultr[table-format=4.1]{6.3}} &
    \multirow{2}{*}{\alignmultr[table-format=4.1]{6.3}} \\
    ~ & $(0.000,4.875,4.875)$ & ~ & ~ & ~ & ~ & ~ \vspace*{.05cm}\\
    \bottomrule\bottomrule
  \end{tabular}
\end{table*}
}

Our results are in good agreement with previous theoretical studies of
\ce{Rb3}. For the quartet ground state Sold\'{a}n~\cite{SoldanPESRb3} found
equilateral bond distances $b$ with $b=\SI{5.450}{\angstrom}$, using the
RHF-RCCSD(T) approach with a [16s13p8d5f3g] basis and the small-core ECP from
Ref.~\cite{SmallECPStoll} (ECP28MDF). The energy of the minimum was determined
at $E_{\text{min}}=-1071\,\mathrm{cm}^{-1}$. Hauser et
al.~\cite{HauserJTK3Rb3,HauserPCCP1} found for the equilateral bond distances
$b=\SI{5.500}{\angstrom}$ with corresponding energy
$E_{\text{min}}=-939\,\mathrm{cm}^{-1}$ and harmonic frequencies
$\lbrace\tilde{\nu}_{D_{3h}},\tilde{\nu}_{C_{2v}},\tilde{\nu}_{C_s}\rbrace =
\lbrace 18,21,21\rbrace\,\mathrm{cm}^{-1}$ using RHF-RCCSD(T) with the
ECP28MDF small-core ECP and the corresponding original basis set augmented by
a $(1s,1p,1d)$ set of diffuse functions. In comparison, our computations give
a binding energy of $-1244\,\mathrm{cm}^{-1}$, equilateral bond distances of
$5.311\,\si{\angstrom}$ and vibrational frequencies of
$\lbrace\tilde{\nu}_{D_{3h}},\tilde{\nu}_{C_{2v}},\tilde{\nu}_{C_s}\rbrace =
\lbrace 23.8, 23.6, 23.6\rbrace\,\mathrm{cm}^{-1}$. In case of the doublet
ground state Hauser et al.~\cite{HauserJTK3Rb3,HauserJTK3Rb3Doublets} obtained
bond distances with $R_{12}=R_{13}=\SI{4.387}{\angstrom}$,
$R_{23}=\SI{5.575}{\angstrom}$ and the equilibrium energy
$E_{\text{min}}=-5321\,\mathrm{cm}^{-1}$ using RHF-UCCSD(T), small-core ECP
and a [14s,11p,6d,3f,1g] uncontracted even-tempered basis set derived from the
ECP28MDF basis. Our calculations result in bond distances with
$R_{12}=R_{13}=\SI{4.379}{\angstrom}$ and $R_{23}=\SI{5.393}{\angstrom}$ with
a corresponding binding energy of $-6017\,\mathrm{cm}^{-1}$. Moreover, we can
extract the vertical transition energy from the quartet ground state to the
high-spin $2{\,}\tensor*[^{4}]{\mathrm{E}}{^\prime}$ manifold from
Fig.~\ref{fig:termscheme} and Tab.~\ref{tab:Rb3statesTriangular} and compare
the result with the one calculated by Hauser et
al.~\cite{HauserJTK3Rb3,HauserPCCP1} using a modified version of CASPT2
(referred to as RS2C in \textsc{Molpro}), the same small-core ECP as well as
the same basis set as described before. Our result is
$E_{2{\,}\tensor*[^{4}]{\mathrm{E}}{^\prime}\leftarrow
  1{\,}\tensor*[^{4}]{\mathrm{A}}{_2^\prime}}=11535\,\mathrm{cm}^{-1}$
compared to the result $11530\,\mathrm{cm}^{-1}$ of Hauser et al. The
corresponding experimental value~\cite{NaglTrimesHeDroplets,HauserPCCP2} is
$11510\,\mathrm{cm}^{-1}$ referring to the lowest-energy maximum band of the
measured band spectra applying laser-induced fluorescence (LIF) spectroscopy
to \ce{Rb3} clusters formed on helium nanodroplets.

In the supplementary material~\cite{Supplementary} in Tabs.~\ref{S-tab:Rb3statesTriangularSupp}
and~\ref{S-tab:Rb3statesLinearSupp} and Fig.~\ref{S-fig:termschemeSupp} we are
providing a more detailed overview on all states, i.e. by including saddle
points, obtained within the energy range up to the $5s+2\cdot 5p$
asymptote. Some of the saddle points define the barrier heights between minima
on PESs. This becomes important for analyzing the JT effect.

\subsection{\label{subsec:SOCSurvey}Survey of Spin-Orbit Coupling Effects}
\begin{figure}[tb]
  \centering
  \includegraphics[width=\columnwidth]{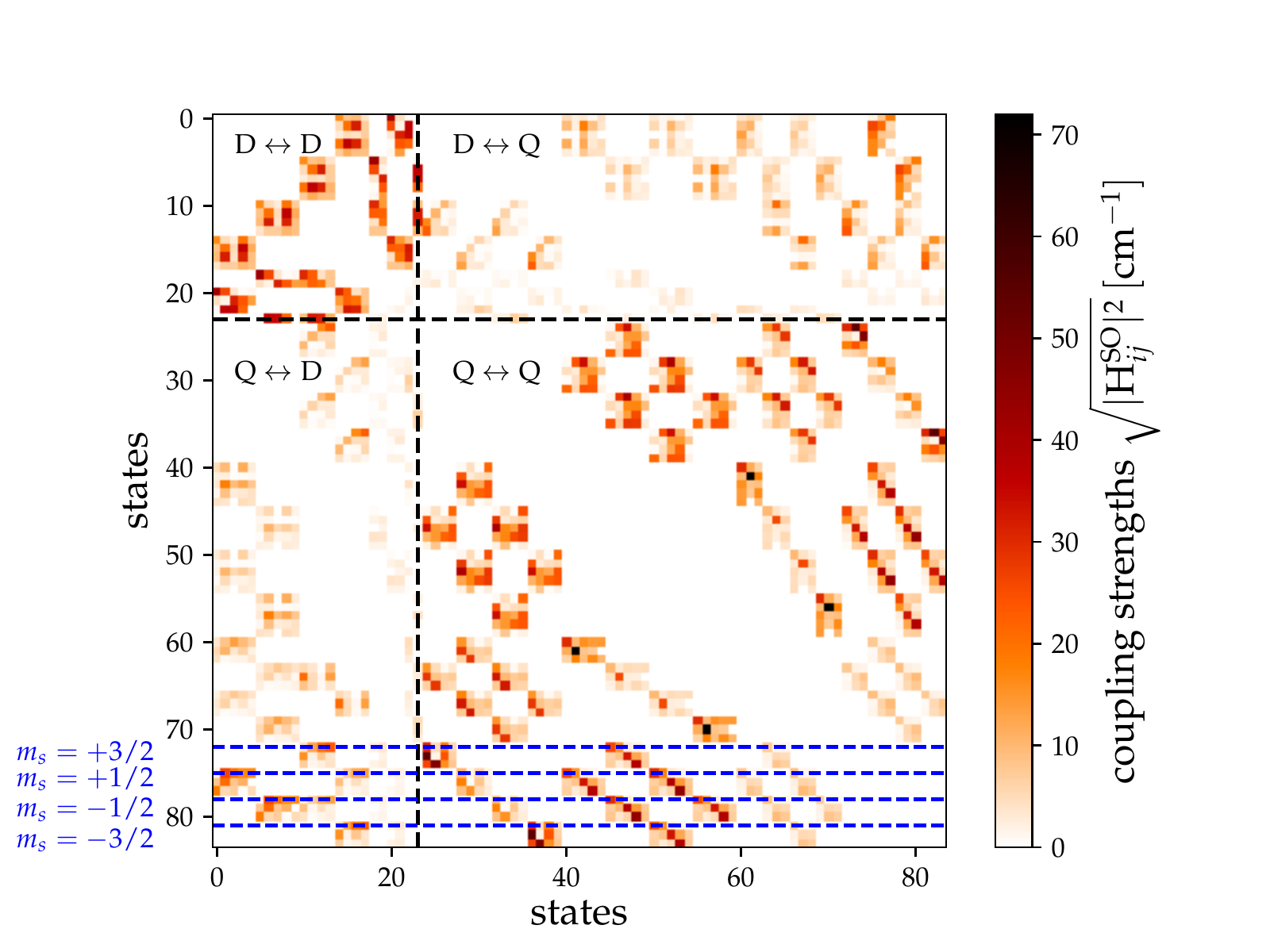}
  \caption{\label{fig:SOHeatMap}(Color online) Heat-map representation of the absolute values
    of the spin-orbit matrix $\sqrt{|\hat{H}^{\text{SO}}_{ij}|^2}$ (without
    diagonal elements). The \ce{Rb3} geometry was fixed to the equilibrium
    configuration of the first excited quartet state
    $1{\,}\tensor*[^4]{\mathrm{A}}{_2}$ (see
    Tab.~\ref{tab:Rb3statesTriangular}). The partitioning separates
    doublet-doublet (D$\leftrightarrow$D), quartet-quartet
    (Q$\leftrightarrow$Q) and quartet-doublet (Q$\leftrightarrow$D)
    couplings. The dashed blue lines mark the rows where the four components
    corresponding to the $1{\,}\tensor*[^4]{\mathrm{A}}{_2}$ state are found
    in the SO-matrix. The SO matrix is sorted according to IRREPs ($C_{2v}$)
    in the sequence
    ($\mathrm{A}_1$/$\mathrm{B}_1$/$\mathrm{B}_2$/$\mathrm{A}_2$) where each
    of them (in zeroth order basis) are ordered with respect to increasing
    energy and are accordingly combined with the $m_s$ spin function starting
    from $m_s=+\nicefrac{1}{2}$ to $m_s=-\nicefrac{1}{2}$ for doublets and
    $m_s=+\nicefrac{3}{2}$ to $m_s=-\nicefrac{3}{2}$ for quartets. This is a
    complete representation with respect to the energetically lowest 12
    doublet (5/4/2/1) and 15 quartet states (4/5/3/3) leading to the $84\times
    84$ SO-matrix.}
\end{figure}
Spin-orbit coupling (SOC) is still a comparatively weak effect for \ce{Rb}
(the SOC induced splitting of the atomic $\tensor*[^2]{\mathrm{P}}{}$ state is
$\approx 240\,\mathrm{cm}^{-1}$) and the classification of states in terms of
their total spin, as in the previous sections, is justified. Nevertheless, in
particular in the vicinity of degeneracies, SOC can lead to a mixing of states
of the same or of different spin. To get an idea of the importance of this
phenomenon we have investigated the size of the couplings for selected nuclear
configurations at the MRCI(ECP+CPP)/UET15 level of theory using the ECP-LS
technique for the corresponding large-core pseudopotential. All important
details about the computation of the corresponding spin-orbit matrix
based on a pseudopotential approach can be found, e.g., in Refs.~\cite{LargeCoreECP,ECPsDolg}. The
computations included 15 quartet (4/5/3/3) and 12 doublet (5/4/2/1) states,
according to the \textsc{Molpro} specific ordering of the IRREPs
($\mathrm{A}_1$/$\mathrm{B}_1$/$\mathrm{B}_2$/$\mathrm{A}_2$). That is, in
total a $84\times 84$ SO-matrix is set up and diagonalized.

To get a qualitative overview we show in Fig.~\ref{fig:SOHeatMap} the absolute
values of the SO-matrix $|\hat{H}^{\text{SO}}_{ij}|$ at the equilibrium
geometry of the first excited quartet state
$1{\,}\tensor*[^4]{\mathrm{A}}{_2}$ as a heat-map representation. It should
look similar for comparable geometrical configurations. The main contributions
come from doublet-doublet ($\mathrm{D}\leftrightarrow\mathrm{D}$),
respectively quartet-quartet ($\mathrm{Q}\leftrightarrow\mathrm{Q}$)
couplings. However, there are also non-vanishing couplings between quartet and
doublet states ($\mathrm{Q}\leftrightarrow\mathrm{D}$ and vice versa). The
corresponding selection rules (for $C_{2v}$ configurations), deduced from
group theory, allow for $\Delta S = 0,\pm 1$ and couplings between all
combinations of IRREPs except the same (a detailed derivation is given in the
supplementary material~\cite{Supplementary}).

The explicit values for resulting shifts and zero-field splittings (i.e. the
lifting of degenerate states in the absence of a magnetic field) are given in
Tabs.~\ref{S-tab:SOCRb3Q} to~\ref{S-tab:SOCRb3DLin} in the supplementary
material~\cite{Supplementary} for the equilibrium states listed in
Tabs.~\ref{tab:Rb3statesTriangular} and~\ref{tab:Rb3statesLinear} together
with the corresponding most dominantly coupling states. Typical coupling
strengths amount to 20 to 70 $\mathrm{cm}^{-1}$, as shown in
Fig.~\ref{fig:SOHeatMap}, but the resulting energy shifts and zero-field
splittings are much smaller. For instance the quartet ground state splits into
the two states $\mathrm{E}_{1/2}$ and $\mathrm{E}_{3/2}$ of the $D_{3h}$ spin
double group~\cite{HauserJTK3Rb3}, but the corresponding zero-field splitting
is less than $0.1\,\mathrm{cm}^{-1}$ and the energy lowering induced by the
SOC is less than $0.2\,\mathrm{cm}^{-1}$. The same observation holds for the
first excited quartet state $1{\,}\tensor*[^4]{\mathrm{A}}{_2}$ for which
these SOC effects are again smaller than $1\,\mathrm{cm}^{-1}$. The reason for
these small values lies in the effective quenching of the orbital angular
momentum in triangular geometries and in the energy separation to other
states. For highly symmetric configurations, in particular for linear
geometries and in the presence of spatial degeneracies, the effects become
larger, e.g. for the $1{\,}\tensor*[^4]{\Pi}{_g}$ state, for which splittings
and energy shifts of up to 200 $\mathrm{cm}^{-1}$ are computed.

The strength of SOC, in particular between the quartet ground state
$1{\,}\tensor*[^4]{\mathrm{B}}{_1}$ and the first excited quartet state
$1{\,}\tensor*[^4]{\mathrm{A}}{_2}$, decays with respect to distortions from
equilateral triangular geometries. Only in the limit of dissociation into both
\ce{Rb2}$+$\ce{Rb} and $3\cdot\ce{Rb}$ SOC effects become larger, since
\ce{Rb2} always has a well-defined $C_{\infty}$ axis. To summarize, we do not
expect significant SOC induced mixing of the states in the vicinity of
equilibrium geometries, in particular for the low-lying states
$1{\,}\tensor*[^4]{\mathrm{B}}{_1}$ and $1{\,}\tensor*[^4]{\mathrm{A}}{_2}$.

\section{\label{sec:PAStatesIdentify}Identifying Appropriate States for
  Photoassociation}

\subsection{\label{subsec:ConfigSpace}Configuration Space Survey}
\begin{figure*}[tb]
  \centering
  \begin{minipage}[t]{.49\textwidth}
    \includegraphics[width=.98\linewidth]{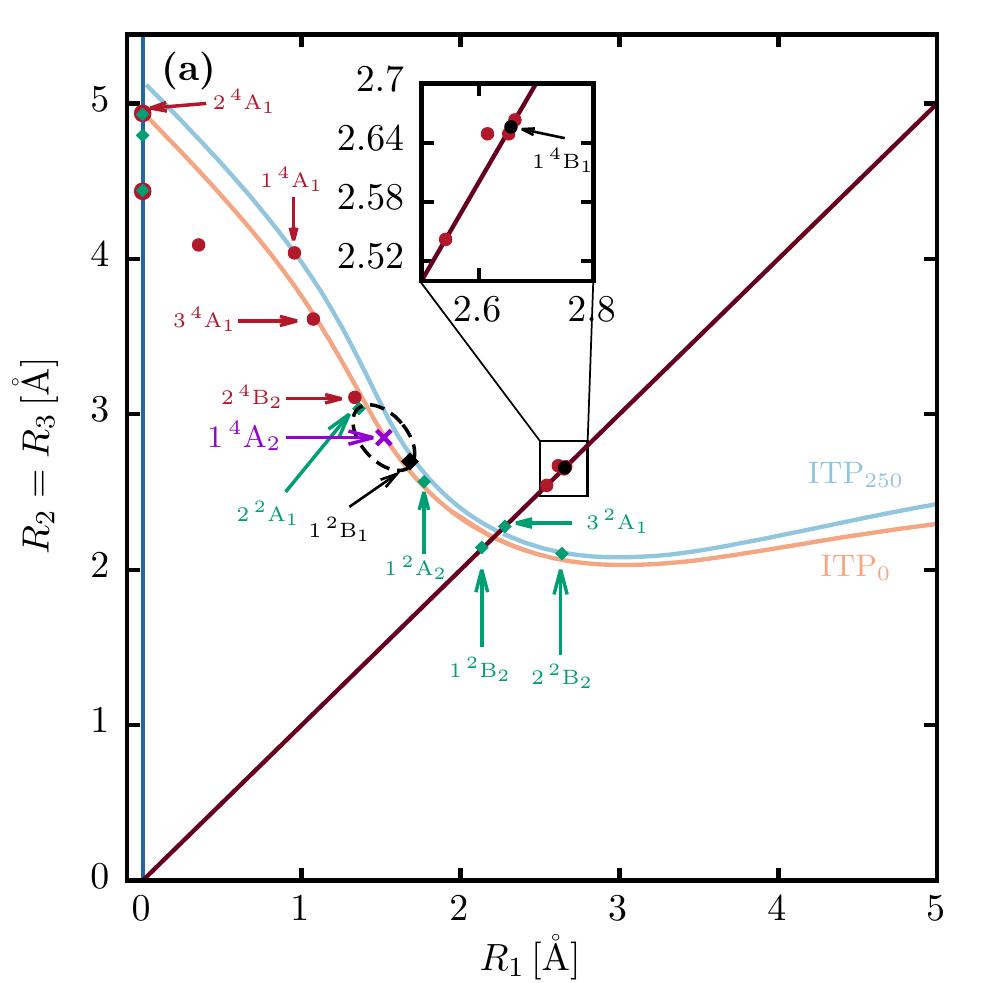}
  \end{minipage}
  \begin{minipage}[t]{.49\textwidth}
    \includegraphics[width=.98\linewidth]{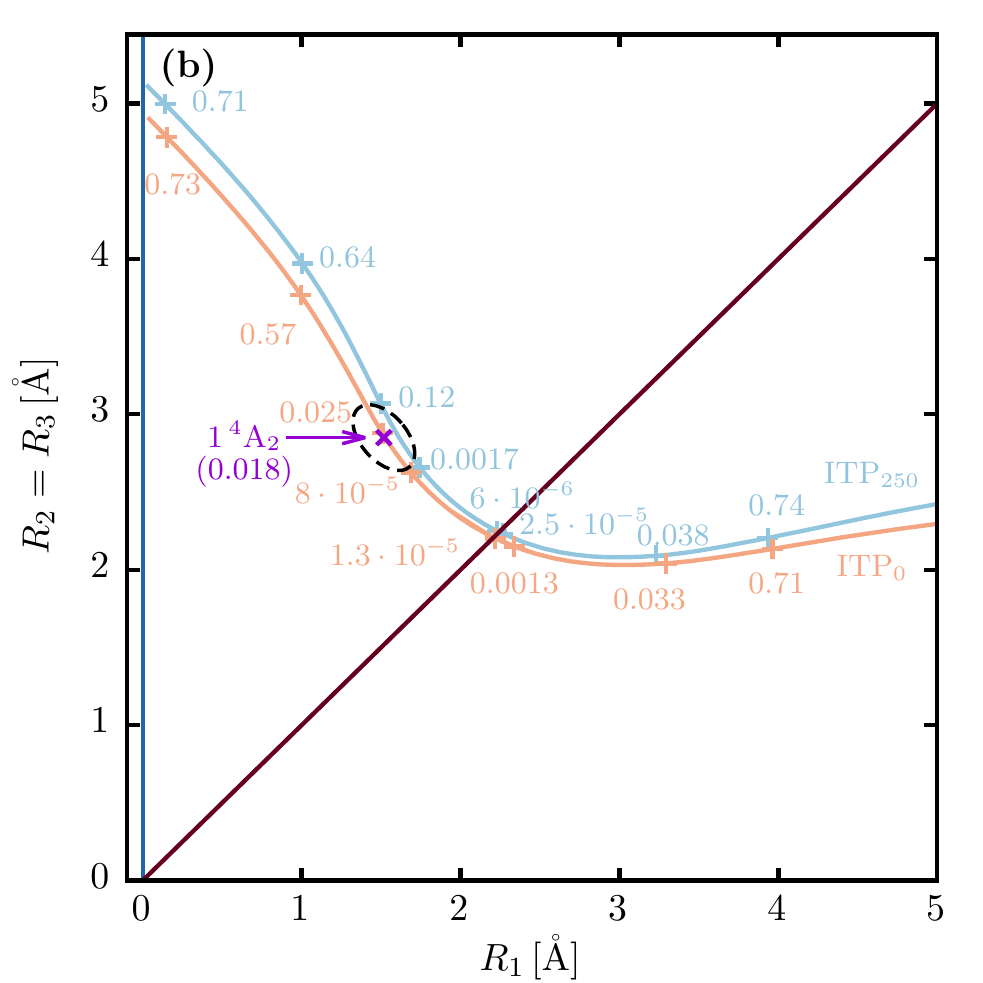}
  \end{minipage}
  \caption{\label{fig:ConfigSpaceSurvey}(Color online) Inner-turning point (ITP) locations
    and locations of equilibrium geometries in the configuration space of
    perimetric coordinates. \textbf{(a)} Since all equilibrium geometries show
    at least $C_{2v}$ symmetry the configuration space survey can be
    restricted to one of the space diagonal surfaces shown in
    Fig.~\ref{fig:coords}~(c). Inner turning points (ITPs) on the quartet
    ground state PES with respect to either the $\ce{Rb2}+\ce{Rb}$ or
    $3\cdot\ce{Rb}$ dissociation scenarios are given in light blue for the
    first one and in light red for the latter. States lying close to this
    lines are promising candidates for showing good Franck-Condon
    factors. Here we focus on the $1{\,}\tensor*[^{4}]{\mathrm{A}}{_2}$ state
    highlighted in purple. The numbers given in \textbf{(b)} along both ITP
    lines represent the electronic dipole transition strengths (in units of
    $[\mathrm{D}^2]$) between the quartet ground state and this
    $1{\,}\tensor*[^{4}]{\mathrm{A}}{_2}$ state at the corrsponding ITP
    locations. Note that the numbers given in \textbf{(b)} do not correlate
    with the equilibrium geometries depicted in \textbf{(a)}. The transition
    dipole strength at the equilibrium geometry of the first excited quartet
    state amounts to $0.018\,\mathrm{D}^2$. The ellipses shown in \textbf{(a)}
    and \textbf{(b)} give an estimate of the size of the vibrational ground
    state wavefunction for the $1{\,}\tensor*[^4]{\mathrm{A}}{_2}$ state.}
\end{figure*}
For the realization of the trimer PA processes, non-vanishing Franck-Condon
factors are required, i.e. a significant overlap of the nuclear scattering
wavefunction of \ce{Rb2}$+$\ce{Rb} or $3\cdot\ce{Rb}$ collisions and the
molecular trimer vibrational wavefunction of the excited state. In this work
we are mostly interested in producing deeply bound trimers close to the
vibrational ground state for reasons of increased stability, lifetime and
simplicity. In fact, as we will show in the following, it turns out that the
equilibrium geometries of a number of excited states are in close proximity to
the inner-turning points (ITPs) of the scattering wavefunction. Since the
scattering wavefunction typically exhibits a local maximum at the ITP this
suggests that favourable Franck-Condon factors might be found for
photoassociating excited trimers in their equilibrium geometry. For trimers,
the ITPs are actually 2D surfaces in the configuration space. They correspond
to those points where the quartet ground state PES equals to the energy of the
scattering state.
For the case of \ce{Rb2}$+$\ce{Rb} this energy is given by the negative
binding energy of the $a{\,}\tensor*[^{3}]{\Sigma}{_u}$ state of \ce{Rb2},
i.e. $\approx -250\,\mathrm{cm}^{-1}$, and for the case of $3\cdot\ce{Rb}$ the
energy is approximately zero. Again, note that PA2 at short distances is
expected to be rather unlikely due to the effective repulsive barrier in the
short-range of the three-body potential,
cf. Ref.~\cite{PhysRevLett.108.263001}. Nevertheless, at large distances PA2
should be possible. The feasibility for PA1 is shown in
Ref.~\cite{Rios2015}. The locations of the ITPs (i.e. $\mathrm{ITP}_{250}$ and
$\mathrm{ITP}_0$) and the positions of the equilibrium geometries are shown in
Fig.~\ref{fig:ConfigSpaceSurvey}~(a). The equilibrium geometries have at least
$C_{2v}$ symmetry and are located on a space diagonal surface, as shown in
Fig.~\ref{fig:coords}~(c) (due to the threefold degeneracy of $C_{2v}$ and
$D_{\infty h}$ configurations, resulting from the indistinguishability of the
three \ce{Rb} atoms, there are three equivalent such representations).

\subsection{Electronic Dipole Transition Moments}
A successful realization of PA processes also requires non-vanishing
electronic dipole transition moments between the initial state and the
corresponding excited state. In $C_{2v}$ symmetry electronic dipole
transitions between all states (with $\Delta S = 0$) are allowed, except
transitions between $\mathrm{A}_1$ and $\mathrm{A}_2$ as well as
$\mathrm{B}_1$ and $\mathrm{B}_2$ (a detailed derivation of this as well as
for the selection rules in $D_{3h}$ is given in the supplementary
material~\cite{Supplementary}). Due to the facts that the density of states increases with
increasing energy and that the transition between the quartet ground state and
the first excited quartet state ($1{\,}\tensor*[^{4}]{\mathrm{A}}{_2}$) is
symmetry-allowed and in close proximity to the ITP lines, we are going to
focus our following investigations on this state.

We study the specific electronic dipole transition strengths (in units of
$[\mathrm{D}^2]$) at ITP configurations in
Fig.~\ref{fig:ConfigSpaceSurvey}~(b). The magnitude of the electronic dipole
transition strengths between the quartet ground state and the first excited
quartet state, $1{\,}\tensor*[^4]{\mathrm{A}}{_2}$, are approximately the same
for $\mathrm{ITP}_{250}$ and $\mathrm{ITP}_0$. In both cases we obtained no
considerable changes in $C_s$ direction. In the vicinity of $D_{3h}$
configurations (diagonal dark red line) we obtain vanishing transition
strengths due to the fact that for $D_{3h}$ geometries the
$1{\,}\tensor*[^{4}]{\mathrm{A}}{_2}$ state forms a degenerate
$1{\,}\tensor*[^{4}]{\mathrm{E}}{^{\prime\prime}}$ JT state (see
Sec.~\ref{subsec:1Epp} for a detailed discussion) where the quartet ground
state is described in terms of the $\mathrm{A}_2^\prime$ IRREP. In the
supplementary material~\cite{Supplementary} we show that electronic dipole transitions between
these states are zero by symmetry. For $C_{2v}$ configurations admixture of
other configurations makes the transition dipole moment non-vanishing, but it
remains rather small.

Using the harmonic vibrational frequencies in
Tab.~\ref{tab:Rb3statesTriangular}, and the topology of the PES in
Fig.~\ref{fig:2DCuts1B1and1A2}~(b) we can estimate the extent of the vibrational
ground state wavefunction for the $1{\,}\tensor*[^4]{\mathrm{A}}{_2}$
state. For each normal mode $i$ the size is approximated by the harmonic
oscillator length. It can be derived from the one-dimensional Schrödinger
equation of a particle of reduced mass $\mu$ (for homonuclear triatomics
$\mu=m/\sqrt{3}$) moving in a harmonic potential, yielding (for
${\,}^{87}\mathrm{Rb}$)
\begin{align}
  \label{OscLengths}
  x_i &= \sqrt{\frac{\hbar}{\mu\omega_i}} =
  \sqrt{\frac{\sqrt{3}\hbar}{100\cdot m({\,}^{87}\mathrm{Rb})\cdot c\tilde{\nu}_i}}\,.
\end{align}
The PES in this region takes on the form of a rotated ellipse with
semi-major axis $a=\SI{0.495}{\angstrom}$ and semi-minor axis
$b=\SI{0.29}{\angstrom}$ calculated from Eq.~\eqref{OscLengths} using
$\tilde{\nu}_{D_{3h}}$ and $\tilde{\nu}_{C_{2v}}$. These findings are
indicated in Fig.~\ref{fig:ConfigSpaceSurvey}. Since there is a
good overlap with the ITPs, a sizeable Franck-Condon factor can be expected.

\subsection{\label{subsec:1Epp}The
  $1{\,}\tensor*[^{4}]{\mathrm{E}}{^{\prime\prime}}$ Jahn-Teller Pair} 
\begin{figure*}[tb]
  \centering
  \begin{minipage}[t]{.49\textwidth}
    \includegraphics[width=\linewidth]{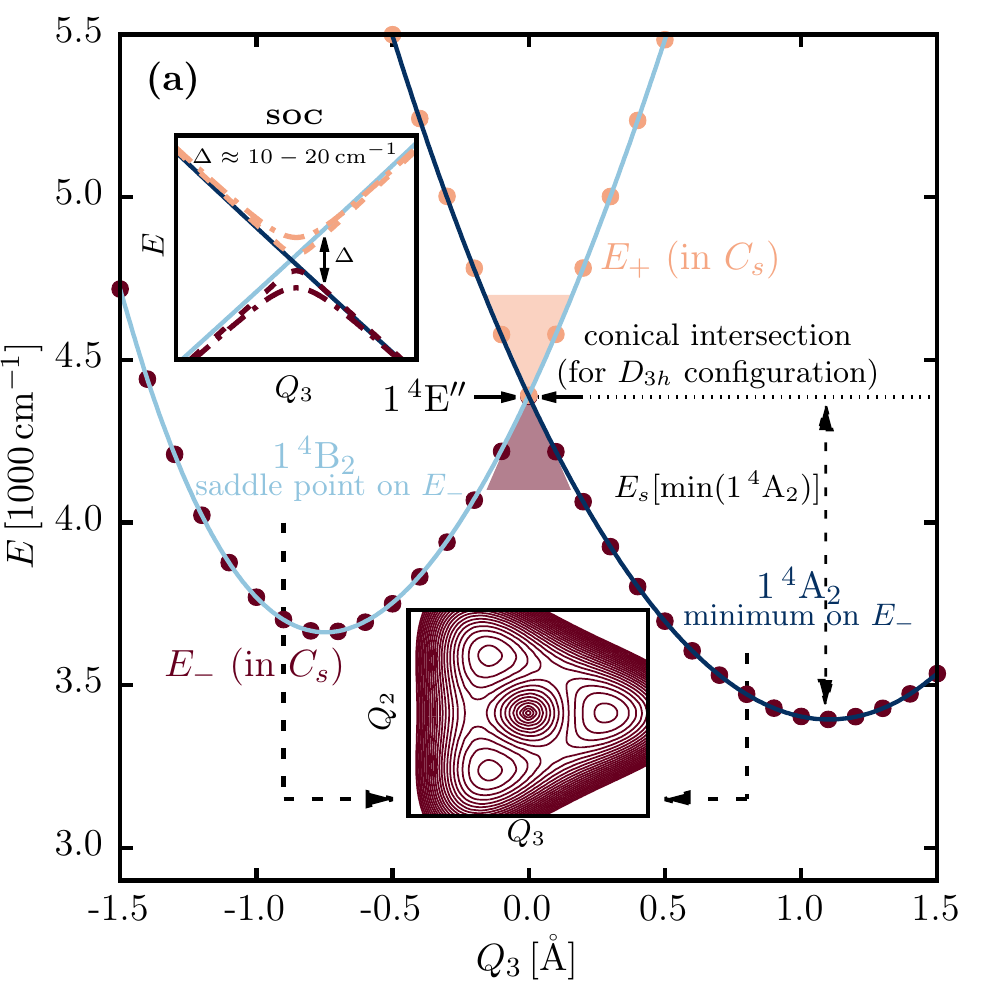}
  \end{minipage}
    \begin{minipage}[t]{.49\textwidth}
    \includegraphics[width=8.6cm,height=8.6cm]{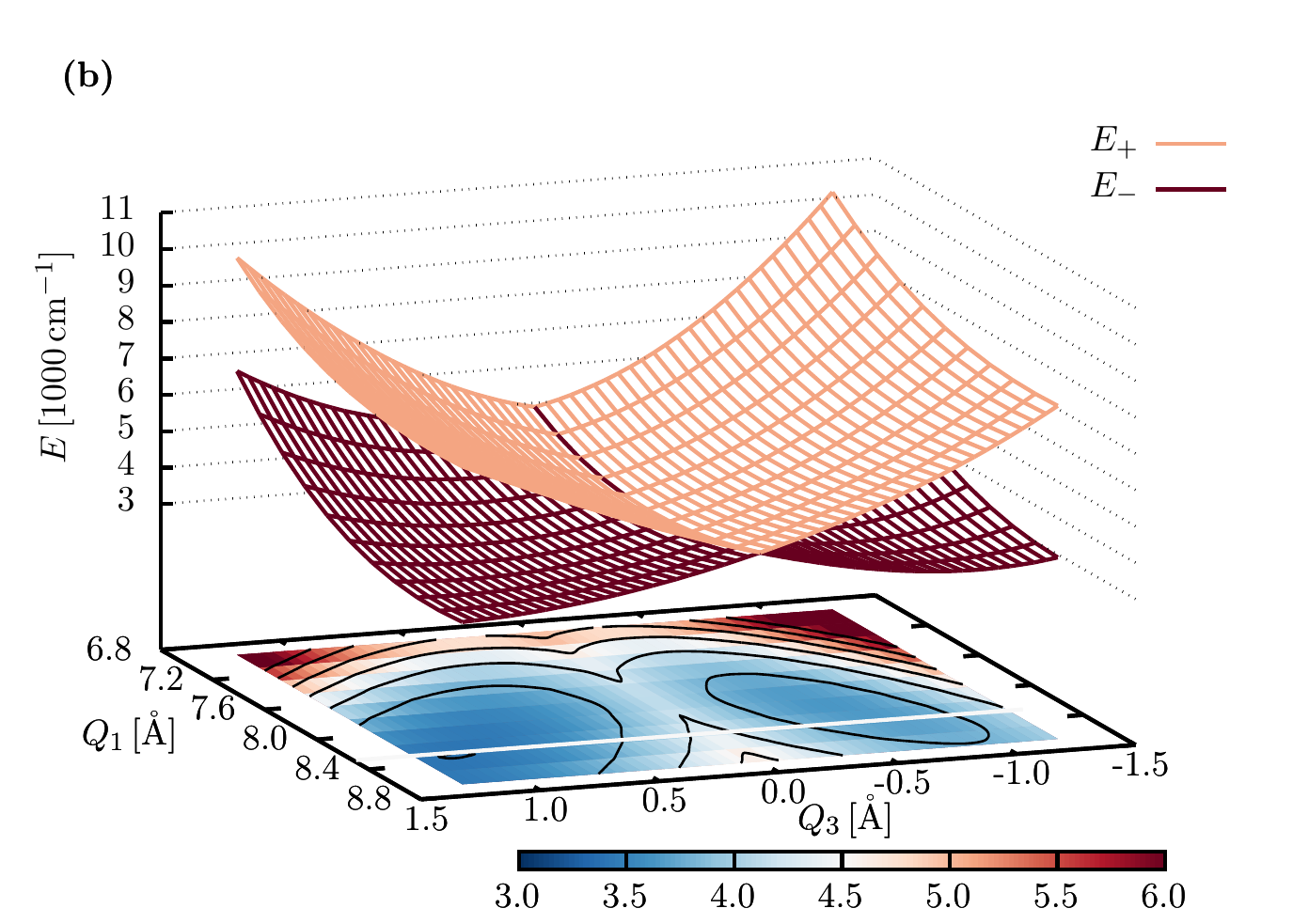}
  \end{minipage}
    \caption{\label{fig:Quartet1EppJT}(Color online) The quartet states (according to
      $C_{2v}$ nomenclature) $1{\,}^4\mathrm{B}_2$ and $1{\,}^4\mathrm{A}_2$
      forming the Jahn-Teller pair $1{\,}^4\mathrm{E}^{\prime\prime}$ in the
      higher symmetry $D_{3h}$ subspace (equilateral triangle). For each
      equilateral triangular configuration the two states show a conical
      intersection (COIN) leading to a COIN seam (one-dimensional curve) in
      the full 3D configuration space. Lowering the symmetry, e.g. by scanning
      along the $C_{2v}$ preserving coordinate $Q_3$, leads to a splitting of
      both states (see \textbf{(a)} for $Q_1=\SI{8.335}{\angstrom}$ and
      $Q_2=\SI{0.0}{\angstrom}$ fixed). Due to the Jahn-Teller character these
      states cannot be viewed separately. The interactions lead to the
      formation of a lower PES sheet $E_-$ showing tricorn topology (three
      equivalent wells alternating regularly with three saddle points,
      separated by the localization energy
      $E_{\text{loc}}=225\,\mathrm{cm}^{-1}$) and a parabolic shaped upper
      surface $E_{+}$. The $1{\,}\tensor*[^4]{\mathrm{A}}{_2}$ state
      represents a global minimum on $E_-$ while the
      $1{\,}\tensor*[^4]{\mathrm{B}}{_2}$ state is a saddle point on this
      surface. This behaviour is illustrated by the lower inset in
      \textbf{(a)} for $Q_1=\SI{8.335}{\angstrom}$. Including spin-orbit
      coupling (SOC) leads to the annihilation of the COIN and to an energy
      splitting $\Delta$ as shown by the inset on the top left in
      \textbf{(a)}. The topology of the PESs in the two-dimensional subspace
      of $Q_1$ and $Q_3$ is shown in \textbf{(b)}. The degenerate line at
      $Q_3=\SI{0.0}{\angstrom}$, where we have $D_{3h}$ symmetry, corresponds
      to the one-dimensional COIN seam. The white line at the bottom
      represents the one-dimensional cut shown in~\textbf{(a)}.}
\end{figure*}
As indicated in Tab.~\ref{tab:Rb3statesTriangular} the first excited quartet
state $1{\,}\tensor*[^{4}]{\mathrm{A}}{_2}$ forms, together with the
$1{\,}\tensor*[^{4}]{\mathrm{B}}{_2}$ state, for equilateral triangular
geometries the JT pair $1{\,}\tensor*[^{4}]{\mathrm{E}}{^{\prime\prime}}$. The
two states are degenerate for every high-symmetry ($D_{3h}$) nuclear
configuration, thus forming a one-dimensional COIN seam in the
full 3D configuration space as already outlined in
Sec.~\ref{subsec:PESCuts}. When lowering the symmetry (scanning along $Q_2$
and/or $Q_3$) both states branch off forming a lower PES sheet $E_-$ revealing
a tricorn topology with three equivalent minima (of
$1{\,}\tensor*[^4]{\mathrm{A}}{_2}$ character) alternating regularly with
three saddle points (of $1{\,}\tensor*[^4]{\mathrm{B}}{_2}$ character) as
illustrated by the lower inset in
Fig.~\ref{fig:Quartet1EppJT}~(a). The upper surface $E_+$ is a
paraboloid of revolution about $Q_2=Q_3=0$~\cite{Cocchini1988Na3}. Spin-orbit
coupling (SOC) removes the COIN with an energy splitting of $\Delta\approx
10-20\,\mathrm{cm}^{-1}$ (i.e. weak SOC) between the corresponding Kramers
pairs of $E_+$ and $E_-$. For details of the underlying (relativistic) JT
theory see, e.g.,
Refs.~\cite{BersukerReview,RochaC3,VibJTExe,RelExEJT,relJTReview,POLUYANOV2008125}
and Refs.~\cite{YarkoniCOINSOC1,YarkoniCOINSOC2,YarkoniCOINSOC3},
respectively, for the effect of SOC on COINs in general. Here it is important
to note that due to the JT interaction the
$1{\,}\tensor*[^{4}]{\mathrm{A}}{_2}$ and
$1{\,}\tensor*[^{4}]{\mathrm{B}}{_2}$ states cannot be viewed separately.
Figure~\ref{fig:Quartet1EppJT}~(b) illustrates the one-dimensional
COIN seam occurring for $Q_2=Q_3=0$ and shows a contour
plot of the trough of the $E_-$--PES in the $D_{3h}$-$C_{2v}$ subspace of
$Q_1$ and $Q_3$. The energetically lowest COIN occurs at
$R_1=R_2=R_3=\SI{2.250}{\angstrom}$ with an energy
$E_{\text{min}(\text{COIN})} = 4146\,\mathrm{cm}^{-1}$ (note that a detailed
overview on all JT pairs is given in Tab.~\ref{S-tab:PJTStates} in the
supplementary material~\cite{Supplementary}). It is convenient~\cite{Cocchini1988Na3} to define a
stabilization energy $E_s$ of the minima on $E_-$ from the COINs as well as a
localization energy $E_{\text{loc}}$ defining the barrier height in the
tricorn potential. In the lower inset of Fig.~\ref{fig:Quartet1EppJT}~(a) this
denotes the energy barrier for transitions between the three equivalent minima
on $E_-$ separated by three saddle points. The stabilization energy
for the cut through the $1{\,}\tensor*[^4]{\mathrm{A}}{_2}$ minimum is
$E_s[\mathrm{min}(1{\,}\tensor*[^4]{\mathrm{A}}{_2})]=991\,\mathrm{cm}^{-1}$
as indicated in Fig.~\ref{fig:Quartet1EppJT}~(a) and clarified in
Fig.~\ref{S-fig:VisualizationEs} of the supplementary material~\cite{Supplementary}. The
localization energy is $E_{\text{loc}}=225\,\mathrm{cm}^{-1}$.

\subsection{\label{subsec:Couplings1Epp}Interactions in the Vicinity of the
  $1{\,}\tensor*[^{4}]{\mathrm{E}}{^{\prime\prime}}$ Global Minimum}
  
 Despite the small transition dipole strengths between the quartet ground state
and the first excited quartet state, discussed above
(cf. Fig.~\ref{fig:ConfigSpaceSurvey}~(b)), we claim that the
$1{\,}\tensor*[^{4}]{\mathrm{A}}{_2}$ state is a promising candidate for PA
experiments. First, its minimum is rather well isolated from intersections
with doublet and quartet states (both in $C_{2v}$ and $C_s$ configuration
space) due to the low density of states. Only the
$2{\,}\tensor*[^{2}]{\mathrm{B}}{_1}$ and
$3{\,}\tensor*[^{2}]{\mathrm{A}}{_1}$ states show intersections, in close
proximity to the $1{\,}\tensor*[^{4}]{\mathrm{A}}{_2}$ minimum, besides the
symmetry-required one with the $1{\,}\tensor*[^{4}]{\mathrm{B}}{_2}$
state. The energetically closest intersection emerges at $C_{2v}$ geometry
with the $3{\,}\tensor*[^2]{\mathrm{A}}{_1}$ state for
$R_2=R_3\approx\SI{2.65}{\angstrom}$ and $R_1\approx\SI{1.7}{\angstrom}$. For
$C_s$ geometries intersections with the $2{\,}\tensor*[^{2}]{\mathrm{B}}{_1}$
state move slightly closer to the minimum of the first excited quartet state
while the $3{\,}\tensor*[^2]{\mathrm{A}}{_1}$ intersections approximately
remain at the same location. However, all intersections are $\gtrsim
60\,\mathrm{cm}^{-1}$ away from the $1{\,}\tensor*[^{4}]{\mathrm{A}}{_2}$
global minimum. The situation is illustrated in
Fig.~\ref{S-fig:IntersecProximity1A2} of the supplementary material~\cite{Supplementary}. As
indicated previously, the $1{\,}\tensor*[^4]{\mathrm{A}}{_2}$ minimum is
stabilized from COINs by
$E_s[\mathrm{min}(1{\,}\tensor*[^4]{\mathrm{A}}{_2})]=991\,\mathrm{cm}^{-1}$.

In the vicinity of the $1{\,}\tensor*[^{4}]{\mathrm{A}}{_2}$ equilibrium
geometry SOC effects are rather small vanishing with $1/r$ for large $r$ (with
$r\propto\sqrt{Q_2^2+Q_3^2}$ measuring the distortion from $D_{3h}$ symmetry
to acute $C_{2v}$ triangular geometries) as follows from relativistic JT
theory~\cite{HauserJTK3Rb3,relJTReview}. Strongest SOCs of the
$1{\,}\tensor*[^4]{\mathrm{A}}{_2}$ state close to its equilibrium geometry are to
the quartet ground state $1{\,}\tensor*[^{4}]{\mathrm{B}}{_1}$ as well as to
the excited state $1{\,}\tensor*[^{4}]{\mathrm{A}}{_1}$ with typical
magnitudes between $30$ to $50\,\mathrm{cm}^{-1}$. The
$1{\,}\tensor*[^4]{\mathrm{A}}{_1}$ state
shows a local minimum at
$E_{\text{min}}(1{\,}\tensor*[^4]{\mathrm{A}}{_1})=6766\,\mathrm{cm}^{-1}$
(cf. Tab.~\ref{tab:Rb3statesTriangular}) and is thus well separated from the
$1{\,}\tensor*[^4]{\mathrm{A}}{_2}$ state. Spin-orbit couplings to doublet
states are slightly weaker with interactions between
$1{\,}\tensor*[^{4}]{\mathrm{A}}{_2}$ and the
$2{\,}\tensor*[^{2}]{\mathrm{A}}{_1}$, $3{\,}\tensor*[^{2}]{\mathrm{A}}{_1}$,
$4{\,}\tensor*[^{2}]{\mathrm{A}}{_1}$, $5{\,}\tensor*[^{2}]{\mathrm{A}}{_1}$,
$3{\,}\tensor*[^{2}]{\mathrm{B}}{_1}$ and
$4{\,}\tensor*[^{2}]{\mathrm{B}}{_1}$ states with orders of $10$ to
$30\,\mathrm{cm}^{-1}$. Those equilibrium states are found either well below
the minimum of the $1{\,}\tensor*[^4]{\mathrm{A}}{_2}$ state at
$229\,\mathrm{cm}^{-1}$ or $1898\,\mathrm{cm}^{-1}$, respectively, or well
above, starting from $5431\,\mathrm{cm}^{-1}$ (the corresponding values are
taken from Tab.~\ref{S-tab:Rb3statesTriangularSupp} of the supplementary
material~\cite{Supplementary}).

\section{\label{sec:Summary}Summary and Outlook}

This work provides a possible roadmap to the experimental realization of PA
processes of single ultracold rubidium trimers. We give a wide-ranging
overview of available states using the MRCI method, together with a large-core
ECP with CPP and a modified even-tempered valence basis-set. By special cuts
through the PESs of both low-- and high-spin species, we revealed their
topology and gave an idea of the mutual position and the expected density of
electronic states. We discussed the prominent feature of the (pseudo)
Jahn-Teller (JT) effect naturally occuring for triangular geometries and outlined
Renner-Teller (combined with pseudo Jahn-Teller) interactions for linear
geometries.
We also provided a survey of SOC effects giving selection rules and showing
that they are weak, particularly for the low-lying states involved in possible
PA schemes.

We studied the equilibrium states as well as the locations of selected inner
turning points (ITPs) on the quartet ground state PES in the configuration
space. Since states lying close to ITPs are promising candidates for good
Franck-Condon factors this analysis helped us to identify suitable states for
PA processes. We focused on the
$1{\,}\tensor*[^{4}]{\mathrm{E}}{^{\prime\prime}}$ state (consisting of the
lowest lying excited states $1{\,}\tensor*[^{4}]{\mathrm{A}}{_2}$ and
$1{\,}\tensor*[^{4}]{\mathrm{B}}{_2}$) for which we investigated the
characteristic JT topology of the corresponding PES and discussed the
consequences of the underlying JT effect. Finally, we investigated the main
coupling effects for the first excited quartet state
($1{\,}\tensor*[^4]{\mathrm{A}}{_2}$), including electronic dipole transition
strengths at ITP geometries, intersections to nearby doublet and quartet
states as well as spin-orbit couplings. This confirms the
$1{\,}\tensor*[^{4}]{\mathrm{A}}{_2}$ state as a promising candidate for PA
experiments.

In a next step we will analyze and fix the breakdown of the Born-Oppenheimer
approximation, connected to the various Jahn-Teller coupling effects, by means
of different diabatization techniques including diabatic PES interpolation
approaches.

Data corresponding to the figures are available in the supplementary material~\cite{Supplementary}.

\begin{acknowledgments}
The research of IQ\textsuperscript{ST} is financially supported by the
Ministry of Science, Research, and Arts Baden-Württemberg. S. R. would like to
acknowledge support from the Deutsche Forschungsgemeinschaft within Project
No. KA 4677/2-1. J. S. and A. K. are grateful to Hermann Stoll for advice on
core-polarization and effective core potentials. The authors would also like
to thank Jos\'{e} P. D'Incao and Paul Julienne for illuminating discussions.
\end{acknowledgments}


\begin{thebibliography}{82}%
\makeatletter
\providecommand \@ifxundefined [1]{%
 \@ifx{#1\undefined}
}%
\providecommand \@ifnum [1]{%
 \ifnum #1\expandafter \@firstoftwo
 \else \expandafter \@secondoftwo
 \fi
}%
\providecommand \@ifx [1]{%
 \ifx #1\expandafter \@firstoftwo
 \else \expandafter \@secondoftwo
 \fi
}%
\providecommand \natexlab [1]{#1}%
\providecommand \enquote  [1]{``#1''}%
\providecommand \bibnamefont  [1]{#1}%
\providecommand \bibfnamefont [1]{#1}%
\providecommand \citenamefont [1]{#1}%
\providecommand \href@noop [0]{\@secondoftwo}%
\providecommand \href [0]{\begingroup \@sanitize@url \@href}%
\providecommand \@href[1]{\@@startlink{#1}\@@href}%
\providecommand \@@href[1]{\endgroup#1\@@endlink}%
\providecommand \@sanitize@url [0]{\catcode `\\12\catcode `\$12\catcode
  `\&12\catcode `\#12\catcode `\^12\catcode `\_12\catcode `\%12\relax}%
\providecommand \@@startlink[1]{}%
\providecommand \@@endlink[0]{}%
\providecommand \url  [0]{\begingroup\@sanitize@url \@url }%
\providecommand \@url [1]{\endgroup\@href {#1}{\urlprefix }}%
\providecommand \urlprefix  [0]{URL }%
\providecommand \Eprint [0]{\href }%
\providecommand \doibase [0]{http://dx.doi.org/}%
\providecommand \selectlanguage [0]{\@gobble}%
\providecommand \bibinfo  [0]{\@secondoftwo}%
\providecommand \bibfield  [0]{\@secondoftwo}%
\providecommand \translation [1]{[#1]}%
\providecommand \BibitemOpen [0]{}%
\providecommand \bibitemStop [0]{}%
\providecommand \bibitemNoStop [0]{.\EOS\space}%
\providecommand \EOS [0]{\spacefactor3000\relax}%
\providecommand \BibitemShut  [1]{\csname bibitem#1\endcsname}%
\let\auto@bib@innerbib\@empty
\bibitem [{\citenamefont {Quemener}\ and\ \citenamefont
  {Julienne}(2012)}]{Quemener2012}%
  \BibitemOpen
  \bibfield  {author} {\bibinfo {author} {\bibfnamefont {G.}~\bibnamefont
  {Quemener}}\ and\ \bibinfo {author} {\bibfnamefont {P.}~\bibnamefont
  {Julienne}},\ }\href {https://pubs.acs.org/doi/abs/10.1021/cr300092g}
  {\bibfield  {journal} {\bibinfo  {journal} {Chem. Rev.}\ }\textbf {\bibinfo
  {volume} {112}},\ \bibinfo {pages} {4949} (\bibinfo {year}
  {2012})}\BibitemShut {NoStop}%
\bibitem [{\citenamefont {Bohn}\ \emph {et~al.}(2017)\citenamefont {Bohn},
  \citenamefont {Rey},\ and\ \citenamefont {Ye}}]{Bohn2017}%
  \BibitemOpen
  \bibfield  {author} {\bibinfo {author} {\bibfnamefont {J.}~\bibnamefont
  {Bohn}}, \bibinfo {author} {\bibfnamefont {A.}~\bibnamefont {Rey}}, \ and\
  \bibinfo {author} {\bibfnamefont {J.}~\bibnamefont {Ye}},\ }\href
  {http://science.sciencemag.org/content/357/6355/1002} {\bibfield  {journal}
  {\bibinfo  {journal} {Science}\ }\textbf {\bibinfo {volume} {357}},\ \bibinfo
  {pages} {1002} (\bibinfo {year} {2017})}\BibitemShut {NoStop}%
\bibitem [{\citenamefont {Balakrishnan}(2016)}]{Balakrishnan2016}%
  \BibitemOpen
  \bibfield  {author} {\bibinfo {author} {\bibfnamefont {N.}~\bibnamefont
  {Balakrishnan}},\ }\href {https://doi.org/10.1063/1.4964096} {\bibfield
  {journal} {\bibinfo  {journal} {J. Chem. Phys.}\ }\textbf {\bibinfo {volume}
  {145}},\ \bibinfo {pages} {150901} (\bibinfo {year} {2016})}\BibitemShut
  {NoStop}%
\bibitem [{\citenamefont {Krems}(2008)}]{Krems2008}%
  \BibitemOpen
  \bibfield  {author} {\bibinfo {author} {\bibfnamefont {R.~V.}\ \bibnamefont
  {Krems}},\ }\href {https://doi.org/10.1039/B802322K} {\bibfield  {journal}
  {\bibinfo  {journal} {Phys. Chem. Chem. Phys.}\ }\textbf {\bibinfo {volume}
  {10}},\ \bibinfo {pages} {4079} (\bibinfo {year} {2008})}\BibitemShut
  {NoStop}%
\bibitem [{\citenamefont {Carr}\ \emph {et~al.}(2009)\citenamefont {Carr},
  \citenamefont {DeMille}, \citenamefont {Krems},\ and\ \citenamefont
  {Ye}}]{Carr2009}%
  \BibitemOpen
  \bibfield  {author} {\bibinfo {author} {\bibfnamefont {L.~D.}\ \bibnamefont
  {Carr}}, \bibinfo {author} {\bibfnamefont {D.}~\bibnamefont {DeMille}},
  \bibinfo {author} {\bibfnamefont {R.~V.}\ \bibnamefont {Krems}}, \ and\
  \bibinfo {author} {\bibfnamefont {J.}~\bibnamefont {Ye}},\ }\href
  {https://doi.org/10.1088/1367-2630/11/5/055049} {\bibfield  {journal}
  {\bibinfo  {journal} {New J. Phys.}\ }\textbf {\bibinfo {volume} {11}},\
  \bibinfo {pages} {055049} (\bibinfo {year} {2009})}\BibitemShut {NoStop}%
\bibitem [{\citenamefont {Doyle}\ \emph {et~al.}(2004)\citenamefont {Doyle},
  \citenamefont {Friedrich}, \citenamefont {Krems},\ and\ \citenamefont
  {Masnou-Seeuws}}]{Doyle2004}%
  \BibitemOpen
  \bibfield  {author} {\bibinfo {author} {\bibfnamefont {J.}~\bibnamefont
  {Doyle}}, \bibinfo {author} {\bibfnamefont {B.}~\bibnamefont {Friedrich}},
  \bibinfo {author} {\bibfnamefont {R.~V.}\ \bibnamefont {Krems}}, \ and\
  \bibinfo {author} {\bibfnamefont {F.}~\bibnamefont {Masnou-Seeuws}},\ }\href
  {https://doi.org/10.1140/epjd/e2004-00151-x} {\bibfield  {journal} {\bibinfo
  {journal} {Eur. Phys. J. D}\ }\textbf {\bibinfo {volume} {31}},\ \bibinfo
  {pages} {149} (\bibinfo {year} {2004})}\BibitemShut {NoStop}%
\bibitem [{\citenamefont {Burt}\ \emph {et~al.}(1997)\citenamefont {Burt},
  \citenamefont {Ghrist}, \citenamefont {Myatt}, \citenamefont {Holland},
  \citenamefont {Cornell},\ and\ \citenamefont {Wieman}}]{Burt1997}%
  \BibitemOpen
  \bibfield  {author} {\bibinfo {author} {\bibfnamefont {E.~A.}\ \bibnamefont
  {Burt}}, \bibinfo {author} {\bibfnamefont {R.~W.}\ \bibnamefont {Ghrist}},
  \bibinfo {author} {\bibfnamefont {C.~J.}\ \bibnamefont {Myatt}}, \bibinfo
  {author} {\bibfnamefont {M.~J.}\ \bibnamefont {Holland}}, \bibinfo {author}
  {\bibfnamefont {E.~A.}\ \bibnamefont {Cornell}}, \ and\ \bibinfo {author}
  {\bibfnamefont {C.~E.}\ \bibnamefont {Wieman}},\ }\href {\doibase
  10.1103/PhysRevLett.79.337} {\bibfield  {journal} {\bibinfo  {journal} {Phys.
  Rev. Lett.}\ }\textbf {\bibinfo {volume} {79}},\ \bibinfo {pages} {337}
  (\bibinfo {year} {1997})}\BibitemShut {NoStop}%
\bibitem [{\citenamefont {Stamper-Kurn}\ \emph {et~al.}(1998)\citenamefont
  {Stamper-Kurn}, \citenamefont {Andrews}, \citenamefont {Chikkatur},
  \citenamefont {Inouye}, \citenamefont {Miesner}, \citenamefont {Stenger},\
  and\ \citenamefont {Ketterle}}]{Stamper1998}%
  \BibitemOpen
  \bibfield  {author} {\bibinfo {author} {\bibfnamefont {D.~M.}\ \bibnamefont
  {Stamper-Kurn}}, \bibinfo {author} {\bibfnamefont {M.~R.}\ \bibnamefont
  {Andrews}}, \bibinfo {author} {\bibfnamefont {A.~P.}\ \bibnamefont
  {Chikkatur}}, \bibinfo {author} {\bibfnamefont {S.}~\bibnamefont {Inouye}},
  \bibinfo {author} {\bibfnamefont {H.-J.}\ \bibnamefont {Miesner}}, \bibinfo
  {author} {\bibfnamefont {J.}~\bibnamefont {Stenger}}, \ and\ \bibinfo
  {author} {\bibfnamefont {W.}~\bibnamefont {Ketterle}},\ }\href {\doibase
  10.1103/PhysRevLett.80.2027} {\bibfield  {journal} {\bibinfo  {journal}
  {Phys. Rev. Lett.}\ }\textbf {\bibinfo {volume} {80}},\ \bibinfo {pages}
  {2027} (\bibinfo {year} {1998})}\BibitemShut {NoStop}%
\bibitem [{\citenamefont {Greene}\ \emph {et~al.}(2017)\citenamefont {Greene},
  \citenamefont {Giannakeas},\ and\ \citenamefont
  {P\'erez-R\'{\i}os}}]{Greene2017}%
  \BibitemOpen
  \bibfield  {author} {\bibinfo {author} {\bibfnamefont {C.~H.}\ \bibnamefont
  {Greene}}, \bibinfo {author} {\bibfnamefont {P.}~\bibnamefont {Giannakeas}},
  \ and\ \bibinfo {author} {\bibfnamefont {J.}~\bibnamefont
  {P\'erez-R\'{\i}os}},\ }\href {\doibase 10.1103/RevModPhys.89.035006}
  {\bibfield  {journal} {\bibinfo  {journal} {Rev. Mod. Phys.}\ }\textbf
  {\bibinfo {volume} {89}},\ \bibinfo {pages} {035006} (\bibinfo {year}
  {2017})}\BibitemShut {NoStop}%
\bibitem [{\citenamefont {Ulmanis}\ \emph {et~al.}(2012)\citenamefont
  {Ulmanis}, \citenamefont {Deiglmayr}, \citenamefont {Repp}, \citenamefont
  {Wester},\ and\ \citenamefont {Weidemüller}}]{Ulmanis2012}%
  \BibitemOpen
  \bibfield  {author} {\bibinfo {author} {\bibfnamefont {J.}~\bibnamefont
  {Ulmanis}}, \bibinfo {author} {\bibfnamefont {J.}~\bibnamefont {Deiglmayr}},
  \bibinfo {author} {\bibfnamefont {M.}~\bibnamefont {Repp}}, \bibinfo {author}
  {\bibfnamefont {R.}~\bibnamefont {Wester}}, \ and\ \bibinfo {author}
  {\bibfnamefont {M.}~\bibnamefont {Weidemüller}},\ }\href {\doibase
  10.1021/cr300215h} {\bibfield  {journal} {\bibinfo  {journal} {Chem. Rev.}\
  }\textbf {\bibinfo {volume} {112}},\ \bibinfo {pages} {4890} (\bibinfo {year}
  {2012})}\BibitemShut {NoStop}%
\bibitem [{\citenamefont {Jones}\ \emph {et~al.}(2006)\citenamefont {Jones},
  \citenamefont {Tiesinga}, \citenamefont {Lett},\ and\ \citenamefont
  {Julienne}}]{Jones2006}%
  \BibitemOpen
  \bibfield  {author} {\bibinfo {author} {\bibfnamefont {K.~M.}\ \bibnamefont
  {Jones}}, \bibinfo {author} {\bibfnamefont {E.}~\bibnamefont {Tiesinga}},
  \bibinfo {author} {\bibfnamefont {P.~D.}\ \bibnamefont {Lett}}, \ and\
  \bibinfo {author} {\bibfnamefont {P.~S.}\ \bibnamefont {Julienne}},\ }\href
  {\doibase 10.1103/RevModPhys.78.483} {\bibfield  {journal} {\bibinfo
  {journal} {Rev. Mod. Phys.}\ }\textbf {\bibinfo {volume} {78}},\ \bibinfo
  {pages} {483} (\bibinfo {year} {2006})}\BibitemShut {NoStop}%
\bibitem [{\citenamefont {K\"ohler}\ \emph {et~al.}(2006)\citenamefont
  {K\"ohler}, \citenamefont {G\'oral},\ and\ \citenamefont
  {Julienne}}]{Koehler2006}%
  \BibitemOpen
  \bibfield  {author} {\bibinfo {author} {\bibfnamefont {T.}~\bibnamefont
  {K\"ohler}}, \bibinfo {author} {\bibfnamefont {K.}~\bibnamefont {G\'oral}}, \
  and\ \bibinfo {author} {\bibfnamefont {P.~S.}\ \bibnamefont {Julienne}},\
  }\href {\doibase 10.1103/RevModPhys.78.1311} {\bibfield  {journal} {\bibinfo
  {journal} {Rev. Mod. Phys.}\ }\textbf {\bibinfo {volume} {78}},\ \bibinfo
  {pages} {1311} (\bibinfo {year} {2006})}\BibitemShut {NoStop}%
\bibitem [{\citenamefont {Chin}\ \emph {et~al.}(2010)\citenamefont {Chin},
  \citenamefont {Grimm}, \citenamefont {Julienne},\ and\ \citenamefont
  {Tiesinga}}]{Chin2010}%
  \BibitemOpen
  \bibfield  {author} {\bibinfo {author} {\bibfnamefont {C.}~\bibnamefont
  {Chin}}, \bibinfo {author} {\bibfnamefont {R.}~\bibnamefont {Grimm}},
  \bibinfo {author} {\bibfnamefont {P.}~\bibnamefont {Julienne}}, \ and\
  \bibinfo {author} {\bibfnamefont {E.}~\bibnamefont {Tiesinga}},\ }\href
  {\doibase 10.1103/RevModPhys.82.1225} {\bibfield  {journal} {\bibinfo
  {journal} {Rev. Mod. Phys.}\ }\textbf {\bibinfo {volume} {82}},\ \bibinfo
  {pages} {1225} (\bibinfo {year} {2010})}\BibitemShut {NoStop}%
\bibitem [{\citenamefont {Dulieu}\ and\ \citenamefont
  {Gabbanini}(2009)}]{Dulieu2009}%
  \BibitemOpen
  \bibfield  {author} {\bibinfo {author} {\bibfnamefont {O.}~\bibnamefont
  {Dulieu}}\ and\ \bibinfo {author} {\bibfnamefont {C.}~\bibnamefont
  {Gabbanini}},\ }\href {\doibase 10.1088/0034-4885/72/8/086401} {\bibfield
  {journal} {\bibinfo  {journal} {Rep. Prog. Phys.}\ }\textbf {\bibinfo
  {volume} {72}},\ \bibinfo {pages} {086401} (\bibinfo {year}
  {2009})}\BibitemShut {NoStop}%
\bibitem [{\citenamefont {Kosloff}(1988)}]{FGH1}%
  \BibitemOpen
  \bibfield  {author} {\bibinfo {author} {\bibfnamefont {R.}~\bibnamefont
  {Kosloff}},\ }\href {https://pubs.acs.org/doi/pdf/10.1021/j100319a003}
  {\bibfield  {journal} {\bibinfo  {journal} {J. Chem. Phys.}\ }\textbf
  {\bibinfo {volume} {92}},\ \bibinfo {pages} {2087} (\bibinfo {year}
  {1988})}\BibitemShut {NoStop}%
\bibitem [{\citenamefont {Dulieu}\ and\ \citenamefont {Julienne}(1995)}]{FGH2}%
  \BibitemOpen
  \bibfield  {author} {\bibinfo {author} {\bibfnamefont {O.}~\bibnamefont
  {Dulieu}}\ and\ \bibinfo {author} {\bibfnamefont {P.~S.}\ \bibnamefont
  {Julienne}},\ }\href {\doibase 10.1063/1.469622} {\bibfield  {journal}
  {\bibinfo  {journal} {J. Chem. Phys.}\ }\textbf {\bibinfo {volume} {103}},\
  \bibinfo {pages} {60} (\bibinfo {year} {1995})}\BibitemShut {NoStop}%
\bibitem [{\citenamefont {Colbert}\ and\ \citenamefont {Miller}(1992)}]{DVR}%
  \BibitemOpen
  \bibfield  {author} {\bibinfo {author} {\bibfnamefont {D.~T.}\ \bibnamefont
  {Colbert}}\ and\ \bibinfo {author} {\bibfnamefont {W.~H.}\ \bibnamefont
  {Miller}},\ }\href {\doibase 10.1063/1.462100} {\bibfield  {journal}
  {\bibinfo  {journal} {J. Chem. Phys.}\ }\textbf {\bibinfo {volume} {96}},\
  \bibinfo {pages} {1982} (\bibinfo {year} {1992})}\BibitemShut {NoStop}%
\bibitem [{\citenamefont {Ferlaino}\ \emph {et~al.}(2011)\citenamefont
  {Ferlaino}, \citenamefont {Zenesini}, \citenamefont {Berninger},
  \citenamefont {Huang}, \citenamefont {N\"agerl},\ and\ \citenamefont
  {Grimm}}]{Ferlaino2011}%
  \BibitemOpen
  \bibfield  {author} {\bibinfo {author} {\bibfnamefont {F.}~\bibnamefont
  {Ferlaino}}, \bibinfo {author} {\bibfnamefont {A.}~\bibnamefont {Zenesini}},
  \bibinfo {author} {\bibfnamefont {M.}~\bibnamefont {Berninger}}, \bibinfo
  {author} {\bibfnamefont {B.}~\bibnamefont {Huang}}, \bibinfo {author}
  {\bibfnamefont {H.-C.}\ \bibnamefont {N\"agerl}}, \ and\ \bibinfo {author}
  {\bibfnamefont {R.}~\bibnamefont {Grimm}},\ }\href {\doibase
  10.1007/s00601-011-0260-7} {\bibfield  {journal} {\bibinfo  {journal}
  {Few-Body Syst.}\ }\textbf {\bibinfo {volume} {51}},\ \bibinfo {pages} {113}
  (\bibinfo {year} {2011})}\BibitemShut {NoStop}%
\bibitem [{\citenamefont {Delacr\'etaz}\ \emph {et~al.}(1986)\citenamefont
  {Delacr\'etaz}, \citenamefont {Grant}, \citenamefont {Whetten}, \citenamefont
  {W\"oste},\ and\ \citenamefont {Zwanziger}}]{Na3ArBeam1986}%
  \BibitemOpen
  \bibfield  {author} {\bibinfo {author} {\bibfnamefont {G.}~\bibnamefont
  {Delacr\'etaz}}, \bibinfo {author} {\bibfnamefont {E.~R.}\ \bibnamefont
  {Grant}}, \bibinfo {author} {\bibfnamefont {R.~L.}\ \bibnamefont {Whetten}},
  \bibinfo {author} {\bibfnamefont {L.}~\bibnamefont {W\"oste}}, \ and\
  \bibinfo {author} {\bibfnamefont {J.~W.}\ \bibnamefont {Zwanziger}},\ }\href
  {\doibase 10.1103/PhysRevLett.56.2598} {\bibfield  {journal} {\bibinfo
  {journal} {Phys. Rev. Lett.}\ }\textbf {\bibinfo {volume} {56}},\ \bibinfo
  {pages} {2598} (\bibinfo {year} {1986})}\BibitemShut {NoStop}%
\bibitem [{\citenamefont {Ernst}\ and\ \citenamefont
  {Rakowsky}(1995)}]{ErnstArBeam1995}%
  \BibitemOpen
  \bibfield  {author} {\bibinfo {author} {\bibfnamefont {W.~E.}\ \bibnamefont
  {Ernst}}\ and\ \bibinfo {author} {\bibfnamefont {S.}~\bibnamefont
  {Rakowsky}},\ }\href {\doibase 10.1002/bbpc.19950990331} {\bibfield
  {journal} {\bibinfo  {journal} {Ber. Bunsenges. Phys. Chem.}\ }\textbf
  {\bibinfo {volume} {99}},\ \bibinfo {pages} {441} (\bibinfo {year}
  {1995})}\BibitemShut {NoStop}%
\bibitem [{\citenamefont {Vituccio}\ \emph {et~al.}(1997)\citenamefont
  {Vituccio}, \citenamefont {Golonzka},\ and\ \citenamefont
  {Ernst}}]{ErnstArBeam1997}%
  \BibitemOpen
  \bibfield  {author} {\bibinfo {author} {\bibfnamefont {D.~T.}\ \bibnamefont
  {Vituccio}}, \bibinfo {author} {\bibfnamefont {O.}~\bibnamefont {Golonzka}},
  \ and\ \bibinfo {author} {\bibfnamefont {W.~E.}\ \bibnamefont {Ernst}},\
  }\href {\doibase https://doi.org/10.1006/jmsp.1997.7272} {\bibfield
  {journal} {\bibinfo  {journal} {J. Mol. Spectrosc.}\ }\textbf {\bibinfo
  {volume} {184}},\ \bibinfo {pages} {237 } (\bibinfo {year}
  {1997})}\BibitemShut {NoStop}%
\bibitem [{\citenamefont {Nagl}\ \emph
  {et~al.}(2008{\natexlab{a}})\citenamefont {Nagl}, \citenamefont {Aub{\"o}ck},
  \citenamefont {Hauser}, \citenamefont {Allard}, \citenamefont {Callegari},\
  and\ \citenamefont {Ernst}}]{NaglTrimesHeDroplets}%
  \BibitemOpen
  \bibfield  {author} {\bibinfo {author} {\bibfnamefont {J.}~\bibnamefont
  {Nagl}}, \bibinfo {author} {\bibfnamefont {G.}~\bibnamefont {Aub{\"o}ck}},
  \bibinfo {author} {\bibfnamefont {A.~W.}\ \bibnamefont {Hauser}}, \bibinfo
  {author} {\bibfnamefont {O.}~\bibnamefont {Allard}}, \bibinfo {author}
  {\bibfnamefont {C.}~\bibnamefont {Callegari}}, \ and\ \bibinfo {author}
  {\bibfnamefont {W.~E.}\ \bibnamefont {Ernst}},\ }\href {\doibase
  10.1063/1.2906120} {\bibfield  {journal} {\bibinfo  {journal} {J. Chem.
  Phys.}\ }\textbf {\bibinfo {volume} {128}},\ \bibinfo {pages} {154320}
  (\bibinfo {year} {2008}{\natexlab{a}})}\BibitemShut {NoStop}%
\bibitem [{\citenamefont {Aub{\"o}ck}\ \emph {et~al.}(2008)\citenamefont
  {Aub{\"o}ck}, \citenamefont {Nagl}, \citenamefont {Callegari},\ and\
  \citenamefont {Ernst}}]{AuboeckRb3K3QuartetsHeDroplets}%
  \BibitemOpen
  \bibfield  {author} {\bibinfo {author} {\bibfnamefont {G.}~\bibnamefont
  {Aub{\"o}ck}}, \bibinfo {author} {\bibfnamefont {J.}~\bibnamefont {Nagl}},
  \bibinfo {author} {\bibfnamefont {C.}~\bibnamefont {Callegari}}, \ and\
  \bibinfo {author} {\bibfnamefont {W.~E.}\ \bibnamefont {Ernst}},\ }\href
  {\doibase 10.1063/1.2976765} {\bibfield  {journal} {\bibinfo  {journal} {J.
  Chem. Phys.}\ }\textbf {\bibinfo {volume} {129}},\ \bibinfo {pages} {114501}
  (\bibinfo {year} {2008})}\BibitemShut {NoStop}%
\bibitem [{\citenamefont {Nagl}\ \emph
  {et~al.}(2008{\natexlab{b}})\citenamefont {Nagl}, \citenamefont {Aub{\"o}ck},
  \citenamefont {Hauser}, \citenamefont {Allard}, \citenamefont {Callegari},\
  and\ \citenamefont {Ernst}}]{NaglHeteroHomoHighSpinTrimersHeDroplets}%
  \BibitemOpen
  \bibfield  {author} {\bibinfo {author} {\bibfnamefont {J.}~\bibnamefont
  {Nagl}}, \bibinfo {author} {\bibfnamefont {G.}~\bibnamefont {Aub{\"o}ck}},
  \bibinfo {author} {\bibfnamefont {A.~W.}\ \bibnamefont {Hauser}}, \bibinfo
  {author} {\bibfnamefont {O.}~\bibnamefont {Allard}}, \bibinfo {author}
  {\bibfnamefont {C.}~\bibnamefont {Callegari}}, \ and\ \bibinfo {author}
  {\bibfnamefont {W.~E.}\ \bibnamefont {Ernst}},\ }\href {\doibase
  10.1103/PhysRevLett.100.063001} {\bibfield  {journal} {\bibinfo  {journal}
  {Phys. Rev. Lett.}\ }\textbf {\bibinfo {volume} {100}},\ \bibinfo {pages}
  {063001} (\bibinfo {year} {2008}{\natexlab{b}})}\BibitemShut {NoStop}%
\bibitem [{\citenamefont {Hauser}\ and\ \citenamefont
  {Ernst}(2011)}]{HauserPCCP1}%
  \BibitemOpen
  \bibfield  {author} {\bibinfo {author} {\bibfnamefont {A.~W.}\ \bibnamefont
  {Hauser}}\ and\ \bibinfo {author} {\bibfnamefont {W.~E.}\ \bibnamefont
  {Ernst}},\ }\href {\doibase 10.1039/C1CP21163C} {\bibfield  {journal}
  {\bibinfo  {journal} {Phys. Chem. Chem. Phys.}\ }\textbf {\bibinfo {volume}
  {13}},\ \bibinfo {pages} {18762} (\bibinfo {year} {2011})}\BibitemShut
  {NoStop}%
\bibitem [{\citenamefont {Giese}\ \emph {et~al.}(2011)\citenamefont {Giese},
  \citenamefont {Stienkemeier}, \citenamefont {Mudrich}, \citenamefont
  {Hauser},\ and\ \citenamefont {Ernst}}]{HauserPCCP2}%
  \BibitemOpen
  \bibfield  {author} {\bibinfo {author} {\bibfnamefont {C.}~\bibnamefont
  {Giese}}, \bibinfo {author} {\bibfnamefont {F.}~\bibnamefont {Stienkemeier}},
  \bibinfo {author} {\bibfnamefont {M.}~\bibnamefont {Mudrich}}, \bibinfo
  {author} {\bibfnamefont {A.~W.}\ \bibnamefont {Hauser}}, \ and\ \bibinfo
  {author} {\bibfnamefont {W.~E.}\ \bibnamefont {Ernst}},\ }\href {\doibase
  10.1039/C1CP21191A} {\bibfield  {journal} {\bibinfo  {journal} {Phys. Chem.
  Chem. Phys.}\ }\textbf {\bibinfo {volume} {13}},\ \bibinfo {pages} {18769}
  (\bibinfo {year} {2011})}\BibitemShut {NoStop}%
\bibitem [{\citenamefont {Martin}\ and\ \citenamefont
  {Davidson}(1978)}]{Davidson1978}%
  \BibitemOpen
  \bibfield  {author} {\bibinfo {author} {\bibfnamefont {R.~L.}\ \bibnamefont
  {Martin}}\ and\ \bibinfo {author} {\bibfnamefont {E.~R.}\ \bibnamefont
  {Davidson}},\ }\href {\doibase 10.1080/00268977800101291} {\bibfield
  {journal} {\bibinfo  {journal} {Mol. Phys.}\ }\textbf {\bibinfo {volume}
  {35}},\ \bibinfo {pages} {1713} (\bibinfo {year} {1978})}\BibitemShut
  {NoStop}%
\bibitem [{\citenamefont {Martins}\ \emph {et~al.}(1983)\citenamefont
  {Martins}, \citenamefont {Car},\ and\ \citenamefont {Buttet}}]{Martins1983}%
  \BibitemOpen
  \bibfield  {author} {\bibinfo {author} {\bibfnamefont {J.~L.}\ \bibnamefont
  {Martins}}, \bibinfo {author} {\bibfnamefont {R.}~\bibnamefont {Car}}, \ and\
  \bibinfo {author} {\bibfnamefont {J.}~\bibnamefont {Buttet}},\ }\href
  {\doibase 10.1063/1.445446} {\bibfield  {journal} {\bibinfo  {journal} {J.
  Chem. Phys.}\ }\textbf {\bibinfo {volume} {78}},\ \bibinfo {pages} {5646}
  (\bibinfo {year} {1983})}\BibitemShut {NoStop}%
\bibitem [{\citenamefont {Thompson}\ \emph {et~al.}(1985)\citenamefont
  {Thompson}, \citenamefont {Izmirlian}, \citenamefont {Lemon}, \citenamefont
  {Truhlar},\ and\ \citenamefont {Mead}}]{Thompson1985Li3Na3K3}%
  \BibitemOpen
  \bibfield  {author} {\bibinfo {author} {\bibfnamefont {T.~C.}\ \bibnamefont
  {Thompson}}, \bibinfo {author} {\bibfnamefont {G.~J.}\ \bibnamefont
  {Izmirlian}}, \bibinfo {author} {\bibfnamefont {S.~J.}\ \bibnamefont
  {Lemon}}, \bibinfo {author} {\bibfnamefont {D.~G.}\ \bibnamefont {Truhlar}},
  \ and\ \bibinfo {author} {\bibfnamefont {C.~A.}\ \bibnamefont {Mead}},\
  }\href {\doibase 10.1063/1.448594} {\bibfield  {journal} {\bibinfo  {journal}
  {J. Chem. Phys.}\ }\textbf {\bibinfo {volume} {82}},\ \bibinfo {pages} {5597}
  (\bibinfo {year} {1985})}\BibitemShut {NoStop}%
\bibitem [{\citenamefont {Cocchini}\ \emph {et~al.}(1988)\citenamefont
  {Cocchini}, \citenamefont {Upton},\ and\ \citenamefont
  {Andreoni}}]{Cocchini1988Na3}%
  \BibitemOpen
  \bibfield  {author} {\bibinfo {author} {\bibfnamefont {F.}~\bibnamefont
  {Cocchini}}, \bibinfo {author} {\bibfnamefont {T.~H.}\ \bibnamefont {Upton}},
  \ and\ \bibinfo {author} {\bibfnamefont {W.}~\bibnamefont {Andreoni}},\
  }\href {\doibase 10.1063/1.454499} {\bibfield  {journal} {\bibinfo  {journal}
  {J. Chem. Phys.}\ }\textbf {\bibinfo {volume} {88}},\ \bibinfo {pages} {6068}
  (\bibinfo {year} {1988})}\BibitemShut {NoStop}%
\bibitem [{\citenamefont {Spiegelmann}\ and\ \citenamefont
  {Pavolini}(1988)}]{Spiegelmann1988NanKn}%
  \BibitemOpen
  \bibfield  {author} {\bibinfo {author} {\bibfnamefont {F.}~\bibnamefont
  {Spiegelmann}}\ and\ \bibinfo {author} {\bibfnamefont {D.}~\bibnamefont
  {Pavolini}},\ }\href {\doibase 10.1063/1.455638} {\bibfield  {journal}
  {\bibinfo  {journal} {J. Chem. Phys.}\ }\textbf {\bibinfo {volume} {89}},\
  \bibinfo {pages} {4954} (\bibinfo {year} {1988})}\BibitemShut {NoStop}%
\bibitem [{\citenamefont {Meiswinkel}\ and\ \citenamefont
  {K\"{o}ppel}(1990)}]{Meiswinkel1990}%
  \BibitemOpen
  \bibfield  {author} {\bibinfo {author} {\bibfnamefont {R.}~\bibnamefont
  {Meiswinkel}}\ and\ \bibinfo {author} {\bibfnamefont {H.}~\bibnamefont
  {K\"{o}ppel}},\ }\href {\doibase
  https://doi.org/10.1016/0301-0104(90)80077-B} {\bibfield  {journal} {\bibinfo
   {journal} {Chem. Phys.}\ }\textbf {\bibinfo {volume} {144}},\ \bibinfo
  {pages} {117} (\bibinfo {year} {1990})}\BibitemShut {NoStop}%
\bibitem [{\citenamefont {de~Vivie-Riedle}\ \emph {et~al.}(1997)\citenamefont
  {de~Vivie-Riedle}, \citenamefont {Gaus}, \citenamefont {Bona{\v
  c}i\'{c}-Kouteck\'{y}}, \citenamefont {Manz}, \citenamefont {Reischl-Lenz},\
  and\ \citenamefont {Saalfrank}}]{TheoreticalStudAbsSpectNa3B1997}%
  \BibitemOpen
  \bibfield  {author} {\bibinfo {author} {\bibfnamefont {R.}~\bibnamefont
  {de~Vivie-Riedle}}, \bibinfo {author} {\bibfnamefont {J.}~\bibnamefont
  {Gaus}}, \bibinfo {author} {\bibfnamefont {V.}~\bibnamefont {Bona{\v
  c}i\'{c}-Kouteck\'{y}}}, \bibinfo {author} {\bibfnamefont {J.}~\bibnamefont
  {Manz}}, \bibinfo {author} {\bibfnamefont {B.}~\bibnamefont {Reischl-Lenz}},
  \ and\ \bibinfo {author} {\bibfnamefont {P.}~\bibnamefont {Saalfrank}},\
  }\href {\doibase https://doi.org/10.1016/S0301-0104(97)00191-2} {\bibfield
  {journal} {\bibinfo  {journal} {Chem. Phys.}\ }\textbf {\bibinfo {volume}
  {223}},\ \bibinfo {pages} {1} (\bibinfo {year} {1997})}\BibitemShut {NoStop}%
\bibitem [{\citenamefont {Hauser}\ \emph {et~al.}(2008)\citenamefont {Hauser},
  \citenamefont {Callegari}, \citenamefont {Sold{\'a}n},\ and\ \citenamefont
  {Ernst}}]{HauserPotassiumDoublets}%
  \BibitemOpen
  \bibfield  {author} {\bibinfo {author} {\bibfnamefont {A.~W.}\ \bibnamefont
  {Hauser}}, \bibinfo {author} {\bibfnamefont {C.}~\bibnamefont {Callegari}},
  \bibinfo {author} {\bibfnamefont {P.}~\bibnamefont {Sold{\'a}n}}, \ and\
  \bibinfo {author} {\bibfnamefont {W.~E.}\ \bibnamefont {Ernst}},\ }\href
  {\doibase 10.1063/1.2956492} {\bibfield  {journal} {\bibinfo  {journal} {J.
  Chem. Phys.}\ }\textbf {\bibinfo {volume} {129}},\ \bibinfo {pages} {044307}
  (\bibinfo {year} {2008})}\BibitemShut {NoStop}%
\bibitem [{\citenamefont {Hauser}\ \emph
  {et~al.}(2010{\natexlab{a}})\citenamefont {Hauser}, \citenamefont
  {Aub{\"o}ck}, \citenamefont {Callegari},\ and\ \citenamefont
  {Ernst}}]{HauserJTK3Rb3}%
  \BibitemOpen
  \bibfield  {author} {\bibinfo {author} {\bibfnamefont {A.~W.}\ \bibnamefont
  {Hauser}}, \bibinfo {author} {\bibfnamefont {G.}~\bibnamefont {Aub{\"o}ck}},
  \bibinfo {author} {\bibfnamefont {C.}~\bibnamefont {Callegari}}, \ and\
  \bibinfo {author} {\bibfnamefont {W.~E.}\ \bibnamefont {Ernst}},\ }\href
  {\doibase 10.1063/1.3394015} {\bibfield  {journal} {\bibinfo  {journal} {J.
  Chem. Phys.}\ }\textbf {\bibinfo {volume} {132}},\ \bibinfo {pages} {164310}
  (\bibinfo {year} {2010}{\natexlab{a}})}\BibitemShut {NoStop}%
\bibitem [{\citenamefont {Hauser}\ \emph
  {et~al.}(2010{\natexlab{b}})\citenamefont {Hauser}, \citenamefont
  {Callegari}, \citenamefont {Sold{\'a}n},\ and\ \citenamefont
  {Ernst}}]{HauserJTK3Rb3Doublets}%
  \BibitemOpen
  \bibfield  {author} {\bibinfo {author} {\bibfnamefont {A.~W.}\ \bibnamefont
  {Hauser}}, \bibinfo {author} {\bibfnamefont {C.}~\bibnamefont {Callegari}},
  \bibinfo {author} {\bibfnamefont {P.}~\bibnamefont {Sold{\'a}n}}, \ and\
  \bibinfo {author} {\bibfnamefont {W.~E.}\ \bibnamefont {Ernst}},\ }\href
  {\doibase https://doi.org/10.1016/j.chemphys.2010.07.025} {\bibfield
  {journal} {\bibinfo  {journal} {Chem. Phys.}\ }\textbf {\bibinfo {volume}
  {375}},\ \bibinfo {pages} {73 } (\bibinfo {year}
  {2010}{\natexlab{b}})}\BibitemShut {NoStop}%
\bibitem [{\citenamefont {Hauser}\ \emph {et~al.}(2015)\citenamefont {Hauser},
  \citenamefont {Pototschnig},\ and\ \citenamefont {Ernst}}]{HauserJTNa3}%
  \BibitemOpen
  \bibfield  {author} {\bibinfo {author} {\bibfnamefont {A.~W.}\ \bibnamefont
  {Hauser}}, \bibinfo {author} {\bibfnamefont {J.~V.}\ \bibnamefont
  {Pototschnig}}, \ and\ \bibinfo {author} {\bibfnamefont {W.~E.}\ \bibnamefont
  {Ernst}},\ }\href {\doibase https://doi.org/10.1016/j.chemphys.2015.07.027}
  {\bibfield  {journal} {\bibinfo  {journal} {Chem. Phys.}\ }\textbf {\bibinfo
  {volume} {460}},\ \bibinfo {pages} {2} (\bibinfo {year} {2015})}\BibitemShut
  {NoStop}%
\bibitem [{\citenamefont {Hauser}\ \emph {et~al.}(2009)\citenamefont {Hauser},
  \citenamefont {Callegari},\ and\ \citenamefont {Ernst}}]{HauserBook2009}%
  \BibitemOpen
  \bibfield  {author} {\bibinfo {author} {\bibfnamefont {A.~W.}\ \bibnamefont
  {Hauser}}, \bibinfo {author} {\bibfnamefont {C.}~\bibnamefont {Callegari}}, \
  and\ \bibinfo {author} {\bibfnamefont {W.~E.}\ \bibnamefont {Ernst}},\
  }\enquote {\bibinfo {title} {Level-structure and magnetic properties from
  one-electron atoms to clusters with delocalized electronic orbitals: Shell
  models for alkali trimers},}\ in\ \href {\doibase
  10.1007/978-90-481-2985-0_10} {\emph {\bibinfo {booktitle} {Advances in the
  Theory of Atomic and Molecular Systems}}},\ \bibinfo {editor} {edited by\ \bibinfo {editor}
  {\bibfnamefont {P.}~\bibnamefont {Piecuch}}, \bibinfo {editor} {\bibfnamefont
  {J.}~\bibnamefont {Maruani}}, \bibinfo {editor} {\bibfnamefont
  {G.}~\bibnamefont {Delgado-Barrio}}, \ and\ \bibinfo {editor} {\bibfnamefont
  {S.}~\bibnamefont {Wilson}}}\ (\bibinfo  {publisher} {Springer Netherlands},\
  \bibinfo {address} {Dordrecht},\ \bibinfo {year} {2009})\ pp.\ \bibinfo
  {pages} {201--215}\BibitemShut {NoStop}%
\bibitem [{\citenamefont {Hauser}(2009)}]{HauserPhD}%
  \BibitemOpen
  \bibfield  {author} {\bibinfo {author} {\bibfnamefont {A.~W.}\ \bibnamefont
  {Hauser}},\ }\emph {\bibinfo {title} {The electronic structure of alkali
  trimers in their doublet and quartet manifolds: shell models and quantum
  chemistry calculations}},\ \href@noop {} {Ph.D. thesis},\ \bibinfo  {school}
  {Technische Universität Graz} (\bibinfo {year} {2009})\BibitemShut {NoStop}%
\bibitem [{\citenamefont {Sold{\'a}n}(2010)}]{SoldanPESRb3}%
  \BibitemOpen
  \bibfield  {author} {\bibinfo {author} {\bibfnamefont {P.}~\bibnamefont
  {Sold{\'a}n}},\ }\href {\doibase 10.1063/1.3455710} {\bibfield  {journal}
  {\bibinfo  {journal} {J. Chem. Phys.}\ }\textbf {\bibinfo {volume} {132}},\
  \bibinfo {pages} {234308} (\bibinfo {year} {2010})}\BibitemShut {NoStop}%
\bibitem [{\citenamefont {P{\'e}rez-R{\'i}os}\ \emph
  {et~al.}(2015)\citenamefont {P{\'e}rez-R{\'i}os}, \citenamefont {Lepers},\
  and\ \citenamefont {Dulieu}}]{Rios2015}%
  \BibitemOpen
  \bibfield  {author} {\bibinfo {author} {\bibfnamefont {J.}~\bibnamefont
  {P{\'e}rez-R{\'i}os}}, \bibinfo {author} {\bibfnamefont {M.}~\bibnamefont
  {Lepers}}, \ and\ \bibinfo {author} {\bibfnamefont {O.}~\bibnamefont
  {Dulieu}},\ }\href {\doibase 10.1103/PhysRevLett.115.073201} {\bibfield
  {journal} {\bibinfo  {journal} {Phys. Rev. Lett.}\ }\textbf {\bibinfo
  {volume} {115}},\ \bibinfo {pages} {073201} (\bibinfo {year}
  {2015})}\BibitemShut {NoStop}%
\bibitem [{\citenamefont {Wang}\ \emph {et~al.}(2012)\citenamefont {Wang},
  \citenamefont {D'Incao}, \citenamefont {Esry},\ and\ \citenamefont
  {Greene}}]{PhysRevLett.108.263001}%
  \BibitemOpen
  \bibfield  {author} {\bibinfo {author} {\bibfnamefont {J.}~\bibnamefont
  {Wang}}, \bibinfo {author} {\bibfnamefont {J.~P.}\ \bibnamefont {D'Incao}},
  \bibinfo {author} {\bibfnamefont {B.~D.}\ \bibnamefont {Esry}}, \ and\
  \bibinfo {author} {\bibfnamefont {C.~H.}\ \bibnamefont {Greene}},\ }\href
  {\doibase 10.1103/PhysRevLett.108.263001} {\bibfield  {journal} {\bibinfo
  {journal} {Phys. Rev. Lett.}\ }\textbf {\bibinfo {volume} {108}},\ \bibinfo
  {pages} {263001} (\bibinfo {year} {2012})}\BibitemShut {NoStop}%
\bibitem [{\citenamefont {Mukherjee}\ and\ \citenamefont
  {Adhikari}(2014)}]{Mukherjee2014}%
  \BibitemOpen
  \bibfield  {author} {\bibinfo {author} {\bibfnamefont {S.}~\bibnamefont
  {Mukherjee}}\ and\ \bibinfo {author} {\bibfnamefont {S.}~\bibnamefont
  {Adhikari}},\ }\href {\doibase
  https://doi.org/10.1016/j.chemphys.2014.05.022} {\bibfield  {journal}
  {\bibinfo  {journal} {Chem. Phys.}\ }\textbf {\bibinfo {volume} {440}},\
  \bibinfo {pages} {106} (\bibinfo {year} {2014})}\BibitemShut {NoStop}%
\bibitem [{\citenamefont {Strauss}\ \emph {et~al.}(2010)\citenamefont
  {Strauss}, \citenamefont {Takekoshi}, \citenamefont {Lang}, \citenamefont
  {Winkler}, \citenamefont {Grimm}, \citenamefont {Hecker~Denschlag},\ and\
  \citenamefont {Tiemann}}]{Strauss2010}%
  \BibitemOpen
  \bibfield  {author} {\bibinfo {author} {\bibfnamefont {C.}~\bibnamefont
  {Strauss}}, \bibinfo {author} {\bibfnamefont {T.}~\bibnamefont {Takekoshi}},
  \bibinfo {author} {\bibfnamefont {F.}~\bibnamefont {Lang}}, \bibinfo {author}
  {\bibfnamefont {K.}~\bibnamefont {Winkler}}, \bibinfo {author} {\bibfnamefont
  {R.}~\bibnamefont {Grimm}}, \bibinfo {author} {\bibfnamefont
  {J.}~\bibnamefont {Hecker~Denschlag}}, \ and\ \bibinfo {author}
  {\bibfnamefont {E.}~\bibnamefont {Tiemann}},\ }\href {\doibase
  10.1103/PhysRevA.82.052514} {\bibfield  {journal} {\bibinfo  {journal} {Phys.
  Rev. A}\ }\textbf {\bibinfo {volume} {82}},\ \bibinfo {pages} {052514}
  (\bibinfo {year} {2010})}\BibitemShut {NoStop}%
\bibitem [{\citenamefont {Salami}\ \emph {et~al.}(2009)\citenamefont {Salami},
  \citenamefont {Bergeman}, \citenamefont {Beser}, \citenamefont {Bai},
  \citenamefont {Ahmed}, \citenamefont {Kotochigova}, \citenamefont {Lyyra},
  \citenamefont {Huennekens}, \citenamefont {Lisdat}, \citenamefont
  {Stolyarov}, \citenamefont {Dulieu}, \citenamefont {Crozet},\ and\
  \citenamefont {Ross}}]{Salami2009}%
  \BibitemOpen
  \bibfield  {author} {\bibinfo {author} {\bibfnamefont {H.}~\bibnamefont
  {Salami}}, \bibinfo {author} {\bibfnamefont {T.}~\bibnamefont {Bergeman}},
  \bibinfo {author} {\bibfnamefont {B.}~\bibnamefont {Beser}}, \bibinfo
  {author} {\bibfnamefont {J.}~\bibnamefont {Bai}}, \bibinfo {author}
  {\bibfnamefont {E.~H.}\ \bibnamefont {Ahmed}}, \bibinfo {author}
  {\bibfnamefont {S.}~\bibnamefont {Kotochigova}}, \bibinfo {author}
  {\bibfnamefont {A.~M.}\ \bibnamefont {Lyyra}}, \bibinfo {author}
  {\bibfnamefont {J.}~\bibnamefont {Huennekens}}, \bibinfo {author}
  {\bibfnamefont {C.}~\bibnamefont {Lisdat}}, \bibinfo {author} {\bibfnamefont
  {A.~V.}\ \bibnamefont {Stolyarov}}, \bibinfo {author} {\bibfnamefont
  {O.}~\bibnamefont {Dulieu}}, \bibinfo {author} {\bibfnamefont
  {P.}~\bibnamefont {Crozet}}, \ and\ \bibinfo {author} {\bibfnamefont {A.~J.}\
  \bibnamefont {Ross}},\ }\href {\doibase 10.1103/PhysRevA.80.022515}
  {\bibfield  {journal} {\bibinfo  {journal} {Phys. Rev. A}\ }\textbf {\bibinfo
  {volume} {80}},\ \bibinfo {pages} {022515} (\bibinfo {year}
  {2009})}\BibitemShut {NoStop}%
\bibitem [{\citenamefont {Drews}\ \emph {et~al.}(2017)\citenamefont {Drews},
  \citenamefont {Dei{\ss}}, \citenamefont {Wolf}, \citenamefont {Tiemann},\ and\
  \citenamefont {Hecker~Denschlag}}]{Drews2017}%
  \BibitemOpen
  \bibfield  {author} {\bibinfo {author} {\bibfnamefont {B.}~\bibnamefont
  {Drews}}, \bibinfo {author} {\bibfnamefont {M.}~\bibnamefont {Dei{\ss}}},
  \bibinfo {author} {\bibfnamefont {J.}~\bibnamefont {Wolf}}, \bibinfo {author}
  {\bibfnamefont {E.}~\bibnamefont {Tiemann}}, \ and\ \bibinfo {author}
  {\bibfnamefont {J.}~\bibnamefont {Hecker~Denschlag}},\ }\href {\doibase
  10.1103/PhysRevA.95.062507} {\bibfield  {journal} {\bibinfo  {journal} {Phys.
  Rev. A}\ }\textbf {\bibinfo {volume} {95}},\ \bibinfo {pages} {062507}
  (\bibinfo {year} {2017})}\BibitemShut {NoStop}%
\bibitem [{\citenamefont {Amiot}\ and\ \citenamefont
  {Verges}(1987)}]{Amiot1987}%
  \BibitemOpen
  \bibfield  {author} {\bibinfo {author} {\bibfnamefont {C.}~\bibnamefont
  {Amiot}}\ and\ \bibinfo {author} {\bibfnamefont {J.}~\bibnamefont {Verges}},\
  }\href {\doibase 10.1080/00268978700100981} {\bibfield  {journal} {\bibinfo
  {journal} {Mol. Phys.}\ }\textbf {\bibinfo {volume} {61}},\ \bibinfo {pages}
  {51} (\bibinfo {year} {1987})}\BibitemShut {NoStop}%
\bibitem [{\citenamefont {Amiot}(1990)}]{Amiot1990}%
  \BibitemOpen
  \bibfield  {author} {\bibinfo {author} {\bibfnamefont {C.}~\bibnamefont
  {Amiot}},\ }\href {\doibase 10.1063/1.459246} {\bibfield  {journal} {\bibinfo
   {journal} {J. Chem. Phys.}\ }\textbf {\bibinfo {volume}
  {93}},\ \bibinfo {pages} {8591} (\bibinfo {year} {1990})}\BibitemShut
  {NoStop}%
\bibitem [{\citenamefont {Amiot}(1986)}]{Amiot1986}%
  \BibitemOpen
  \bibfield  {author} {\bibinfo {author} {\bibfnamefont {C.}~\bibnamefont
  {Amiot}},\ }\href {\doibase 10.1080/00268978600101491} {\bibfield  {journal}
  {\bibinfo  {journal} {Mol. Phys.}\ }\textbf {\bibinfo {volume} {58}},\
  \bibinfo {pages} {667} (\bibinfo {year} {1986})}\BibitemShut {NoStop}%
\bibitem [{\citenamefont {Silberbach}\ \emph {et~al.}(1986)\citenamefont
  {Silberbach}, \citenamefont {Schwerdtfeger}, \citenamefont {Stoll},\ and\
  \citenamefont {Preuss}}]{LargeCoreECP}%
  \BibitemOpen
  \bibfield  {author} {\bibinfo {author} {\bibfnamefont {H.}~\bibnamefont
  {Silberbach}}, \bibinfo {author} {\bibfnamefont {P.}~\bibnamefont
  {Schwerdtfeger}}, \bibinfo {author} {\bibfnamefont {H.}~\bibnamefont
  {Stoll}}, \ and\ \bibinfo {author} {\bibfnamefont {H.}~\bibnamefont
  {Preuss}},\ }\href {\doibase 10.1088/0022-3700/19/5/011} {\bibfield
  {journal} {\bibinfo  {journal} {J. Phys. B: At. Mol. Phys.}\ }\textbf
  {\bibinfo {volume} {19}},\ \bibinfo {pages} {501} (\bibinfo {year}
  {1986})}\BibitemShut {NoStop}%
\bibitem [{Sup()}]{Supplementary}%
  \BibitemOpen
  \href@noop {} {}\bibinfo {note} {See Supplemental Material at \url{} for
  details on technical aspects, which includes Refs.~[52-57], or for more detailed
  tables and figures. Raw data corresponding to
  all figures shown in this work can be found there as well}\BibitemShut {NoStop}%
\bibitem [{\citenamefont {Huzinaga}\ and\ \citenamefont
  {Miguel}(1990)}]{HigherlExp1990}%
  \BibitemOpen
  \bibfield  {author} {\bibinfo {author} {\bibfnamefont {S.}~\bibnamefont
  {Huzinaga}}\ and\ \bibinfo {author} {\bibfnamefont {B.}~\bibnamefont
  {Miguel}},\ }\href {\doibase https://doi.org/10.1016/0009-2614(90)80112-Q}
  {\bibfield  {journal} {\bibinfo  {journal} {Chem. Phys. Lett.}\ }\textbf
  {\bibinfo {volume} {175}},\ \bibinfo {pages} {289} (\bibinfo {year}
  {1990})}\BibitemShut {NoStop}%
\bibitem [{\citenamefont {Huzinaga}\ and\ \citenamefont
  {Klobukowski}(1993)}]{HigherlExp1993}%
  \BibitemOpen
  \bibfield  {author} {\bibinfo {author} {\bibfnamefont {S.}~\bibnamefont
  {Huzinaga}}\ and\ \bibinfo {author} {\bibfnamefont {M.}~\bibnamefont
  {Klobukowski}},\ }\href {\doibase
  https://doi.org/10.1016/0009-2614(93)89323-A} {\bibfield  {journal} {\bibinfo
   {journal} {Chem. Phys. Lett.}\ }\textbf {\bibinfo {volume} {212}},\ \bibinfo
  {pages} {260} (\bibinfo {year} {1993})}\BibitemShut {NoStop}%
\bibitem [{\citenamefont {Tew}\ and\ \citenamefont
  {Klopper}(2006)}]{TewKlopper2006}%
  \BibitemOpen
  \bibfield  {author} {\bibinfo {author} {\bibfnamefont {D.~P.}\ \bibnamefont
  {Tew}}\ and\ \bibinfo {author} {\bibfnamefont {W.}~\bibnamefont {Klopper}},\
  }\href {\doibase 10.1063/1.2338037} {\bibfield  {journal} {\bibinfo
  {journal} {J. Chem. Phys.}\ }\textbf {\bibinfo {volume} {125}},\ \bibinfo
  {pages} {094302} (\bibinfo {year} {2006})}\BibitemShut {NoStop}%
\bibitem [{\citenamefont {Cherkes}\ \emph {et~al.}(2009)\citenamefont
  {Cherkes}, \citenamefont {Klaiman},\ and\ \citenamefont
  {Moiseyev}}]{EvenTemperedBasisSet}%
  \BibitemOpen
  \bibfield  {author} {\bibinfo {author} {\bibfnamefont {I.}~\bibnamefont
  {Cherkes}}, \bibinfo {author} {\bibfnamefont {S.}~\bibnamefont {Klaiman}}, \
  and\ \bibinfo {author} {\bibfnamefont {N.}~\bibnamefont {Moiseyev}},\ }\href
  {\doibase 10.1002/qua.22090} {\bibfield  {journal} {\bibinfo  {journal} {Int.
  J. Quantum Chem.}\ }\textbf {\bibinfo {volume} {109}},\ \bibinfo {pages}
  {2996} (\bibinfo {year} {2009})}\BibitemShut {NoStop}%
\bibitem [{\citenamefont {Linstrom}\ and\ \citenamefont
  {Mallard}(2019)}]{NIST}%
  \BibitemOpen
  \bibfield  {author} {\bibinfo {author} {\bibfnamefont {P.~J.}\ \bibnamefont
  {Linstrom}}\ and\ \bibinfo {author} {\bibfnamefont {W.~G.}\ \bibnamefont
  {Mallard}},\ }\href@noop {} {\enquote {\bibinfo {title} {{NIST Chemistry
  WebBook,} {NIST Standard Reference Database Number 69}},}\ } (\bibinfo {year}
  {2019}),\ \bibinfo {note} {data retrieved
  \url{https://doi.org/10.18434/T4D303}}\BibitemShut {NoStop}%
\bibitem [{\citenamefont {Bishop}(1993)}]{bishop1993group}%
  \BibitemOpen
  \bibfield  {author} {\bibinfo {author} {\bibfnamefont {D.}~\bibnamefont
  {Bishop}},\ }\href {https://books.google.de/books?id=EdoRzQEACAAJ} {\emph
  {\bibinfo {title} {Group Theory and Chemistry}}},\ Dover books on physics and
  chemistry\ (\bibinfo  {publisher} {Dover},\ \bibinfo {year}
  {1993})\BibitemShut {NoStop}%
\bibitem [{\citenamefont {Werner}\ and\ \citenamefont {Knowles}(1988)}]{MRCI1}%
  \BibitemOpen
  \bibfield  {author} {\bibinfo {author} {\bibfnamefont {H.-J.}\ \bibnamefont
  {Werner}}\ and\ \bibinfo {author} {\bibfnamefont {P.~J.}\ \bibnamefont
  {Knowles}},\ }\href {\doibase 10.1063/1.455556} {\bibfield  {journal}
  {\bibinfo  {journal} {J. Chem. Phys.}\ }\textbf {\bibinfo {volume} {89}},\
  \bibinfo {pages} {5803} (\bibinfo {year} {1988})}\BibitemShut {NoStop}%
\bibitem [{\citenamefont {Knowles}\ and\ \citenamefont {Werner}(1988)}]{MRCI2}%
  \BibitemOpen
  \bibfield  {author} {\bibinfo {author} {\bibfnamefont {P.~J.}\ \bibnamefont
  {Knowles}}\ and\ \bibinfo {author} {\bibfnamefont {H.-J.}\ \bibnamefont
  {Werner}},\ }\href {\doibase https://doi.org/10.1016/0009-2614(88)87412-8}
  {\bibfield  {journal} {\bibinfo  {journal} {Chem. Phys. Lett.}\ }\textbf
  {\bibinfo {volume} {145}},\ \bibinfo {pages} {514} (\bibinfo {year}
  {1988})}\BibitemShut {NoStop}%
\bibitem [{\citenamefont {Knowles}\ and\ \citenamefont {Werner}(1992)}]{MRCI3}%
  \BibitemOpen
  \bibfield  {author} {\bibinfo {author} {\bibfnamefont {P.~J.}\ \bibnamefont
  {Knowles}}\ and\ \bibinfo {author} {\bibfnamefont {H.-J.}\ \bibnamefont
  {Werner}},\ }\href {\doibase 10.1007/BF01117405} {\bibfield  {journal}
  {\bibinfo  {journal} {Theor. Chim. Acta}\ }\textbf {\bibinfo {volume} {84}},\
  \bibinfo {pages} {90} (\bibinfo {year} {1992})}\BibitemShut {NoStop}%
\bibitem [{\citenamefont {Werner}\ and\ \citenamefont {Reinsch}(1982)}]{MRCI4}%
  \BibitemOpen
  \bibfield  {author} {\bibinfo {author} {\bibfnamefont {H.-J.}\ \bibnamefont
  {Werner}}\ and\ \bibinfo {author} {\bibfnamefont {E.~A.}\ \bibnamefont
  {Reinsch}},\ }\href {\doibase 10.1063/1.443357} {\bibfield  {journal}
  {\bibinfo  {journal} {J. Chem. Phys.}\ }\textbf {\bibinfo {volume} {76}},\
  \bibinfo {pages} {3144} (\bibinfo {year} {1982})}\BibitemShut {NoStop}%
\bibitem [{\citenamefont {Werner}(1987)}]{MRCI5}%
  \BibitemOpen
  \bibfield  {author} {\bibinfo {author} {\bibfnamefont {H.-J.}\ \bibnamefont
  {Werner}},\ }\href {\doibase 10.1002/9780470142943.ch1} {\bibfield  {journal}
  {\bibinfo  {journal} {Adv. Chem. Phys.}\ }\textbf {\bibinfo {volume} {69}},\
  \bibinfo {pages} {1} (\bibinfo {year} {1987})}\BibitemShut {NoStop}%
\bibitem [{\citenamefont {Werner}\ \emph {et~al.}(2018)\citenamefont {Werner},
  \citenamefont {Knowles}, \citenamefont {Knizia}, \citenamefont {Manby},
  \citenamefont {{Sch\"{u}tz}} \emph {et~al.}}]{MOLPRO_brief}%
  \BibitemOpen
  \bibfield  {author} {\bibinfo {author} {\bibfnamefont {H.-J.}\ \bibnamefont
  {Werner}}, \bibinfo {author} {\bibfnamefont {P.~J.}\ \bibnamefont {Knowles}},
  \bibinfo {author} {\bibfnamefont {G.}~\bibnamefont {Knizia}}, \bibinfo
  {author} {\bibfnamefont {F.~R.}\ \bibnamefont {Manby}}, \bibinfo {author}
  {\bibfnamefont {M.}~\bibnamefont {{Sch\"{u}tz}}},  \emph {et~al.},\
  }\href@noop {} {\enquote {\bibinfo {title} {Molpro, version 2018.2, a package
  of ab initio programs},}\ } (\bibinfo {year} {2018}),\ \bibinfo {note} {see
  http://www.molpro.net}\BibitemShut {NoStop}%
\bibitem [{\citenamefont {Jeung}(1997)}]{CoreCoreRepulsionDiatomics}%
  \BibitemOpen
  \bibfield  {author} {\bibinfo {author} {\bibfnamefont {G.-H.}\ \bibnamefont
  {Jeung}},\ }\href {\doibase https://doi.org/10.1006/jmsp.1996.7209}
  {\bibfield  {journal} {\bibinfo  {journal} {J. Mol. Spectrosc.}\ }\textbf
  {\bibinfo {volume} {182}},\ \bibinfo {pages} {113} (\bibinfo {year}
  {1997})}\BibitemShut {NoStop}%
\bibitem [{\citenamefont {Gu{\'e}rout}\ \emph {et~al.}(2009)\citenamefont
  {Gu{\'e}rout}, \citenamefont {Sold{\'a}n}, \citenamefont {Aymar},
  \citenamefont {Deiglmayr},\ and\ \citenamefont
  {Dulieu}}]{SoldanCoreRepulsion}%
  \BibitemOpen
  \bibfield  {author} {\bibinfo {author} {\bibfnamefont {R.}~\bibnamefont
  {Gu{\'e}rout}}, \bibinfo {author} {\bibfnamefont {P.}~\bibnamefont
  {Sold{\'a}n}}, \bibinfo {author} {\bibfnamefont {M.}~\bibnamefont {Aymar}},
  \bibinfo {author} {\bibfnamefont {J.}~\bibnamefont {Deiglmayr}}, \ and\
  \bibinfo {author} {\bibfnamefont {O.}~\bibnamefont {Dulieu}},\ }\href
  {\doibase 10.1002/qua.22304} {\bibfield  {journal} {\bibinfo  {journal} {Int.
  J. Quant. Chem.}\ }\textbf {\bibinfo {volume} {109}},\ \bibinfo {pages}
  {3387} (\bibinfo {year} {2009})}\BibitemShut {NoStop}%
\bibitem [{\citenamefont {Suno}\ \emph {et~al.}(2002)\citenamefont {Suno},
  \citenamefont {Esry}, \citenamefont {Greene},\ and\ \citenamefont
  {Burke}}]{PhysRevA.65.042725}%
  \BibitemOpen
  \bibfield  {author} {\bibinfo {author} {\bibfnamefont {H.}~\bibnamefont
  {Suno}}, \bibinfo {author} {\bibfnamefont {B.~D.}\ \bibnamefont {Esry}},
  \bibinfo {author} {\bibfnamefont {C.~H.}\ \bibnamefont {Greene}}, \ and\
  \bibinfo {author} {\bibfnamefont {J.~P.}\ \bibnamefont {Burke}},\ }\href
  {\doibase 10.1103/PhysRevA.65.042725} {\bibfield  {journal} {\bibinfo
  {journal} {Phys. Rev. A}\ }\textbf {\bibinfo {volume} {65}},\ \bibinfo
  {pages} {042725} (\bibinfo {year} {2002})}\BibitemShut {NoStop}%
\bibitem [{\citenamefont {Coolidge}\ and\ \citenamefont
  {James}(1937)}]{PhysRev.51.855}%
  \BibitemOpen
  \bibfield  {author} {\bibinfo {author} {\bibfnamefont {A.~S.}\ \bibnamefont
  {Coolidge}}\ and\ \bibinfo {author} {\bibfnamefont {H.~M.}\ \bibnamefont
  {James}},\ }\href {\doibase 10.1103/PhysRev.51.855} {\bibfield  {journal}
  {\bibinfo  {journal} {Phys. Rev.}\ }\textbf {\bibinfo {volume} {51}},\
  \bibinfo {pages} {855} (\bibinfo {year} {1937})}\BibitemShut {NoStop}%
\bibitem [{\citenamefont {Pekeris}(1958)}]{PhysRev.112.1649}%
  \BibitemOpen
  \bibfield  {author} {\bibinfo {author} {\bibfnamefont {C.~L.}\ \bibnamefont
  {Pekeris}},\ }\href {\doibase 10.1103/PhysRev.112.1649} {\bibfield  {journal}
  {\bibinfo  {journal} {Phys. Rev.}\ }\textbf {\bibinfo {volume} {112}},\
  \bibinfo {pages} {1649} (\bibinfo {year} {1958})}\BibitemShut {NoStop}%
\bibitem [{\citenamefont {Pekeris}(1959)}]{PhysRev.115.1216}%
  \BibitemOpen
  \bibfield  {author} {\bibinfo {author} {\bibfnamefont {C.~L.}\ \bibnamefont
  {Pekeris}},\ }\href {\doibase 10.1103/PhysRev.115.1216} {\bibfield  {journal}
  {\bibinfo  {journal} {Phys. Rev.}\ }\textbf {\bibinfo {volume} {115}},\
  \bibinfo {pages} {1216} (\bibinfo {year} {1959})}\BibitemShut {NoStop}%
\bibitem [{\citenamefont {Pekeris}\ \emph {et~al.}(1962)\citenamefont
  {Pekeris}, \citenamefont {Schiff},\ and\ \citenamefont
  {Lifson}}]{PhysRev.126.1057}%
  \BibitemOpen
  \bibfield  {author} {\bibinfo {author} {\bibfnamefont {C.~L.}\ \bibnamefont
  {Pekeris}}, \bibinfo {author} {\bibfnamefont {B.}~\bibnamefont {Schiff}}, \
  and\ \bibinfo {author} {\bibfnamefont {H.}~\bibnamefont {Lifson}},\ }\href
  {\doibase 10.1103/PhysRev.126.1057} {\bibfield  {journal} {\bibinfo
  {journal} {Phys. Rev.}\ }\textbf {\bibinfo {volume} {126}},\ \bibinfo {pages}
  {1057} (\bibinfo {year} {1962})}\BibitemShut {NoStop}%
\bibitem [{\citenamefont {Pekeris}(1962)}]{PhysRev.127.509}%
  \BibitemOpen
  \bibfield  {author} {\bibinfo {author} {\bibfnamefont {C.~L.}\ \bibnamefont
  {Pekeris}},\ }\href {\doibase 10.1103/PhysRev.127.509} {\bibfield  {journal}
  {\bibinfo  {journal} {Phys. Rev.}\ }\textbf {\bibinfo {volume} {127}},\
  \bibinfo {pages} {509} (\bibinfo {year} {1962})}\BibitemShut {NoStop}%
\bibitem [{\citenamefont {Accad}\ \emph {et~al.}(1971)\citenamefont {Accad},
  \citenamefont {Pekeris},\ and\ \citenamefont {Schiff}}]{PhysRevA.4.516}%
  \BibitemOpen
  \bibfield  {author} {\bibinfo {author} {\bibfnamefont {Y.}~\bibnamefont
  {Accad}}, \bibinfo {author} {\bibfnamefont {C.~L.}\ \bibnamefont {Pekeris}},
  \ and\ \bibinfo {author} {\bibfnamefont {B.}~\bibnamefont {Schiff}},\ }\href
  {\doibase 10.1103/PhysRevA.4.516} {\bibfield  {journal} {\bibinfo  {journal}
  {Phys. Rev. A}\ }\textbf {\bibinfo {volume} {4}},\ \bibinfo {pages} {516}
  (\bibinfo {year} {1971})}\BibitemShut {NoStop}%
\bibitem [{\citenamefont {Schiff}\ \emph {et~al.}(1971)\citenamefont {Schiff},
  \citenamefont {Pekeris},\ and\ \citenamefont {Accad}}]{PhysRevA.4.885}%
  \BibitemOpen
  \bibfield  {author} {\bibinfo {author} {\bibfnamefont {B.}~\bibnamefont
  {Schiff}}, \bibinfo {author} {\bibfnamefont {C.~L.}\ \bibnamefont {Pekeris}},
  \ and\ \bibinfo {author} {\bibfnamefont {Y.}~\bibnamefont {Accad}},\ }\href
  {\doibase 10.1103/PhysRevA.4.885} {\bibfield  {journal} {\bibinfo  {journal}
  {Phys. Rev. A}\ }\textbf {\bibinfo {volume} {4}},\ \bibinfo {pages} {885}
  (\bibinfo {year} {1971})}\BibitemShut {NoStop}%
\bibitem [{\citenamefont {Davidson}(1977)}]{Davidson1977}%
  \BibitemOpen
  \bibfield  {author} {\bibinfo {author} {\bibfnamefont {E.~R.}\ \bibnamefont
  {Davidson}},\ }\href {\doibase 10.1021/ja00444a015} {\bibfield  {journal}
  {\bibinfo  {journal} {J. Am. Chem. Soc.}\ }\textbf {\bibinfo {volume} {99
  (2)}},\ \bibinfo {pages} {397} (\bibinfo {year} {1977})}\BibitemShut
  {NoStop}%
\bibitem [{\citenamefont {Bersuker}(2001)}]{BersukerReview}%
  \BibitemOpen
  \bibfield  {author} {\bibinfo {author} {\bibfnamefont {I.~B.}\ \bibnamefont
  {Bersuker}},\ }\href {\doibase 10.1021/cr0004411} {\bibfield  {journal}
  {\bibinfo  {journal} {Chem. Rev.}\ }\textbf {\bibinfo {volume} {101}},\
  \bibinfo {pages} {1067} (\bibinfo {year} {2001})}\BibitemShut {NoStop}%
\bibitem [{\citenamefont {Rocha}\ and\ \citenamefont
  {Varandas}(2016)}]{RochaC3}%
  \BibitemOpen
  \bibfield  {author} {\bibinfo {author} {\bibfnamefont {C.~M.~R.}\
  \bibnamefont {Rocha}}\ and\ \bibinfo {author} {\bibfnamefont {A.~J.~C.}\
  \bibnamefont {Varandas}},\ }\href {\doibase 10.1063/1.4941382} {\bibfield
  {journal} {\bibinfo  {journal} {J. Chem. Phys.}\ }\textbf {\bibinfo {volume}
  {144}},\ \bibinfo {pages} {064309} (\bibinfo {year} {2016})}\BibitemShut
  {NoStop}%
\bibitem [{\citenamefont {Domcke}\ \emph {et~al.}(2006)\citenamefont {Domcke},
  \citenamefont {Mishra},\ and\ \citenamefont {Poluyanov}}]{RelExEJT}%
  \BibitemOpen
  \bibfield  {author} {\bibinfo {author} {\bibfnamefont {W.}~\bibnamefont
  {Domcke}}, \bibinfo {author} {\bibfnamefont {S.}~\bibnamefont {Mishra}}, \
  and\ \bibinfo {author} {\bibfnamefont {L.~V.}\ \bibnamefont {Poluyanov}},\
  }\href {\doibase https://doi.org/10.1016/j.chemphys.2005.09.009} {\bibfield
  {journal} {\bibinfo  {journal} {Chem. Phys.}\ }\textbf {\bibinfo {volume}
  {322}},\ \bibinfo {pages} {405 } (\bibinfo {year} {2006})}\BibitemShut
  {NoStop}%
\bibitem [{\citenamefont {Thapaliya}\ \emph {et~al.}(2015)\citenamefont
  {Thapaliya}, \citenamefont {Dawadi}, \citenamefont {Ziegler},\ and\
  \citenamefont {Perry}}]{VibJTExe}%
  \BibitemOpen
  \bibfield  {author} {\bibinfo {author} {\bibfnamefont {B.~P.}\ \bibnamefont
  {Thapaliya}}, \bibinfo {author} {\bibfnamefont {M.~B.}\ \bibnamefont
  {Dawadi}}, \bibinfo {author} {\bibfnamefont {C.}~\bibnamefont {Ziegler}}, \
  and\ \bibinfo {author} {\bibfnamefont {D.~S.}\ \bibnamefont {Perry}},\ }\href
  {\doibase https://doi.org/10.1016/j.chemphys.2015.07.017} {\bibfield
  {journal} {\bibinfo  {journal} {Chem. Phys.}\ }\textbf {\bibinfo {volume}
  {460}},\ \bibinfo {pages} {31 } (\bibinfo {year} {2015})}\BibitemShut
  {NoStop}%
\bibitem [{\citenamefont {Barckholtz}\ and\ \citenamefont
  {Miller}(1998)}]{relJTReview}%
  \BibitemOpen
  \bibfield  {author} {\bibinfo {author} {\bibfnamefont {T.~A.}\ \bibnamefont
  {Barckholtz}}\ and\ \bibinfo {author} {\bibfnamefont {T.~A.}\ \bibnamefont
  {Miller}},\ }\href {\doibase 10.1080/014423598230036} {\bibfield  {journal}
  {\bibinfo  {journal} {Int. Rev. in Phys. Chem.}\ }\textbf {\bibinfo {volume}
  {17}},\ \bibinfo {pages} {435} (\bibinfo {year} {1998})}\BibitemShut
  {NoStop}%
\bibitem [{\citenamefont {Poluyanov}\ and\ \citenamefont
  {Domcke}(2008)}]{POLUYANOV2008125}%
  \BibitemOpen
  \bibfield  {author} {\bibinfo {author} {\bibfnamefont {L.~V.}\ \bibnamefont
  {Poluyanov}}\ and\ \bibinfo {author} {\bibfnamefont {W.}~\bibnamefont
  {Domcke}},\ }\href {\doibase https://doi.org/10.1016/j.chemphys.2008.05.020}
  {\bibfield  {journal} {\bibinfo  {journal} {Chem. Phys.}\ }\textbf {\bibinfo
  {volume} {352}},\ \bibinfo {pages} {125 } (\bibinfo {year}
  {2008})}\BibitemShut {NoStop}%
\bibitem [{\citenamefont {Liu}\ \emph {et~al.}(2010)\citenamefont {Liu},
  \citenamefont {Bersuker}, \citenamefont {Zou},\ and\ \citenamefont
  {Boggs}}]{PJTEvsRTE}%
  \BibitemOpen
  \bibfield  {author} {\bibinfo {author} {\bibfnamefont {Y.}~\bibnamefont
  {Liu}}, \bibinfo {author} {\bibfnamefont {I.~B.}\ \bibnamefont {Bersuker}},
  \bibinfo {author} {\bibfnamefont {W.}~\bibnamefont {Zou}}, \ and\ \bibinfo
  {author} {\bibfnamefont {J.~E.}\ \bibnamefont {Boggs}},\ }\href {\doibase
  https://doi.org/10.1016/j.chemphys.2010.07.029} {\bibfield  {journal}
  {\bibinfo  {journal} {Chem. Phys.}\ }\textbf {\bibinfo {volume} {376}},\
  \bibinfo {pages} {30 } (\bibinfo {year} {2010})}\BibitemShut {NoStop}%
\bibitem [{\citenamefont {Bersuker}(2013)}]{PJTEReview}%
  \BibitemOpen
  \bibfield  {author} {\bibinfo {author} {\bibfnamefont {I.~B.}\ \bibnamefont
  {Bersuker}},\ }\href {\doibase 10.1021/cr300279n} {\bibfield  {journal}
  {\bibinfo  {journal} {Chem. Rev.}\ }\textbf {\bibinfo {volume} {113}},\
  \bibinfo {pages} {1351} (\bibinfo {year} {2013})}\BibitemShut {NoStop}%
\bibitem [{\citenamefont {Wolf}\ \emph {et~al.}(2017)\citenamefont {Wolf},
  \citenamefont {Dei{\ss}}, \citenamefont {Kr{\"u}kow}, \citenamefont
  {Tiemann}, \citenamefont {Ruzic}, \citenamefont {Wang}, \citenamefont
  {D{\textquoteright}Incao}, \citenamefont {Julienne},\ and\ \citenamefont
  {Denschlag}}]{Wolf921}%
  \BibitemOpen
  \bibfield  {author} {\bibinfo {author} {\bibfnamefont {J.}~\bibnamefont
  {Wolf}}, \bibinfo {author} {\bibfnamefont {M.}~\bibnamefont {Dei{\ss}}},
  \bibinfo {author} {\bibfnamefont {A.}~\bibnamefont {Kr{\"u}kow}}, \bibinfo
  {author} {\bibfnamefont {E.}~\bibnamefont {Tiemann}}, \bibinfo {author}
  {\bibfnamefont {B.~P.}\ \bibnamefont {Ruzic}}, \bibinfo {author}
  {\bibfnamefont {Y.}~\bibnamefont {Wang}}, \bibinfo {author} {\bibfnamefont
  {J.~P.}\ \bibnamefont {D{\textquoteright}Incao}}, \bibinfo {author}
  {\bibfnamefont {P.~S.}\ \bibnamefont {Julienne}}, \ and\ \bibinfo {author}
  {\bibfnamefont {J.~H.}\ \bibnamefont {Denschlag}},\ }\href {\doibase
  10.1126/science.aan8721} {\bibfield  {journal} {\bibinfo  {journal}
  {Science}\ }\textbf {\bibinfo {volume} {358}},\ \bibinfo {pages} {921}
  (\bibinfo {year} {2017})}\BibitemShut {NoStop}%
\bibitem [{\citenamefont {Lim}\ \emph {et~al.}(2005)\citenamefont {Lim},
  \citenamefont {Schwerdtfeger}, \citenamefont {Metz},\ and\ \citenamefont
  {Stoll}}]{SmallECPStoll}%
  \BibitemOpen
  \bibfield  {author} {\bibinfo {author} {\bibfnamefont {I.~S.}\ \bibnamefont
  {Lim}}, \bibinfo {author} {\bibfnamefont {P.}~\bibnamefont {Schwerdtfeger}},
  \bibinfo {author} {\bibfnamefont {B.}~\bibnamefont {Metz}}, \ and\ \bibinfo
  {author} {\bibfnamefont {H.}~\bibnamefont {Stoll}},\ }\href {\doibase
  10.1063/1.1856451} {\bibfield  {journal} {\bibinfo  {journal} {J. Chem.
  Phys.}\ }\textbf {\bibinfo {volume} {122}},\ \bibinfo {pages} {104103}
  (\bibinfo {year} {2005})}\BibitemShut {NoStop}%
\bibitem [{\citenamefont {Dolg}(2000)}]{ECPsDolg}%
  \BibitemOpen
  \bibfield  {author} {\bibinfo {author} {\bibfnamefont {M.}~\bibnamefont
  {Dolg}},\ }\enquote {\bibinfo {title} {Effective core potentials},}\ in\
  \href {https://core.ac.uk/download/pdf/35009884.pdf} {\emph {\bibinfo
  {booktitle} {Modern Methods and Algorithms of Quantum Chemistry}}},\ \bibinfo
  {editor} {edited by\ \bibinfo {editor} {\bibfnamefont {J.}~\bibnamefont
  {Grotendorst}}}\ (\bibinfo  {publisher} {NIC-Directors},\ \bibinfo {address}
  {Jülich},\ \bibinfo {year} {2000})\ pp.\ \bibinfo {pages}
  {507--540}\BibitemShut {NoStop}%
\bibitem [{\citenamefont {Matsika}\ and\ \citenamefont
  {Yarkony}(2001{\natexlab{a}})}]{YarkoniCOINSOC1}%
  \BibitemOpen
  \bibfield  {author} {\bibinfo {author} {\bibfnamefont {S.}~\bibnamefont
  {Matsika}}\ and\ \bibinfo {author} {\bibfnamefont {D.~R.}\ \bibnamefont
  {Yarkony}},\ }\href {\doibase 10.1063/1.1378324} {\bibfield  {journal}
  {\bibinfo  {journal} {J. Chem. Phys.}\ }\textbf {\bibinfo {volume} {115}},\
  \bibinfo {pages} {2038} (\bibinfo {year} {2001}{\natexlab{a}})}\BibitemShut
  {NoStop}%
\bibitem [{\citenamefont {Matsika}\ and\ \citenamefont
  {Yarkony}(2001{\natexlab{b}})}]{YarkoniCOINSOC2}%
  \BibitemOpen
  \bibfield  {author} {\bibinfo {author} {\bibfnamefont {S.}~\bibnamefont
  {Matsika}}\ and\ \bibinfo {author} {\bibfnamefont {D.~R.}\ \bibnamefont
  {Yarkony}},\ }\href {\doibase 10.1063/1.1391444} {\bibfield  {journal}
  {\bibinfo  {journal} {J. Chem. Phys.}\ }\textbf {\bibinfo {volume} {115}},\
  \bibinfo {pages} {5066} (\bibinfo {year} {2001}{\natexlab{b}})}\BibitemShut
  {NoStop}%
\bibitem [{\citenamefont {Matsika}\ and\ \citenamefont
  {Yarkony}(2002)}]{YarkoniCOINSOC3}%
  \BibitemOpen
  \bibfield  {author} {\bibinfo {author} {\bibfnamefont {S.}~\bibnamefont
  {Matsika}}\ and\ \bibinfo {author} {\bibfnamefont {D.~R.}\ \bibnamefont
  {Yarkony}},\ }\href {\doibase 10.1063/1.1427914} {\bibfield  {journal}
  {\bibinfo  {journal} {J. Chem. Phys.}\ }\textbf {\bibinfo {volume} {116}},\
  \bibinfo {pages} {2825} (\bibinfo {year} {2002})}\BibitemShut {NoStop}%
\end{thebibliography}
\providecommand{\noopsort}[1]{}\providecommand{\singleletter}[1]{#1}%

\end{document}